\documentclass[journal,onecolumn]{IEEEtran}
\usepackage{graphicx}
\usepackage{float}
\usepackage{amsmath}
\usepackage{amssymb}
\usepackage{tabularx}
\usepackage{array}

\usepackage{url}

\AtBeginEnvironment{acronym}{%
  }
\usepackage[nolist]{acronym}

\usepackage[justification=centering]{caption}
\usepackage{hyperref}
\usepackage{xcolor}
\usepackage{booktabs}
\usepackage{multirow}

\begin{document}

\title{Technical Report on Resilient and Secure Large-Scale Energy Internet Systems}
\vspace{2cm}

\author{
        Ioannis~Zografopoulos$^{1}$,
        Karen~Largman$^{1}$,
        Isaac~Ortega~Romero$^{1}$,
        S~M~Zia~Ur~Rashid$^{2}$,
        Yexiang~Chen$^{3}$,
        George~Fragkos$^{4}$,
        Charalambos~Konstantinou$^{5}$,
        Subhash~Lakshminarayana$^{3}$,
        Juan~Ospina$^{6}$,
        Airin~Rahman$^{7}$,
        Suman~Rath$^{2}$,
        Vivek~Kumar~Singh$^{8}$,
        Mucun~Sun$^{9}$,
        and~Wei~Sun$^{7}$
\thanks{$^{1}$University of Massachusetts Boston, Boston, MA, USA.}
\thanks{$^{2}$The University of Tulsa, Tulsa, OK, USA.}
\thanks{$^{3}$University of Warwick, Coventry, UK.}
\thanks{$^{4}$Sandia National Laboratories, Albuquerque, NM, USA.}
\thanks{$^{5}$King Abdullah University of Science and Technology (KAUST), Thuwal, KSA.}
\thanks{$^{6}$Orennia, USA.}
\thanks{$^{7}$University of Central Florida, Orlando, FL, USA.}
\thanks{$^{8}$National Laboratory of the Rockies, USA.}
\thanks{$^{9}$Idaho National Laboratory, Idaho Falls, ID, USA.\\} 

\thanks{This Technical Report draft was prepared by the \href{https://cmte.ieee.org/pes-rsei/}{Task Force on Resilient and Secure Large-Scale Energy Internet Systems}.}%
\thanks{Preprint prepared August, 2026.}
}

\markboth{Technical Report on Resilient and Secure Large-Scale Energy Internet Systems, August~2026}%
{This Technical Report draft was prepared by the Task Force on Resilient and Secure Large-Scale Energy Internet Systems}

\maketitle

\begin{abstract}
This IEEE PES Task Force report examines the security and resilience of large-scale Energy Internet (EI) systems, in which electricity, information, and market layers are tightly coupled through pervasive digitalization. The report characterizes the EI cyber-physical threat landscape and surveys detection, assurance, and mitigation techniques; presents modeling, control, and decision-making frameworks that capture cyber-physical interdependencies, including storage integration, multi-dimensional resilience, and electricity price forecasting; examines adversarial risks and trustworthy deployment of artificial intelligence; and introduces graph-based, attack-resilient information routing. The report closes with recommendations for research, standardization, and regulatory efforts needed to realize a resilient and secure large-scale EI.
\end{abstract}
\vspace{0.5cm}

\vspace{0.5cm}

\begin{center}
\small
    {\textbf{{Acknowledgements}}}\\
\end{center}
\small{

\indent \indent The Task Force gratefully acknowledges the effort of our Task Force Meetings participants, who contributed through their discussions 
\indent \indent and hard work in compiling this report. 
}

\newpage
\tableofcontents
\newpage
\listoffigures
\listoftables
\newpage

\IEEEpeerreviewmaketitle

\begin{acronym}[CPES-QSM] 
\acro{5G}{Fifth Generation}
\acro{A-FDIA}{Adversarial False Data Injection Attack}
\acro{ACOPF}{AC Optimal Power Flow}
\acro{ADMS}{Advanced Distribution Management Systems}
\acro{AGC}{Automatic Generation Control}
\acro{AI}{Artificial Intelligence}
\acro{AMI}{Advanced Metering Infrastructure}
\acro{AOC}{Avoided Outage Cost}
\acro{API}{Application Programming Interface}
\acro{BDD}{Bad Data Detection}
\acro{CAI}{Change of Attack Intensity}
\acro{CESER}{Cybersecurity, Energy Security, and Emergency Response}
\acro{CGMES}{Common Grid Model Exchange Standard}
\acro{CHIL}{Controller Hardware-in-the-Loop}
\acro{CIM}{Common Information Model}
\acro{CIP}{Critical Infrastructure Protection}
\acro{CNN}{Convolutional Neural Network}
\acro{CNT}{Complex Network Theory}
\acro{CPES-QSM}{Cyber-Physical Energy System Quantitative Security Metric}
\acro{CPPS}{Cyber-Physical Power System}
\acro{CPS}{Cyber-Physical System}
\acro{CRR}{Congestion Revenue Right}
\acro{CW}{Carlini \& Wagner}
\acro{DERs}{Distributed Energy Resources}
\acro{DIG}{Decision Integrity Gap}
\acro{DLMP}{Distributed Locational Marginal Price}
\acro{DL}{Deep Learning}
\acro{DM}{Delivery and Management}
\acro{DNN}{Deep Neural Network}
\acro{DoS}{Denial-of-Service}
\acro{DOE}{Department of Energy}
\acro{DRL}{Deep Reinforcement Learning}
\acro{DSE}{Dynamic State Estimation}
\acro{DSO}{Distribution System Operator}
\acro{DT}{Digital Twins}
\acro{EI}{Energy Internet}
\acro{EMS}{Energy Management System}
\acro{EMT}{Electromagnetic Transient}
\acro{ENTSO-E}{European Network of Transmission System Operators for Electricity}
\acro{EPF}{Electricity Price Forecasting}
\acro{EPS}{Electric Power Systems}
\acro{ESDAR}{Energy-State-Driven Adaptive Restoration}
\acro{EUE}{Expected Unserved Energy}
\acro{EV}{Electric Vehicles}
\acro{FACTS}{Flexible Alternating Current Transmission Systems}
\acro{FDIA}{False Data Injection Attack}
\acro{FERC}{Federal Energy Regulatory Commission}
\acro{FGSM}{Fast Gradient Sign Method}
\acro{FNCS}{Framework for Network Co-simulation}
\acro{FREE}{Future Renewable Electric Energy}
\acro{FREEDM}{Future Renewable Electric Energy Delivery and Management}
\acro{FTR}{Financial Transmission Right}
\acro{GAT}{Graph Attention Network}
\acro{GOOSE}{Generic Object Oriented Substation Event}
\acro{GNN}{Graph Neural Network}
\acro{GW}{Gigawatt}
\acro{HELICS}{Hierarchical Engine for Large-scale Infrastructure Co-simulation}
\acro{HILP}{High-Impact, Low-Probability}
\acro{HIL}{Hardware-in-the-Loop}
\acro{HPC}{High Performance Computing}
\acro{HV}{High-Voltage}
\acro{HVDC}{High-Voltage Direct Current}
\acro{IBR}{Inverter-based Resources}
\acro{ICS}{Industrial Control System}
\acro{ICT}{Information and Communication Technology}
\acro{IEC}{International Electrotechnical Commission}
\acro{IEEE}{Institute of Electrical and Electronics Engineers}
\acro{IED}{Intelligent Electronic Device}
\acro{IESO}{Independent Electricity System Operator}
\acro{IoT}{Internet of Things}
\acro{IP}{Internet Protocol}
\acro{ISO}{Independent System Operator}
\acro{IT}{Information Technology}
\acro{KNN}{K-nearest Neighbor}
\acro{LAA}{Load-Altering Attack}
\acro{LDES}{Long-Duration Energy Storage}
\acro{LFC}{Load Frequency Control}
\acro{LLM}{Large Language Model}
\acro{LMP}{Locational Marginal Price}
\acro{LSTM}{Long Short-Term Memory}
\acro{MDP}{Markov Decision Process}
\acro{MG}{Microgrid}
\acro{MITM}{Man-in-the-Middle}
\acro{ML}{Machine Learning}
\acro{MLP}{Multilayer Perceptron}
\acro{MPS}{Mobile Power Resources}
\acro{MST}{Minimum Spanning Tree}
\acro{MTD}{Moving Target Defenses}
\acro{MV/LV}{Medium-to-low Voltage}
\acro{NASPI}{North American Synchrophasor Initiative}
\acro{NERC}{North American Electric Reliability Corporation}
\acro{NIST}{National Institute of Standards and Technology}
\acro{NNSA}{National Nuclear Security Administration}
\acro{OPF}{Optimal Power Flow}
\acro{OSI}{Open Systems Interconnection}
\acro{OT}{Operational Technology}
\acro{PCA}{Principal Component Analysis}
\acro{PCC}{Point of Common Coupling}
\acro{PEPS}{Power Electronic-dominated Systems}
\acro{PF}{Power Flow}
\acro{PHIL}{Power Hardware-in-the-loop}
\acro{PLL}{Phase-Locked Loop}
\acro{PLC}{Power Line Communication}
\acro{PMU}{Phasor Measurement Units}
\acro{PNNL}{Pacific Northwest National Laboratory}
\acro{PINN}{Physics-Informed Neural Network}
\acro{PSS}{Power System Stabilizer}
\acro{PV}{Photovoltaic}
\acro{QoS}{Quality of Service}
\acro{RI}{Resilience Index}
\acro{RL}{Reinforcement Learning}
\acro{RO}{Robust Optimization}
\acro{RTO}{Regional Transmission Organization}
\acro{SAI}{Storage Adequacy Index}
\acro{SBOM}{Software Bill of Materials}
\acro{SCADA}{Supervisory Control and Data Acquisition}
\acro{SCED}{Security-Constrained Economic Dispatch}
\acro{SCOPF}{Security-Constrained Optimal Power Flow}
\acro{SCUC}{Security-Constrained Unit Commitment}
\acro{SDN}{Software-defined Network}
\acro{SE}{State Estimation}
\acro{SGAM}{Smart Grid Architecture Model}
\acro{SHAP}{SHapley Additive exPlanations}
\acro{SoC}{State of Charge}
\acro{SSE}{Static State Estimation}
\acro{SSIM}{Structural Similarity Index Measure}
\acro{SSSR}{Steady-State Security-Region}
\acro{STATCOM}{Static Synchronous Compensator}
\acro{SVM}{Support Vector Machine}
\acro{SV}{Sample Values}
\acro{t-SNE}{t-distributed Stochastic Neighbor Embedding}
\acro{TCN}{Temporal Convolutional Network}
\acro{TEE}{Trusted Execution Environment}
\acro{TF}{Task Force}
\acro{TnD}{Transmission and Distribution}
\acro{TTP}{Tactics, Techniques, and Procedures}
\acro{TSO}{Transmission System Operator}
\acro{VPP}{Virtual Power Plant}
\acro{WAMPAC}{Wide-area Monitoring, Protection, and Control}
\acro{WAMS}{Wide-area Measurement Systems}
\acro{WADC}{Wide-area Damping Control}
\acro{WAN}{Wide-area Network}
\end{acronym}

\section{Task Force Scope} \label{s:Scope}
%
%
%
%

\IEEEPARstart{T}{}his \ac{TF} report aims to analyze the application of the \ac{EI} paradigm in studying and modeling the security and resilience of modern energy systems by fostering collaboration among engineers, scientists, industry personnel, system operators, and regulators. The goal is not only to reduce the likelihood of adverse events, but also to limit their scope and impact, drawing lessons to improve future responses. In light of how Internet integration is reshaping energy networks through the adoption of new Internet and \ac{IT} techniques, as well as emerging energy policies and markets, the report’s objective is to investigate the behavior of interconnected multi-energy systems. To this end, the report examines the landscape of emerging technologies and studies that can address disruptive cyber-physical events, leverage lessons learned from the Internet vertical and apply this knowledge to define methods for enhancing \ac{EI} security and resilience.
Beyond establishing this overarching vision, the \ac{TF} report covers a broad spectrum of interrelated topics that collectively span the cyber, physical, and decision-making layers of large-scale \ac{EI} systems. These include the cybersecurity threat landscape and assurance of \ac{EI} deployments, the modeling and control of \ac{TnD} systems through \ac{EI} architectures, and resilience considerations across multiple dimensions, including storage-integrated and decision-aware operation. The report further addresses the market and forecasting implications of the \ac{EI} era, the opportunities and adversarial risks introduced by the integration of \ac{AI} into energy systems, and the algorithmic foundations, such as attack-resilient information routing, that underpin trustworthy \ac{EI} operation. Each of these topics is developed in detail in the corresponding chapters of this report, as outlined in the following subsection.
\section{Background and Motivation}
\subsection{The Energy Internet Definition}
The \ac{EI} is formally defined as ``a next-generation energy infrastructure that integrates
advanced power electronics, \ac{ICT}, and intelligent
management to interconnect a large number of \ac{DERs}, energy storage
systems, and flexible loads, enabling bidirectional flows of both energy and information in a
manner analogous to the routing of data packets on the Internet'' \cite{ref1,ref2}. First popularized through
Rifkin's vision of a ``Third Industrial Revolution'' \cite{ref1} and given a concrete engineering realization
in the \ac{FREE} \ac{DM} system \cite{ref2}, the
\ac{EI} is architected around key enabling elements: energy routers and solid-state transformers that
manage bidirectional power and information exchange, plug-and-play interfaces for \ac{DERs} and
storage, transactive and peer-to-peer energy markets, and distributed intelligence for autonomous
coordination and control \cite{ref3}.

Conceptually, the \ac{EI} is described through a multi-layer model comprising: i) a physical
(energy) layer, encompassing generation, storage, conversion, and delivery assets across
interconnected multi-energy carriers, ii) a cyber (information) layer, responsible for sensing,
communication, and control data exchange among millions of grid-edge devices, and iii) a business
(market and decision) layer, where energy transactions, pricing, and policy mechanisms operate
\cite{ref3,ref4}. The tight vertical coupling among these layers and the horizontal coupling across energy
cells, microgrids, and interconnected \ac{TnD} systems, are precisely what distinguish the \ac{EI} from the
traditional, hierarchically operated power grid, and what motivates the security and resilience
investigation undertaken by this \ac{TF}.

\subsection{Motivation}
\ac{EPS} form the foundation of critical infrastructure, with national
security and economic stability heavily dependent on their safe, secure, and resilient operation. As
the most complex machine ever constructed, the electric grid has become a prime target for cyber-threats \cite{ref5,ref6}. Its numerous vulnerabilities mean it can no longer reliably provide cyber-secure and
disaster-resilient power to businesses and households, posing a significant and urgent threat to
society and the economy. Real-world incidents, most notably the coordinated attacks against the
Ukrainian power grid in 2015 and 2016 \cite{ref7}, as well as demonstrated attack vectors targeting \ac{DERs},
inverter-based microgrids, and their supporting control and communication infrastructure
\cite{ref8,ref9,ref10} underscore that adversaries can weaponize the cyber layer to induce physical
consequences at scale.

The power system community has recently devoted increasing attention to the concepts of
security and resilience, their definitions, and their application and interdependence in the context
of power grid operation and planning \cite{ref1,ref11}. However, with the advent of the \ac{EI}, there is a
pressing need for deeper exploration. Specifically, the challenge is to re-examine, analyze, and
understand the security and resilience of the newly formed, internet-dependent, and highly
distributed power systems. This includes investigating how these critical concepts are addressed
within the multi-layer model of the \ac{EI}, considering the complexity and interconnections of modern
energy systems. Resilience, in particular, can no longer be treated as a single-dimensional attribute:
recent studies demonstrate that physical, operational, and digital-cyber dimensions of resilience
interact nonlinearly, and that assessments confined to any one dimension fail to capture the
degradation produced by simultaneous cross-dimensional failures \cite{ref4}.

The security and resilience of the \ac{EI} pose significant challenges in controlling and
maintaining access to critical system resources and services at the physical layer, as well as ensuring
the confidentiality, availability, integrity, and non-repudiation of information exchanged at the
cyber and business layers \cite{ref4,ref12}. Recent advancements in artificial intelligence, \ac{5G}
communication \cite{islam2023resource}, blockchain technologies, modern transport-layer security mechanisms for \ac{IP}-based industrial protocols \cite{ref13}, industrial and consumer \ac{IoT} \cite{zografopoulos2022time}, transportation electrification, and the
high penetration of \ac{DERs} \cite{ref6,ref10} necessitate the creation of new approaches, mechanisms, toolsets,
and best practices \cite{liu2021faster}. These innovations are essential to comprehending the multi-layer interaction of
commands and data and to better characterizing the flexibility and adaptation of the \ac{EI} in response
to high-impact security and resilience events, for example, by leveraging \ac{DERs} and storage
flexibility to mitigate ongoing attacks \cite{ref14}, or by employing event-triggered islanding to preserve
stable operation under adverse conditions \cite{ref15}. Equally important is the ability to validate
such mechanisms under realistic conditions, using real-time co-simulation, controller-in-the-loop,
and \ac{HIL} testbeds that accurately reproduce the cyber-physical coupling of
\ac{EI} deployments \cite{ref9,ref16,ref17}.

Although several active \ac{IEEE} \ac{TF}s address power system security and resilience, they
primarily focus on cybersecurity communications, resilience definitions, and
attributes, predominantly targeting bulk power system operations. The proposed
\ac{TF} report advances the security and resilience within the \ac{EI} domain, examining how the multi-network, multi-level, and multi-layer nature of the \ac{EI} influences the cross-section of its Energy and
Internet components. This report illuminates the intersection of Energy and Internet networks,
providing a platform for the academic and industrial communities to explore state-of-the-art
approaches, methods, and systems at the confluence of \ac{EI} and security/resilience, thereby
advancing their joint applications, in direct alignment with the scope, contributions, and report
organization outlined in Section \ref{s:Scope}.
\section{Cybersecurity and Assurance of Large-Scale Energy Internet Systems: Threat Landscape, Detection, and Resilience}
\subsection{Introduction}
\label{sec:cyber_intro}

The \ac{EI} extends the smart grid into a wide-area, multi-carrier system of systems in which
electricity, heat, gas, hydrogen, transportation, and information networks are tightly coupled
and operated through pervasive digitalization. \ac{DERs}, \ac{IBR}, demand-side
flexibility, peer-to-peer energy trading, and software-defined coordination are no longer
peripheral features but the operational backbone of this paradigm. The same digital interfaces that
enable real-time, decentralized coordination also dissolve the clear perimeter that once separated
 \ac{OT} from \ac{IT}, and they multiply the points at which an adversary
can observe, deceive, or manipulate the physical system. Cybersecurity is therefore not an add-on
to the \ac{EI}; it is a precondition for its safe and economically efficient operation.

This section reviews the cyber-physical threat landscape of large-scale \ac{EI} systems and the
state of the art in assessment, detection, and resilience. The central premise, consistent with a
decade of work by the present authors and the broader community, is that confidentiality-,
integrity-, and availability-centric \ac{IT} security is necessary but insufficient for the \ac{EI}. Because
attacks ultimately express themselves through physical quantities (frequency, voltage, power
flows, and market prices), defense must be cyber-physical and must be evaluated against physical
and economic impact rather than against information-layer indicators alone \cite{ref6,ref16}. The
interdependence between the cyber and physical layers, formalized by the activity on cyber-physical interdependence for power system operation and control \cite{ref16}, is the unifying lens adopted
throughout.

The remainder of the section is organized as follows. Subsection~\ref{sec:threat_taxonomy} develops a layered
threat taxonomy and surveys representative attack classes and real-world incidents. Subsection~\ref{sec:ibr_der_vuln}
examines the vulnerabilities specific to \ac{IBR}- and \ac{DERs}-rich systems. Subsection~\ref{sec:impact_modeling} discusses
quantitative impact modeling and the testbeds used to obtain it. Subsection~\ref{sec:detection_localization} reviews detection
and localization. Subsection~\ref{sec:mitigation_defense} addresses mitigation, defense, and resilience, including economic
mechanisms. Subsection~\ref{sec:emerging_challenges} identifies emerging challenges, most prominently the integration of
large, dynamic \ac{AI} data-center loads, and outlines research directions, before Subsection~\ref{sec:cyber_conclusion}
concludes.

\subsection{Threat Landscape and Attack Taxonomy}
\label{sec:threat_taxonomy}
A useful way to organize \ac{EI} threats is by the layer of the cyber-physical stack at which the
adversary acts: the device and edge layer (smart inverters, protective relays, \ac{IED}, \ac{EV} chargers, controllable loads, and their firmware); the
communication layer (field-bus and wide-area protocols such as IEC 61850, DNP3, Modbus, and
\ac{IEEE} 2030.5, together with timing services); the control and coordination layer (local controls,
microgrid secondary/tertiary control, state estimation, and \ac{DERs} aggregation); and the market and
data layer (price formation, demand response signals, and the data pipelines that feed analytics and
optimization). An attack at any single layer can propagate across layers, which is precisely what
distinguishes \ac{EI} security from conventional \ac{IT} security \cite{ref6,ref16}.

\subsubsection{False Data Injection and Measurement-Integrity Attacks}
\ac{FDIA} attacks corrupt the measurements or estimates on
which monitoring and control depend, and can be crafted to evade conventional bad-data detection
in state estimation. Defending against them has been an active research line, including data-driven
and prior-information-based resilient estimation that exploits the redundancy and physics of the
system to expose inconsistencies that a stealthy attacker cannot fully reconcile \cite{ref4}. Surveys of
machine-learning methods for \ac{FDIA} detection map the rapidly growing body of model-based,
learning-based, and hybrid approaches and their respective blind spots \cite{ref18}.

\subsubsection{Load-Altering and Demand-Side Attacks}
Among the most consequential \ac{EI}-specific threats are \ac{LAA}s, in
which an adversary compromises a population of high-wattage, internet-connected devices
(electric-vehicle chargers, heat pumps, smart thermostats, and similar loads) and synchronously
manipulates demand to perturb frequency and voltage. The feasibility of such attacks was first
demonstrated conceptually for \ac{IoT} botnets of high-wattage devices \cite{ref19}, and modern high-wattage
botnet variants have since been studied; comprehensive treatment of the threat model, transmission-and distribution-level impact, market effects, and the spectrum of detection and localization
schemes is provided in a recent survey by the present group \cite{ref20}. Dynamic \ac{LAA}s, which modulate
demand in response to observed frequency, are especially dangerous under low-inertia
conditions, a regime that became measurable during pandemic-era demand depressions and that is
now structural as conventional generation is displaced by converter-interfaced resources \cite{ref21}. The
attack surface continues to expand as more high-wattage, internet-connected loads (e.g., \ac{EV}-charging infrastructure) come online, and the threat has begun to be studied in the \ac{TSO}-\ac{DSO} coordinated setting, where the
interaction of transmission- and distribution-level responses changes both impact and observability
\cite{ref22}.

\subsubsection{Firmware, Hardware, and Supply-Chain Attacks}
Because \ac{EI} field devices are embedded systems with long lifecycles and frequent remote
updates, firmware modification and supply-chain compromise are persistent risks. Detecting
firmware tampering at the device level, including on inverter-based microgrid controllers, has
motivated hardware-assisted methods that monitor low-level execution signatures rather than
trusting software self-reports \cite{ref23}. Supply-chain assurance, increasingly framed around \ac{SBOM}s,
remains immature in practice, and a recent empirical study highlights
the gap between \ac{SBOM}-based vulnerability management as promised and as it actually performs in
the field \cite{ref24}.

\subsubsection{Communication, Timing, and Availability Attacks}
Wide-area monitoring, protection, and control rely on accurate time and timely delivery.
Spoofing of satellite timing can corrupt phase-angle measurements and the controls that depend on
them, an effect that has been reproduced experimentally in real-time \ac{HIL} environments \cite{ref25}. \ac{DoS} and time-delay attacks degrade the availability and
latency guarantees of microgrid and \ac{DERs} coordination, motivating event-triggered and delay-tolerant control designs in which a compromised segment can be islanded before a disturbance
propagates \cite{ref15}.

\subsubsection{Economic and Market-Layer Attacks}
As markets penetrate deeper into distribution and into the behavior of flexible loads, price
formation itself becomes an attack surface. Adversaries can manipulate demand to move
locational prices or exploit price-responsive automation to convert an economic signal into a
physical disturbance. The coupling of market and physical layers (and the corresponding need to
price and hedge residual risk) is examined in work on cyber-insurance for distribution grids \cite{ref26}
and, most pointedly, in grid-security analysis of price-responsive data-center workloads
(Subsection~\ref{sec:emerging_challenges}) \cite{ref27}.

\subsubsection{Real-World Incidents}
The threat model is not hypothetical. The 2015-2016 attacks on the Ukrainian
grid established that coordinated intrusions can cause physical outages. More recently,
ransomware and intrusion campaigns have repeatedly struck utilities and energy companies, and
the rapid digitalization of renewable assets has introduced systemic device-level weaknesses:
vendor-spanning vulnerability disclosures in widely deployed solar inverters showed that a large
fraction of disclosed issues were rated high or critical severity \cite{ref28}. In 2025, a wave of cyberattacks
in Poland was associated with a grid outage affecting on the order of half a million people and
disrupted conventional, solar, and wind assets, underscoring that inverter-dominated infrastructure
is squarely within adversaries' reach. These incidents motivate the device-, system- and market-level analyses that follow.

\subsection{Vulnerabilities of Inverter-Based and DER-Rich Systems}
\label{sec:ibr_der_vuln}
The defining structural change of the \ac{EI} is the displacement of synchronous machines by
\ac{IBR}. This transition reduces system inertia and short circuit strength, narrows stability margins,
and pushes dynamics into faster, control-defined timescales. It also changes the cybersecurity
calculus, because the behavior of an \ac{IBR}-dominated grid is determined by software control loops
that are, in principle, reachable and modifiable by an adversary \cite{ref6}.

\subsubsection{Smart-inverter and synchronization-loop attacks}
Smart inverters expose configurable grid-support functions (volt-VAR, volt-watt,
frequency-watt) and remote-management interfaces that, if manipulated, can be weaponized to
inject disturbances under the guise of legitimate control \cite{ref29}. Of particular concern are stealthy,
parameter-based attacks on the synchronization loops, e.g., \ac{PLL}, that allow
inverters to track grid angle and frequency. Because such manipulations can be small, slow, and
physically plausible, they can evade threshold-based detection while gradually degrading stability
or steering the system toward an unsafe operating point. At the same time, the distributed, control-rich nature of inverter-dominated microgrids confers a degree of inherent cyber resilience: properly
designed local controls can blunt the impact of \ac{PLL} attacks even before a dedicated detector
intervenes, and quantifying this inherent resilience is itself a useful design objective \cite{ref30}.

\subsubsection{DER Coordination and Aggregation}
Large-scale \ac{DERs} coordination, whether through aggregators, virtual power plants, or
distribution-level markets, multiplies the number of stakeholders and trust boundaries. A
comprehensive \ac{DERs} cybersecurity outlook catalogs the vulnerabilities, attack vectors, impacts, and
mitigations across this ecosystem and stresses the need for defense-in-depth that spans device,
communication, and coordination layers rather than securing any one of them in isolation \cite{ref6}. The
heterogeneity of \ac{DERs} ownership and the limited cyber-hardening of consumer-grade equipment
make assumptions of a trusted edge untenable; security architectures must assume partial
compromise and degrade gracefully.

\subsubsection{Standards and Interoperability as a Security Surface}
Interoperability standards are simultaneously a security asset and a security surface. \ac{IEEE}
1547 and \ac{IEEE} 2800 define interconnection and ride-through behavior for \ac{DERs} and \ac{IBR}; \ac{IEEE}
2030.5 and IEC 61850 define communication; and IEC 62443 and the \ac{NIST} Cybersecurity
Framework provide process and control baselines. Where these standards specify security
extensions, for example secure \ac{DERs} communication profiles, adoption lags deployment, leaving
fielded devices reliant on perimeter assumptions that the \ac{EI} no longer satisfies \cite{ref20}. Fragmentation
across regional grid codes and standard versions further complicates assurance, since a device
certified under one regime may behave differently, or expose different interfaces, under another.
Securing the \ac{EI} thus requires not only better devices but harmonized, security-by-design
standardization.

\subsection{Impact Modeling and Quantitative Assessment}
\label{sec:impact_modeling}
Because \ac{EI} threats are judged by their physical and economic consequences, rigorous
impact assessment is central to defense. The field has matured from qualitative threat enumeration
toward quantitative, reproducible evaluation using high-fidelity co-simulation and hardware-in-the-loop testbeds that capture the interaction of power electronics, protection, communication
latency, and control.

Quantitative security metrics translate attack scenarios into operational risk. The \ac{CPES-QSM} methodology, for instance, provides a quantitative basis for reasoning about the secure
operation of cyber-physical energy systems, linking attack feasibility to measurable system-level
outcomes \cite{ref31}. Comprehensive modeling references consolidate the threats, modeling abstractions,
available resources, and metrics, along with worked case studies, into a single framework for
practitioners \cite{ref4}.

On the tooling side, real-time digital simulation and multi-vendor co-simulation make it
possible to subject controls and protections to realistic attacks before deployment, capturing the
communication impairments and device-level dynamics that pure phasor-domain models omit.
Such environments have been used to reproduce and bound the impact of load-altering and false-data attacks across transmission and distribution, and faster-than-real-time variants extend the
capability toward look-ahead operational decision support, allowing operators to evaluate the
trajectory of a developing incident faster than it unfolds in the field \cite{ref20}. These environments are
also the natural proving ground for the detectors and mitigations discussed next, because validation
under realistic latency and protection logic is what determines whether a method transfers to
practice.

\subsection{Detection and Localization}
\label{sec:detection_localization}
Detection methods for \ac{EI} attacks fall into three broad families, with increasing use of
hybrids that combine their complementary strengths.

Model-based methods exploit explicit physics and control models to form residuals or
observers whose deviations signal an attack. Residual-based detection for inverter-based
microgrids, for example, uses the mismatch between observed and model-predicted behavior to
flag malicious manipulation while remaining robust to benign disturbances \cite{ref32}. Observer-based
designs additionally enable localization, attributing an anomaly to a specific node or device.

Data-driven methods learn normal and adversarial patterns from data and are effective
where accurate models are unavailable or where attacks are subtle and high-dimensional. Surveys
of machine-learning \ac{FDIA} detection chart this design space and its pitfalls, including the
vulnerability of detectors themselves to adversarial manipulation \cite{ref18}. The dependence of purely
data-driven detectors on representative training data, and their fragility under distribution shift,
remain open concerns.

Physics-informed and hybrid methods embed governing equations into learning models,
improving data efficiency, generalization, and trustworthiness. Physics-informed neural networks
for robust state estimation, for instance, retain the interpretability and consistency of model-based
methods while gaining the flexibility of learning, and they have been shown to harden estimation
against \ac{FDIA} \cite{ref33}. Hybrid pipelines that pair a physics-based residual generator with a learned
classifier are an increasingly common and pragmatic compromise \cite{ref20}.

A cross-cutting lesson is that detection must be co-designed with the physical and communication
layers. Detectors validated only in idealized phasor-domain models often fail when confronted with
real protection logic, measurement noise, and latency; conversely, detectors evaluated in \ac{HIL}
environments transfer far more reliably to practice. Hardware-assisted detection,
e.g., monitoring device execution at the firmware level, complements network- and measurement-layer detection by catching compromises that never manifest as anomalous grid behavior until it is
too late \cite{ref23}.

\subsection{Mitigation, Defense, and Resilience}
\label{sec:mitigation_defense}
Prevention and detection are necessary but cannot be assumed complete. The organizing
principle adopted here is that resilience, the ability to anticipate, withstand, recover from, and adapt
to adverse events, is the master key for the \ac{EI}, because it does not depend on detecting
every attack in advance \cite{ref34}. Several complementary mechanisms operationalize this principle.

\subsubsection{Resilient and Event-Triggered Control}
Control designs that tolerate corrupted or delayed information by construction reduce
dependence on perfect cyber hygiene. Event-triggered and discontinuous-communication
schemes maintain stable microgrid operation while limiting the communication an adversary can
exploit, and event-triggered islanding can isolate a compromised segment before a disturbance
propagates \cite{ref15}. Designing controls to degrade gracefully under partial compromise, rather than to
fail when any assumption is violated, is the practical expression of resilience-by-design.

\subsubsection{Moving Target Defense and Deception}
\ac{MTD} deliberately and unpredictably varies system parameters or
topology, for example, distributed flexible AC transmission settings, so that an attacker's
reconnaissance is invalidated and stealthy attacks are exposed. A recent survey of \ac{MTD} in power
grids consolidates its design principles, the tradeoff between security gain and operational cost, and
open directions, and clarifies where \ac{MTD} is and is not appropriate \cite{ref18}. \ac{MTD} is most attractive
precisely against the stealthy, slow attacks that defeat threshold detection.

\subsubsection{Digital Twins for Defense}
High-fidelity \ac{DT} enable continuous comparison between observed and expected
behavior, safe what-if testing of attack and response scenarios, and operator situational awareness
without disturbing the live system. The fundamentals, challenges, and prospects of \ac{DT} for
large-scale power systems (including their use for cyber-security defense) are addressed in
dedicated task-force activity \cite{ref35}. For \ac{IBR}-rich systems specifically, digital-twin defenses against
stealthy sensor and parameter attacks on grid-following inverters are an active and promising
line. The principal challenges are model fidelity, synchronization latency, and the security of the
twin itself, which becomes a high-value target.

\subsubsection{Hardware-Rooted Trust and Authentication}
Because edge compromise cannot be excluded, anchoring trust in hardware is essential.
Battery-based authentication for \ac{DERs}, which derives device identity from intrinsic physical
characteristics, exemplifies lightweight, hard-to-clone authentication suited to resource-constrained field devices \cite{ref36}. Combined with hardware-assisted firmware monitoring \cite{ref23}, such
mechanisms raise the cost of edge compromise and shorten the time to detect it.

\subsubsection{Economic and Risk-Transfer Mechanisms}
Not all residual risk is best handled by engineering controls; some is best priced and
transferred. Cyber-insurance instruments tailored to energy systems, for example, policies that
hedge against load-altering attacks and extreme load variations in distribution grids \cite{ref26}, internalize
cyber risk into operational and investment decisions and create incentives for better security
postures. The recent association of a battery-storage-related grid incident with gaps in cyber-insurance coverage illustrates both the relevance and the current immaturity of these instruments.
Market-aware mitigation, which accounts for how defensive actions interact with price formation
and participant behavior, is a natural extension.

\subsection{Emerging Challenges and Research Directions}
\label{sec:emerging_challenges}
\subsubsection{Large, Dynamic AI Data-Center Loads}
The most consequential near-term development for \ac{EI} security and resilience is the rapid
interconnection of hyperscale \ac{AI} data centers. These large dynamic digital loads exhibit multi-hundred-megawatt ramps over very short timescales as training and inference workloads start and
stop, and grid operators have already observed unplanned events, including the near-instantaneous
loss of roughly 1.5 \ac{GW} of data-center load in a single 2024 event, that resemble the signatures of
contingencies and of load-altering attacks. Reliability authorities have flagged emerging large loads
as among the most pressing risks to bulk-system reliability, with dedicated \ac{TF} and white-paper
activity characterizing their behavior and risks \cite{ref37}. From a security standpoint, two features are
critical. First, the volatility of these loads narrows the margin between a benign workload swing
and a destabilizing disturbance, which both raises the stakes of, and provides cover for, an attack.
Second, because data-center demand is increasingly price-responsive, the market layer becomes a
control channel into a very large physical load. Recent work models exactly this coupling and
analyzes the grid-security implications of market-driven data-center workload scheduling, together
with mitigations \cite{ref27}. Treating flexible data-center load as a grid-interactive asset (capable of
providing fast response and even synthetic support) is promising, but it must be secured as critical
control infrastructure rather than assumed benign.

\subsubsection{Coupled Markets and Physics Across TSO-DSO Boundaries}
As coordination deepens across the transmission-distribution interface and across energy
carriers, attacks and defenses must be analyzed in this coupled setting. Impact assessment of load-altering attacks in \ac{TSO}-\ac{DSO}-coordinated systems shows that cross-boundary interactions
materially change both impact and detectability \cite{ref22}. Security analysis that stops at a single voltage
level or a single market will miss cross-layer propagation, and coordinated defense across operators
is needed to match the reach of coordinated attacks.

\subsubsection{Quantum Threats and Post-Quantum Assurance}
Long-lived energy infrastructure must anticipate cryptographically relevant quantum
computing, which threatens the public-key cryptography underpinning device authentication and
secure messaging. Quantum-secure digital-signature schemes for standardized substation
messaging, such as IEC 61850 R-GOOSE and R-SV, demonstrate that post-quantum protection
can be made compatible with the stringent latency budgets of protection communications \cite{ref38}.
Migration planning, e.g., crypto-agility, inventory, and prioritization of long-lifecycle
assets, should begin now.

\subsubsection{AI as Both Shield and Weapon, and Supply-Chain Assurance}
\ac{ML} strengthens detection and operational decision-making, but the
same techniques enable more capable, adaptive, and evasive attacks, and learned
components introduce their own attack surface through poisoning, evasion, and model theft.
Robust, physics-grounded, and verifiable \ac{AI} is therefore a security requirement, not merely a
performance one \cite{ref33}. In parallel, software and firmware supply chains, and the SBOM-based
practices meant to secure them, need substantial strengthening to match deployment realities \cite{ref24}.
Across all these directions, the unifying need is for security-by-design, cyber-physical evaluation,
and resilience that does not presuppose perfect detection.

\subsection{Conclusion}
\label{sec:cyber_conclusion}
The \ac{EI} inherits the vulnerabilities of every domain it integrates and adds new ones at their
interfaces. Its defining features (inverter-dominated dynamics, pervasive \ac{DERs}, deep market
penetration, and increasingly large and volatile digital loads) make it both more capable and more
exposed. The evidence reviewed here supports three conclusions. First, \ac{EI} security must be cyber-physical: threats are realized through physical and economic quantities, and so must be modeled,
detected, and judged. Second, defense must be layered and resilient: with edge compromise
impossible to exclude, systems should be designed to detect quickly, degrade gracefully, and
recover, supported by hardware-rooted trust, moving target defense, digital twins, and economic
risk transfer. Third, the research and standards communities must move ahead of deployment,
particularly on securing large dynamic loads such as \ac{AI} data centers, harmonizing security-by-design standards, and migrating to post-quantum assurance. Realizing a resilient and secure large-scale \ac{EI} is achievable, but only if cybersecurity is treated as foundational infrastructure rather than
as a feature added after the fact.
\section{Modeling and Control of Transmission and Distribution Networks via Energy-Internet Systems}
\subsection{Evolution Toward Energy-Internet Enabled Transmission and Distribution Networks}
Over the last few years, \ac{EPS} have quickly transitioned from centralized infrastructures
toward highly distributed cyber-physical energy systems. This transformation is driven by three
primary factors: the accelerated deployment of renewable energy resources, widespread
electrification across transportation and industry, and the rapid growth of large electricity
consumers such as hyperscale data centers \cite{ref39}. These developments have significantly altered both
short-term, i.e., real-time and day-ahead, operations and long-term planning strategies of \ac{TnD}
networks, forcing utilities, regulators, and system operators to reconsider conventional approaches
to grid expansion planning and operations.

Unlike traditional power systems, where generation is predominantly supplied by large
synchronous power plants connected to the transmission network, modern `smart' grids
accommodate highly variable generations connected at the distribution system-level, widely known
\ac{DERs}, which include \ac{PV}, battery energy storage systems, \ac{EV}, and flexible demand-side assets. Thus, power and energy systems are no longer managed exclusively through a top-down hierarchical structure but through coordinated interactions between \ac{TSO}, \ac{DSO}, aggregators,
and distributed intelligent devices operating at the network edge; an evolution that requires
substantially greater observability, controllability, and interoperability across all voltage levels
while preserving system reliability and operational security \cite{ref40}.

The rapid increase in deployment of energy-internet systems has exposed fundamental
limitations within existing transmission infrastructures. Across North America, thousands of
gigawatts of generation and storage projects remain stalled in interconnection queues due to
transmission congestion, lengthy engineering studies, and insufficient network capacity.
Recognizing these challenges, the U.S. \ac{DOE} released the Transmission
Interconnection Roadmap in 2024 \cite{ref41}, which advocates for modernizing interconnection
procedures, adopting standardized technical requirements, and improving coordination among
utilities, \ac{RTO}, \ac{ISO}, regulators,
and project developers to accelerate the integration of new energy resources into the bulk \ac{EPS}.
Instead of relying solely on greenfield transmission expansion, the roadmap emphasizes improving
the utilization of existing infrastructure through enhanced planning methodologies, and advanced
operational tools, that allow for more efficient interconnection processes that take into consideration
the distribution system's ability to serve as active resources.

However, the increasing complexity of the integration of energy-internet systems into \ac{TnD}
networks extends beyond transmission-distribution planning. The \ac{DOE}'s Future of Resource
Adequacy Report \cite{ref42} further highlights that maintaining long-term system reliability will require
coordinated deployment of diverse energy resources while simultaneously addressing the growing
uncertainty associated with climate variability, extreme weather events, and rapidly changing
electricity demand. The report emphasizes that future resource adequacy cannot be evaluated
independently at either the transmission or distribution level; instead, it requires integrated planning
frameworks capable of capturing the mutual dependencies between bulk system operation and
active distribution networks.

On the other hand, the proliferation of \ac{IED}, \ac{AMI},
\ac{PMU}, distributed sensors, edge computing platforms, and cloud-based
energy management systems has created an unprecedented volume of operational data coming from
electrical systems. As the grid becomes increasingly decentralized, reliable operation depends not
only on physical infrastructure but also on secure and reliable communication architectures capable
of exchanging measurements, control commands, market information, and operational constraints
across multiple domains. Consequently, the modernization of \ac{TnD} systems has evolved from a
purely electrical infrastructure challenge into a multidisciplinary cyber-physical systems problem
requiring close integration between communication networks, data analytics, optimization, and
control.

In essence, the convergence of electrical infrastructure and digital communication has
given rise to the concept of the Energy-Internet, in which \ac{DERs}, network operators, market
participants, and stakeholders interact through standardized communication protocols and
interoperable information models. Rather than viewing \ac{TnD} networks as separate operational
entities, the \ac{EI} promotes continuous bidirectional exchange of electrical information, enabling
coordinated monitoring, control, and real-time decision making across the entire energy
infrastructure. As this modernization increases, the paradigm is rapidly changing, providing the
technological basis upon which advanced modeling, optimization, and cyber-resilient control
strategies are (or need to be) developed.

\subsection{Energy Internet: Regulatory Frameworks and Data Exchange Across Transmission and Distribution Systems}
The modernization of \ac{TnD} infrastructures extends beyond the deployment of new
electrical devices. It requires a corresponding transformation in the way operational information is
generated, exchanged, and utilized. As \ac{DERs} continue to proliferate throughout distribution
networks, making them `active', traditional communication and operational paradigms, where
transmission operators possessed near-complete system observability while distribution systems
remained largely passive, are no longer sufficient. Effective operation of modern \ac{TnD} systems will
depend on continuous bidirectional information exchange among \ac{TSO}s, \ac{DSO}s, aggregators,
market participants, and intelligent control devices deployed in the `edge' network. This digital
ecosystem is widely recognized as the \ac{EI}, where electricity and information flow simultaneously
through tightly integrated cyber-physical communication infrastructures.

Different from conventional \ac{SCADA}
architectures, the Energy-Internet is made up of heterogeneous communication technologies
capable of supporting real-time monitoring, distributed optimization, power markets, and
autonomous control across multiple voltage levels. Data exchanged throughout this architecture
extends well beyond conventional measurements of voltage, current, and frequency. Modern \ac{TnD}
coordination requires the continuous transmission of other type of network information, such as
feeder topology, asset characteristics, \ac{DERs} capabilities, and protection settings, as well as dynamic
operational data such as active and reactive power injections, state-of-charge estimates for battery
energy storage systems, inverter operating modes, flexibility bids, locational constraints, and high-resolution telemetry generated by \ac{PMU}; information that enables system operators to model
accurate system-wide representations used for coordinated monitoring, \ac{SE},
congestion management, and \ac{OPF} calculations.

In terms of regulations, the growing dependence on shared operational \ac{TnD} data has
reshaped regulatory policies governing the interaction between \ac{TnD} networks. In the United
States, the \ac{FERC} issued Order No. 2222, requiring \ac{RTO}s
and \ac{ISO}s to allow aggregated \ac{DERs}, or aggregators, to participate directly in wholesale electricity
markets \cite{ref43}. This regulation order comes with benefits and drawbacks, which
require distribution system operators and aggregators to expand their operational responsibilities in
a secure and standardized way. New services such as ancillary, frequency regulation, reserve
capacity, and market participation depend heavily on accurate information, standardized
communication interfaces, and synchronized operational awareness across both \ac{TnD} network-levels.

One example is the adoption of standardized communication protocols and interoperable
information models that enable secure information exchange among heterogeneous devices
manufactured by different vendors, e.g., the IEC 61850 standard \cite{ref44}, which provides object-oriented communication models for substation automation and \ac{IED}, and the IEC 62351 \cite{ref45}, which
defines risk management processes through authentication, encryption, data integrity, and secure
management mechanisms specifically designed for power system communications.

At the operational level, protocols including DNP3 Secure Authentication, IEC 60870-5-104, and \ac{IEEE} 2030.5 support interoperability between \ac{DERs}, energy management systems,
aggregators, and utility control centers. Together, all these standards establish the digital
infrastructure required for interoperable Energy-Internet deployments while ensuring compatibility
across \ac{TnD} environments.

Recent regulatory developments further emphasize that data sharing itself has become a
critical operational asset. For example, the Ontario Energy Board's proposed amendments to the
Distribution System Code require local distribution companies to provide standardized static and
operational telemetry associated with \ac{DERs} to the \ac{IESO}
\cite{ref46}. These initiatives reflect an international trend toward treating operational data as an
essential component of grid reliability rather than simply an auxiliary communication service. By
improving system-wide observability, standardized data exchange enables
more accurate forecasting, enhanced situational awareness, faster contingency analysis, and
coordinated control.

Despite all these developments and regulations, there are still many questions that need to
be answered. For instance, the same communication infrastructure that enables advanced
coordination between \ac{TnD} systems substantially expands the cyber-physical attack surface of
modern \ac{EPS}s. Every additional communication channel, intelligent sensor, edge controller, edge-device, or market interface represents a potential entry point for malicious threat actors seeking to
manipulate measurements, disrupt communications, or compromise operational decision-making
processes. \ac{FDIA} and \ac{DoS} attacks, GPS spoofing, and coordinated attacks targeting \ac{OT} networks
can exploit vulnerabilities within the digital communication layer to propagate physical
consequences throughout interconnected \ac{TnD} systems. Thus, maintaining the reliability of
Energy-Internet systems will require not only standardized communication protocols but also
advanced computational techniques capable of continuously validating, interpreting, and securing
the massive streams of data exchanged across the \ac{TnD} boundaries.

\subsection{Mathematical Modeling and Intelligent Control for Energy-Internet-Enabled Transmission and Distribution Networks}
The continuous need to exchange operational data throughout the \ac{EI} systems will change
fundamentally how \ac{TnD} networks are monitored and controlled. Unlike conventional power
systems, where supervisory control is largely based on centralized and homogeneous
measurements, modern \ac{TnD} networks generate heterogeneous data streams that originate from
\ac{PMU}, \ac{IED}, \ac{AMI}, \ac{DERs}, controllers, and edge computing platforms. Heterogeneous measurements
that describe a power system that is both electrically interconnected and digitally distributed, make
traditional deterministic control strategies increasingly inefficient for maintaining reliable
operations.

To transform large volumes of data into actionable operational decisions, system operators
rely on mathematical models that estimate the current state of the electrical network, predict future
operating conditions, and determine optimal control actions under physical and operational
constraints. Among these methods, \ac{SE} remains one of the most fundamental components of system
operations. By combining redundant measurements from geographically dispersed sensors with
network topology and equipment models, \ac{SE} reconstructs the most `probable' operating condition
of the power system even when measurements are incomplete, delayed, or corrupted. The estimated
system state subsequently serves as the foundation for studies such as contingency analysis, voltage
stability assessment, congestion management, and/or \ac{SCOPF}.

Building upon system observability, optimization algorithms determine how
generation resources, storage systems, controllable loads, and flexible \ac{DERs} aggregations need to
be coordinated to satisfy both physical constraints and economic objectives. Classical formulations
based on \ac{PF} and \ac{OPF} minimize generation costs and/or network losses while
respecting power balance equations, voltage limits, thermal constraints, and equipment operating
boundaries. However, the increasing variability introduced by renewable generation and demand
response at both \ac{TnD} levels has motivated the development of stochastic, robust, and distributed
optimization techniques capable of explicitly incorporating uncertainty into operational \ac{OPF}
problem specifications. These have become particularly important for integrated \ac{TnD} operation,
where decisions made at one voltage level directly influence operating conditions throughout lower
voltage levels of the electrical network.

Alternative ways to manage uncertainty in these systems are based on \ac{AI} and \ac{ML}, whose
proposal has been accelerated by the rapid growth and adoption of sensing infrastructure and
computational capabilities. Rather than relying exclusively on physics-based mathematical models,
data-driven methods, i.e., \ac{AI} and \ac{ML}, learn complex nonlinear relationships directly from
historical and real-time operational measurements. Deep neural networks, graph neural networks,
support vector machines, ensemble learning algorithms, and transformer-based architectures are
examples of \ac{AI} models increasingly employed for power grid applications including load
forecasting, renewable generation prediction, fault location, topology identification, voltage
stability assessment, and optimal control.

Among recent developments, \ac{RL} and \ac{DRL}
have emerged as powerful approaches for sequential decision-making under uncertainty.
Instead of optimizing a single operating point, \ac{RL} agents continuously interact with simulated
electrical networks, learning control policies that maximize long-term operational performance
through repeated observation and feedback. These approaches have shown significant potential for
real-time voltage regulation, coordinated \ac{DERs} dispatch, battery energy storage management,
congestion mitigation, microgrid energy management, and adaptive network reconfiguration.

Beyond operational optimization, latest developments have focused on mathematical
frameworks that provide an essential layer of cyber-physical resilience in \ac{TnD} networks. Because
Energy-Internet architectures depend heavily on continuous information exchange, malicious
manipulation of measurement data can propagate directly into operational decision-making
processes. Threat actors can target a system with the objective of significantly degrading \ac{SE}
accuracy, mislead optimization algorithms, and trigger inappropriate control actions.
Consequently, modern control architectures increasingly integrate anomaly detection, statistical
estimation, and \ac{AI}-assisted cyber threat intelligence directly into the operational workflow. Rather
than functioning as isolated cybersecurity tools, these algorithms continuously validate incoming
measurements, quantify uncertainty, identify abnormal system behavior, and enable resilient
control decisions even under compromised operating conditions.

The convergence of physics-based power system models, optimization theory, artificial
intelligence, and cybersecurity is establishing a new paradigm for operating modern \ac{TnD} systems.
Within the Energy-Internet domain, mathematical modeling is no longer limited
to representing electrical behavior; it also serves as the computational foundation that transforms
information into reliable situational awareness, autonomous control, market coordination, and
cyber-resilient operation. The following subsections build upon these concepts by examining the
specific modeling and control methodologies that enable integrated \ac{TnD} networks
to operate securely, efficiently, and autonomously under increasingly strained operating conditions.

\begin{figure}[H]
\centering
\includegraphics[width=0.45\linewidth]{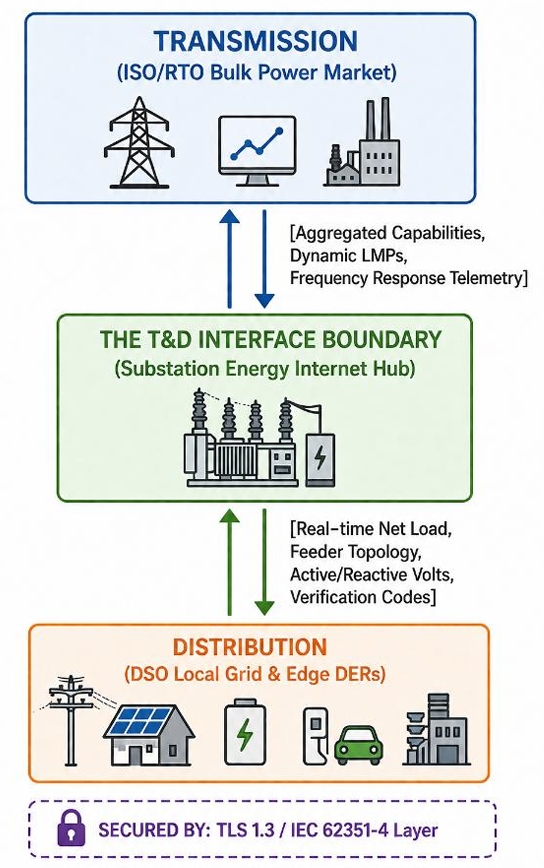}
\caption{Transmission and Distribution Interface Boundary. Data Exchange between TSOs and DSOs via Energy-Internet Hub.}
\label{fig:td_boundary}
\end{figure}

\subsection{Control and Optimization Across the Transmission and Distribution Boundary}
Within the Energy-Internet domain, the \ac{TnD} boundary represents far more than a physical
boundary between voltage levels. This boundary functions as the principal exchange point through
which operational information, market signals, and control commands are continuously shared
between \ac{TSO}s, \ac{DSO}s, \ac{DERs} aggregators, and customer-level energy management systems, see
Fig.~\ref{fig:td_boundary}. Therefore, the data exchanged across this interface possesses significant operational and
economic value, directly influencing network reliability, market efficiency, and infrastructure
utilization.

Historically, transmission operators maintained conservative operating margins because
distribution systems offered limited visibility into distributed generation and consumption. As
renewable generation and \ac{DERs} continue to proliferate, this lack of observability forces \ac{TSO}s to
commit additional reserve generation to compensate for uncertainty at the distribution level. These
reserves frequently consist of expensive fast-ramping generation units that increase operational
costs while reducing the effective utilization of renewable resources. Conversely, distribution
operators require continuous awareness of transmission operating conditions to prevent excessive
reverse power flow, voltage excursions, transformer overloading, and equipment malfunction
associated with high penetrations of inverter-based generation. Improving situational awareness
across the \ac{TnD} interface therefore enables both operational reliability and more
efficient utilization of existing infrastructure.

The realization of these benefits depends upon the continuous exchange of high-quality
operational data, e.g.:
\begin{enumerate}
\item Coming from the distribution network:
\begin{enumerate}
\item Aggregated active and reactive power capability curves that describe the flexibility
available from \ac{PV} systems, battery energy storage systems, and controllable loads.
\item Real-time net load measurements at the substation boundary that provide transmission
operators with an accurate representation of local demand after accounting for generation.
\item Network topology updates and contingency notifications that communicate changes
that may influence network security or operational flexibility.
\item \ac{DLMP}s.
\end{enumerate}
\item Coming from the transmission network:
\begin{enumerate}
\item Dispatch instructions.
\item Boundary voltage references.
\item Frequency regulation signals.
\item \ac{LMP}s that enable distributed resources to participate in
wholesale electricity markets while simultaneously supporting system-wide reliability
objectives.
\end{enumerate}
\end{enumerate}

Rather than serving merely as monitoring information, these data streams can become
decision variables of modern optimization algorithms. Integrated \ac{TnD} operation increasingly
relies on \ac{TnD} \ac{OPF} and \ac{TnD} \ac{SCOPF} formulations that coordinate multiple network operators
without requiring complete disclosure of proprietary distribution-level information. Instead of
transmitting every customer measurement to a centralized control center, local distribution
management systems independently solve internal optimization problems while exchanging only
boundary information, such as active and reactive power injections, boundary voltages, and other
boundary variables, with the transmission operator. Iterative decomposition methods as well as
integrated approaches allow optimization problems to converge toward a global (or local, in case
of nonconvex \ac{ACOPF} optimal operating point while preserving computational scalability and
organizational independence.

Because every optimization and control decision depends upon the integrity of exchanged
data, the mathematical accuracy of these data streams becomes equally important as the
communication infrastructure that transports them. Errors introduced through sensor failures,
communication delays, synchronization problems, or malicious cyberattacks propagate directly
into the optimization models responsible for controlling the \ac{TnD} network. Consequently,
understanding how corrupted measurements influence \ac{SE} and \ac{OPF} optimizations is essential for
developing resilient Energy-Internet architectures capable of maintaining reliable operation under
both uncertain and adversarial conditions.

\subsubsection{Mathematical Vulnerabilities in State Estimation and Optimal Power Flow}
Every optimization routine, dispatch decision, market-clearing process, and control action
relies on data collected from distributed sensors, \ac{IED},\ac{PMU}, \ac{AMI}, and accurate representative
models. Consequently, corrupted measurements or models do not simply degrade situational
awareness, but they directly alter the mathematical models used to estimate the operating state of
the power system and compute optimal control actions. Whether introduced through sensor
degradation, communication failures, or malicious attacks, compromised information propagates
into \ac{SE} and subsequently into \ac{TnD} \ac{OPF} solutions, potentially driving the network toward
economically inefficient or physically unsafe operating conditions.

The first computational stage affected by corrupted data is \ac{SE}. \ac{SE} combines redundant
field measurements with the physical network model to reconstruct the most probable operating
condition of the electrical grid. The measurement model is expressed as:
\begin{equation}
z = h(x) + e
\label{eq:se_model}
\end{equation}
where $z$ is the measurement vector containing voltage, current, power flow, and power injection
measurements; $x$ is the unknown system state composed of bus voltage magnitudes and voltage
phase angles; $h(x)$ represents the nonlinear measurement function; and $e$ is the measurement
error vector accounting for sensor inaccuracies and measurement noise.

Since practical measurements always contain uncertainty, \ac{SE} determines the operating
point that minimizes the discrepancy between measured and calculated quantities
while remaining consistent with the physical network model.

To prevent erroneous measurements from influencing operational decisions, many \ac{SE}
models implement \ac{BDD} algorithms, commonly used residual-based
detectors that evaluate the consistency between incoming measurements and the estimated system
state according to:
\begin{equation}
\lVert z - h(\hat{x}) \rVert \leq \tau
\label{eq:bdd}
\end{equation}
where $\hat{x}$ is the estimated system state and $\tau$ is a statistically determined
detection threshold. Measurements producing residuals greater than this threshold are classified as
inconsistent and removed before optimization routines are performed.

However, one of the defining characteristics of \ac{FDIA} attacks is that an adversary can
manipulate measurements while preserving the statistical properties expected by the estimator. If
the attacker constructs a corrupted measurement vector:
\begin{equation}
z_{attack} = z + c
\label{eq:zattack}
\end{equation}
where
\begin{equation}
c = h(\hat{x} + \Delta) - h(\hat{x}),
\label{eq:cdef}
\end{equation}
the resulting measurements satisfy:
\begin{equation}
\lVert z_{attack} - h(\hat{x}+\Delta) \rVert = \lVert z - h(\hat{x}) \rVert \leq \tau,
\label{eq:stealthy}
\end{equation}
allowing the corrupted data to bypass conventional residual-based detection. Rather than increasing
the estimation error, the attack shifts the estimated operating point itself,
causing subsequent optimization algorithms to operate on an incorrect representation of the
physical \ac{TnD} system.

Because \ac{SE} provides critical information used in \ac{OPF}, estimation
errors immediately propagate into network control and economic dispatch. For instance, the
classical \ac{ACOPF} problem, in polar coordinates, seeks the operating point that minimizes the total
generation cost:
\begin{equation}
\min \sum_i f(P_{Gi})
\label{eq:acopf}
\end{equation}
subject to the electrical network constraints. The active power balance equation is expressed as:
\begin{equation}
P_{Gi} - P_{Di} = V_i \sum_j V_j (G_{ij}\cos\theta_{ij} + B_{ij}\sin\theta_{ij}), \; \forall i \in N
\label{eq:pbalance}
\end{equation}
where $P_{Gi}$ and $P_{Di}$ denote generated and demanded active power, respectively, $V_i$ and $V_j$ are bus
voltage magnitudes, and $G_{ij}$ and $B_{ij}$ are the conductance and susceptance elements of the network
admittance matrix.

The optimization must also satisfy voltage magnitude limits,
\begin{equation}
V_{min} \leq V_i \leq V_{max}, \; \forall i \in N
\label{eq:vlimits}
\end{equation}
and transmission line thermal limits,
\begin{equation}
|S_{ij}| \leq S_{ij,max}, \; \forall (i,j) \in L
\label{eq:slimits}
\end{equation}
A full \ac{TnD} \ac{ACOPF} formulation can be found in \cite{ref47}. Within this optimization framework,
the data exchanged across the \ac{TnD} interface defines many of the parameters appearing in both the
objective function and the physical network constraints. Consequently, an attacker does not
(necessarily) need to alter the physical electrical infrastructure to influence system operation.
Manipulating measurements associated with boundary loads, distributed generation, voltage
magnitudes, transformer tap positions, \ac{DLMP}s, or transmission line ratings is sufficient
to modify the feasible operating region explored by the \ac{OPF} solver.

The consequences of these manipulations are both operational and economic. Artificially
reducing the reported thermal capacity of a transmission corridor creates nonexistent congestion,
forcing the optimization engine to dispatch more expensive generation while increasing \ac{LMP}s.
Conversely, underestimating actual line loading may cause the solver to schedule excessive power
transfers through already congested transmission paths, resulting in thermal overloads, accelerated
equipment degradation, and, under severe operating conditions, cascading failures. Similar effects
occur at the distribution level, where falsified voltage or reactive power measurements may cause
\ac{IBR} to inject inappropriate reactive power, destabilizing Volt-VAR control and increasing the
probability of unnecessary protection operations or localized voltage problems.

These observations demonstrate that protecting modern Energy-Internet-based
infrastructures requires more than securing communication channels alone. Since optimization and
control decisions are fundamentally mathematical processes, preserving the integrity of \ac{SE} and
\ac{OPF} has become a critical component of cyber-physical resilience. Consequently, recent research
has shifted toward resilient optimization frameworks that integrate distributed \ac{SE}, robust
optimization, physics-informed machine learning, and \ac{AI} to detect corrupted measurements before
they influence system operation and market decisions \cite{ref47}.

\subsubsection{Optimization Mitigation Frameworks in Energy-Internet Environments}
To protect the mathematical integrity of optimization methods from corrupted datasets, modern
systems integrate mathematical safeguards directly into the optimization pipeline. Some examples
are presented below:
\begin{enumerate}
\item \textbf{\ac{RO}}: Instead of relying on a deterministic \ac{OPF} solver that trusts
all input variables blindly, modern interfaces use \ac{RO}. The solver treats
incoming data parameters as uncertain values bound within a mathematical ``uncertainty
set''. The algorithm optimizes for the worst-case scenario within that set, rendering the
solution immune to micro-manipulations \cite{ref48}.
\item \textbf{Homomorphic Encryption for Multi-Area OPF}: When executing decentralized \ac{TnD}
\ac{ACOPF}, the optimization problem can be divided into a master problem, representing the
transmission network, and sub-problems representing each distribution system attached.
By using homomorphic encryption, \ac{DSO}s can exchange
boundary node Lagrangian multipliers $\lambda_i$
and consensus variables with the \ac{TSO} in a completely encrypted format. The global \ac{OPF}
problem converges mathematically to the lowest-cost configuration without the
transmission solver ever seeing or having the opportunity to ingest unencrypted, easily
corrupted data.
\item \textbf{Feasibility-Restricted Slack Variables}: Modern resilient \ac{ACOPF} solvers append
dynamic penalty slacks $x_{ii}$ to the optimization constraints. If highly corrupted data points
make the optimization problem mathematically impossible to solve (causing optimizer
divergence or an ``Infeasible'' error), these slack variables trigger immediately. They relax
the corrupted constraints mathematically and alert control operators, triggering corrected
resolve processes.
\end{enumerate}
\section{Decision-Aware Cyber-Physical Resilience for Storage-Integrated Energy Internet Systems}
\subsection{Motivation: From Storage Availability to Decision-Aware Resilience}
The \ac{EI} represents a structural evolution of the power grid from a centrally dispatched
physical network to a distributed, digitized, multi-energy system. In this architecture, electricity,
transportation loads, \ac{DERs}, storage, communication infrastructure, and market signals are
coordinated through bidirectional energy and information flows. The \ac{FREEDM} architecture
provided an early engineering foundation for this vision through distributed intelligence and power-electronics-based energy routers \cite{ref2}. Later \ac{EI} frameworks extended this concept toward multi-energy coordination, communication-enabled distributed control, and flexible grid operation
\cite{ref3,ref49,ref50}.

This evolution changes resilience from a primarily physical-restoration problem to a cyber-physical decision problem. Conventional resilience analysis emphasizes the ability to withstand,
absorb, and recover from high-impact disturbances, with performance commonly measured
through service loss, restoration progress, and recovery capability \cite{ref11,ref51}. In storage-integrated
\ac{EI} systems, this framing remains necessary but is no longer sufficient. Resilience also depends on
whether usable stored energy is available at the disruption onset, whether reported storage states
reflect true physical inventories, and whether candidate recovery actions satisfy voltage, current,
inverter, and network feasibility limits.

Thus, installed storage capacity alone does not guarantee resilience. A storage resource
may be physically intact but depleted by prior market dispatch, incorrectly reported because of
sensor error or adversarial manipulation, or unable to execute a recovery action because inverter or
network limits are violated. In these cases, resilience degradation originates in the decision process
rather than in asset availability.

The cyber-information layer further amplifies this dependency. \ac{EI} operation relies on
sensing, communication, estimation, computation, and control to coordinate distributed resources
in real time \cite{ref16,ref52}. Measurement error, communication delay, forecast corruption, or cyber
manipulation can distort the system state observed by the decision layer, causing recovery actions
to be planned for operating conditions that do not physically exist. This motivates a decision-aware
resilience framework that jointly evaluates storage adequacy, information integrity, and physical
feasibility before resilience actions are executed.

This decision-layer vulnerability is formally captured by the decision-induced resilience
deficit:
\begin{equation}
\Delta R_t^{dec} = R_t^{exp}(\hat{z}_t, u_t) - R_t^{real}(z_t, u_t)
\label{eq:deltaR}
\end{equation}
where $z_t$ denotes the true decision-relevant system state, $\hat{z}_t$ denotes the reported, estimated, or
perceived state available to the decision layer, and $u_t$ is the selected resilience action.
A positive $\Delta R_t^{dec}$ indicates that the decision layer overestimated the achievable resilience because
the action was selected using imperfect information, such as depleted but unobserved storage
energy, distorted storage-state measurements, communication delay, forecast error, or ignored
physical feasibility limits.

The main principle of this section is therefore that \ac{EI} resilience actions should be selected
only after jointly checking storage adequacy, information integrity, and physical
feasibility. Accordingly, the contribution of this section is a decision-aware admissibility
framework for storage-integrated \ac{EI} resilience. Rather than treating storage capacity as a standalone
resilience resource, the framework evaluates whether a candidate resilience action is energy-ready,
information-trustworthy, and physically feasible before execution. This shifts resilience
assessment from post-event outcome measurement to pre-action screening.

\subsection{Storage-Integrated Energy Internet Resilience: Key Challenges}
Storage-integrated \ac{EI} systems require resilience assessment beyond conventional
restoration metrics. Network hardening, generation availability, emergency control, and restoration
speed remain important in resilience studies \cite{ref11}, while quantitative resilience assessment
commonly relies on metrics such as load served, unserved energy, and recovery time \cite{ref51}. Natural-disaster resilience reviews also show that restoration performance depends on both infrastructure
preparedness and operational recovery strategies \cite{ref53}. In storage-integrated \ac{EI} systems, these
conventional concerns must be extended to include stored-energy availability, storage-state
information integrity, decision relevance, and physical feasibility.

First, storage introduces cross-horizon energy coupling. Storage may remain physically
intact yet provide negligible emergency support if pre-event dispatch depleted its inventory. The
energy state at disturbance onset is inherited from normal-operation decisions, not independently
set at recovery time. This intertemporal dependency is most consequential for \ac{LDES}, whose multi-hour to multi-day discharge capability \cite{ref54,ref55} means pre-event
depletion accumulates over the same timescale that determines resilience value. For hydrogen
systems, pre-event production and inventory scheduling directly govern post-event recovery
capability \cite{ref56}.

Second, storage readiness is cyber-physical. Operators usually act on estimated or
communicated storage states, such as battery \ac{SoC}, hydrogen inventory, pressure,
flow, electrolyzer status, or fuel-cell availability. \ac{FDIA} can systematically bias these estimates and
mislead operational decisions \cite{ref57}, with the attack surface spanning measurement, communication,
and control layers \cite{ref4,ref16,ref52}. The resulting impact appears at the decision layer when corrupted
state information causes storage dispatch, load pickup, or restoration sequencing to be selected for
a physical state that is not actually available.

Third, resilience can degrade through decision error, even without additional physical
damage. Premature storage depletion, excessive load pickup, poorly coordinated demand response,
or an unsustainable recovery sequence can widen the gap between planned and realized resilience.
This risk is amplified in \ac{DERs}-rich and hydrogen-integrated systems because distributed control,
inverter interfaces, and monitored process variables increase cyber-physical dependency \cite{ref58,ref59}.

Fourth, resilience actions require source-grid-load-storage coordination. Storage dispatch,
\ac{DERs} support, islanding, demand response, topology reconfiguration, and recovery sequencing
must be coordinated under limited energy and network headroom. Restoration planning therefore
requires resource availability, sequencing logic, operating constraints, and security checks to be
considered together \cite{ref60}.

Fifth, stored energy does not guarantee physical feasibility. In inverter-rich \ac{EI} systems,
current limits, modulation-voltage limits, voltage stability, and network transfer constraints may
bind before energy is exhausted. Security-region and grid-forming inverter studies show that
inverter constraints can restrict feasible restoration actions \cite{ref61,ref62}.

Together, these challenges show that storage-integrated \ac{EI} resilience is not only an outcome-measurement problem. It is an action-screening problem in which
each candidate resilience action must be checked for storage adequacy, information integrity, and physical feasibility before execution. These challenges motivate an action-screening view of \ac{EI} resilience in which each candidate action
is evaluated using storage adequacy, information integrity, and physical feasibility before
execution.

\begin{table}[!t]
\caption{Decision-Aware Resilience Challenges in Storage-Integrated EI Systems}
\label{tab:decision_challenges}
\centering
\small
\begin{tabular}{p{3.5cm}p{4.5cm}p{4.5cm}p{4cm}}
\toprule
\textbf{Challenge} & \textbf{Main issue} & \textbf{Resilience consequence} & \textbf{Needed decision support} \\
\midrule
Cross-horizon energy coupling & Pre-event storage use affects disturbed-period energy & Low energy at disruption onset & Storage-aware reserve scheduling \\
Cyber-physical storage state & Storage state is measured, estimated, and communicated & Wrong dispatch or underuse of storage & Information validation and DIG \\
Decision-induced resilience deficit & Reported state differs from true state & Planned RI exceeds actual RI & Decision-aware resilience metrics \\
Source-grid-load-storage coordination & Multiple resources must act together & Misallocation of limited energy & Coordinated EI operation \\
Physical feasibility limits & IBR/network constraints restrict actions & Energetically possible but infeasible operation & Security-margin screening \\
\bottomrule
\end{tabular}
\end{table}

\subsection{Decision-Aware Framework for Storage-Integrated Energy Internet Resilience}
A storage-integrated \ac{EI} should be analyzed as a coupled cyber-physical decision system
rather than as a collection of independent energy assets. The proposed framework introduces an
admissibility layer between \ac{SE} and resilience action execution. This layer screens
each candidate action using three coupled conditions: storage adequacy, information integrity, and
physical feasibility. Cyber-physical energy-security studies emphasize that physical components,
communication infrastructure, control systems, and decision logic must be considered jointly when
assessing risk and resilience \cite{ref4}. Cyber-physical interdependence studies also show that power-system operation increasingly depends on the interaction of sensing, communication, monitoring,
control, and physical network dynamics \cite{ref16}.

The physical layer provides energy and grid support; the cyber layer measures, estimates,
and communicates system states; the decision layer selects resilience actions; and the performance
layer evaluates realized outcomes and feeds information back for adaptation. The key distinction
is that the decision layer is treated as a resilience-critical component. A resilience action may fail
not only because of physical damage, but also because the decision layer acts on depleted storage,
corrupted state information, delayed communication, or ignored inverter/network feasibility limits.

In this framework, resilience is evaluated as a decision-aware process rather than as asset
availability alone. The physical energy state must be accurately represented by the cyber-information layer, translated into reliable operational decisions, screened for physical admissibility,
and reflected in realized resilience outcomes.

\begin{figure}[H]
\centering
\includegraphics[width=0.65\linewidth]{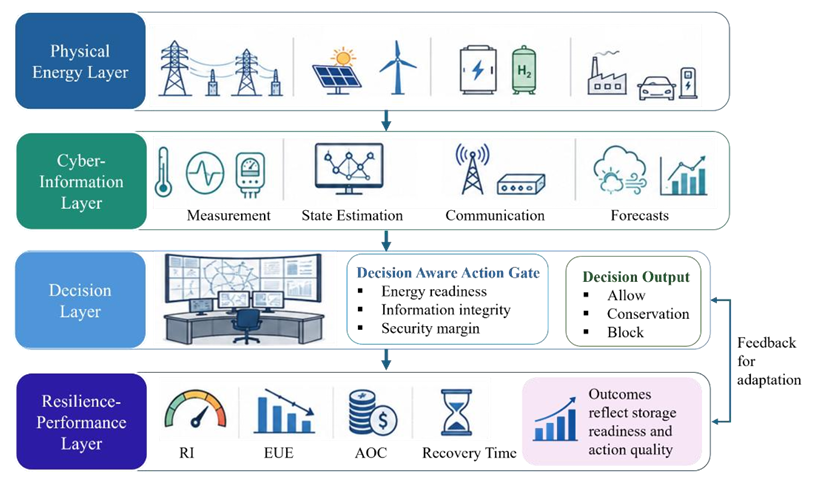}
\caption{Decision-aware storage-integrated EI resilience framework.}
\label{fig:decision_framework}
\end{figure}

Fig.~\ref{fig:decision_framework} illustrates the proposed decision-aware storage-integrated \ac{EI} resilience
framework. The physical energy layer supplies the true operating and storage states, while the
cyber-information layer measures, estimates, communicates, and forecasts the information used by
the decision layer. Candidate resilience actions are then screened through the decision-aware action
gate using storage adequacy, information integrity, and security-margin checks. The resulting
actions affect resilience-performance outcomes, including the \ac{RI}, \ac{EUE}, \ac{AOC}, and recovery time, which provide
feedback for adaptive operation.

Storage energy state is the central variable connecting these layers. Unlike conventional
grid assets, storage may be physically intact after a disturbance but unable to support emergency
operation if it was depleted beforehand. Therefore, the energy available at the disturbance onset is
not an independent restoration input; it is inherited from pre-event operation:
\begin{equation}
E_{s,t_0} = E_{s,t_0^-}, \; \forall s \in \mathcal{S}
\label{eq:energy_inherit}
\end{equation}
where $t_0^-$ is the final pre-disturbance time step and $t_0$ is the disturbance onset. This condition
prevents artificial energy reset and makes storage readiness a cross-horizon resilience state.

Because storage state is measured, estimated, and communicated through cyber
infrastructure, the decision layer may observe a reported state rather than the true physical state:
\begin{equation}
\hat{E}_{s,t} = E_{s,t} + a_{s,t}^E + \epsilon_{s,t}^E
\label{eq:reported_state}
\end{equation}
where $a_{s,t}^E$ represents malicious or disruptive bias and $\epsilon_{s,t}^E$ represents measurement, estimation, or
communication uncertainty. Overestimated storage energy can authorize aggressive restoration
actions that cannot be sustained, while underestimated energy may leave usable resilience
capability unused.

Three compact metrics connect storage readiness, information integrity, and physical
feasibility. The \ac{SAI} is:
\begin{equation}
SAI_t = \frac{\sum_{s\in\mathcal{S}} E_{s,t}}{\sum_{s\in\mathcal{S}} E_s^{max}}
\label{eq:sai}
\end{equation}
the system-level \ac{DIG} is:
\begin{equation}
DIG_t = \frac{\sum_{s\in\mathcal{S}} (\hat{E}_{s,t} - E_{s,t})}{\sum_{s\in\mathcal{S}} E_s^{max}}
\label{eq:dig}
\end{equation}
and the physical security margin is:
\begin{equation}
\Delta_t^{sec} = \min\{\Delta_t^{PF}, \Delta_t^{OC}, \Delta_t^{MOD}\}
\label{eq:secmargin}
\end{equation}
where $\Delta_t^{PF}$, $\Delta_t^{OC}$, and $\Delta_t^{MOD}$ denote margins to power-flow feasibility, inverter overcurrent limits,
and modulation-voltage limits, respectively.

The proposed decision-aware admissible-action set is:
\begin{equation}
\mathcal{A}_t^{allow} = \{a_t \in \mathcal{A}_t : SAI_t \geq SAI^{min}, |DIG_t| \leq DIG^{max}, \Delta_t^{sec} \geq \Delta^{min}\}
\label{eq:actiongate}
\end{equation}

This action gate is the central contribution of the framework. It converts storage-integrated
\ac{EI} resilience from an outcome-only assessment into a decision-admissibility problem. The first
condition, $SAI_t$, checks whether sufficient usable stored energy exists. The second
condition, $DIG_t$, checks whether the information used by the decision layer is trustworthy. The
third condition, $\Delta_t^{sec}$, checks whether the selected action remains physically feasible under inverter
and network limits.

The framework separates resilience assessment into decision-side readiness
metrics and performance-side outcome metrics. Readiness metrics determine whether
a candidate action should be attempted: $SAI_t$ measures stored-energy readiness, $DIG_t$ quantifies
cyber-to-decision distortion, and $\Delta_t^{sec}$ screens inverter and network feasibility. Outcome metrics
evaluate the realized consequence after the action is executed: $RI$ measures weighted load
service, $EUE$ quantifies unserved energy, $AOC$ converts reduced outage energy into economic
value, and $T^{rec}$ measures recovery speed. This separation is important because conventional
resilience assessment mainly reports what happened after a disturbance, whereas storage-integrated \ac{EI} resilience also requires metrics that explain whether the system was energy-ready,
information-trustworthy, and physically capable of executing the selected action.

\subsection{Storage Readiness and Cross-Horizon Resilience Coupling}
This subsection supports the first term of the action gate, $SAI_t$, by showing that storage
readiness depends on the energy carried from pre-event operation into the disturbance period.
Storage readiness links pre-event operation with post-event resilience performance. In a storage-integrated \ac{EI}, the energy available for emergency support is not an independent restoration input;
it is determined by prior dispatch, charging opportunity, renewable availability, and reserve-preservation decisions. Therefore, storage capacity contributes to resilience only when usable
energy is carried into the disturbance period.

This cross-horizon dependency is demonstrated through the unified \ac{LDES} co-optimization
study \cite{ref63}, where normal market operation and blackout restoration are solved in a single
framework with shared storage-state variables. Because stored energy is transferred from market
operation into restoration, the model captures whether each technology can sustain critical-load
service after disruption. The \ac{IEEE} 39-bus results are summarized in Table~\ref{tab:ldes_impact}.

\begin{table}[!t]
\caption{Storage Technology Impact on Blackout Restoration Performance \cite{ref63}}
\label{tab:ldes_impact}
\centering
\small
\begin{tabular}{p{4cm}p{2.8cm}p{2.2cm}p{3.3cm}p{2.7cm}}
\toprule
\textbf{Configuration} & \textbf{EUE (MWh)} & \textbf{RI (\%)} & \textbf{Restoration time (h)} & \textbf{AOC (\$M/yr)} \\
\midrule
No storage & 27,488.70 & 79.5 & 7 & 0.00 \\
Li-ion & 18,062.40 & 86.8 & 6 & 47.13 \\
Flow battery LDES & 14,853.50 & 90.5 & 3 & 63.18 \\
Hydrogen LDES & 5,435.76 & 96.0 & 2 & 110.26 \\
\bottomrule
\end{tabular}
\end{table}

Table~\ref{tab:ldes_impact} shows that storage-enabled restoration reduces unserved energy, improves \ac{RI},
shortens restoration time, and increases avoided outage cost. The improvement is strongest for
long-duration technologies. Li-ion storage improves performance relative to the no-storage case,
but its shorter discharge duration limits sustained restoration support. Flow battery and hydrogen
\ac{LDES} provide longer support, with hydrogen \ac{LDES} achieving the lowest \ac{EUE}, highest \ac{RI}, and
shortest restoration time.

\begin{figure}[H]
\centering
\includegraphics[width=0.7\linewidth]{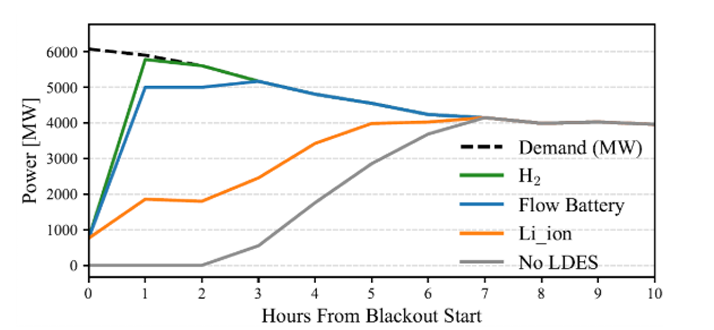}
\caption{Cross-horizon storage readiness and blackout resilience \cite{ref63}.}
\label{fig:cross_horizon}
\end{figure}

The corresponding restoration trajectories are shown in Fig.~\ref{fig:cross_horizon}. The figure illustrates that
the no-storage case restores load slowly, while \ac{LDES}-supported cases provide earlier and more
sustained service. Hydrogen \ac{LDES} follows the demand trajectory most closely during the early
restoration period, indicating stronger readiness for immediate post-blackout support.

These findings support the storage-adequacy condition in the proposed action gate. A
candidate resilience action should not be evaluated only by installed MW capacity or restoration
priority; it should also depend on whether sufficient stored MWh is available at the disturbance
onset. In practical terms, a system with large storage capacity but low energy readiness may still
experience delayed recovery, high unserved energy, and limited avoided outage benefit.

\subsection{Energy-State-Driven Adaptive Resilience Operation}
This subsection extends the storage-readiness concept from a static pre-event check to a
real-time adaptive control signal. Storage energy should not only be a constraint; it should also act
as a real-time supervisory signal. Fixed restoration sequences are energy-blind: they may
continue load pickup even when reserves are low or proceed too conservatively when storage is
abundant. In a storage-integrated \ac{EI}, the restoration pace should adapt to the remaining energy
state. Hydrogen-based resilience resources are particularly suitable for this role because hydrogen
storage can provide long-duration support across restoration horizons \cite{ref64}. For hydrogen-integrated restoration, the normalized energy-state signal is:
\begin{equation}
\phi_t = \frac{\sum_{h\in\mathcal{H}} H_{h,t}^{st}}{\sum_{h\in\mathcal{H}} H_h^{st,max}}, \quad 0 \leq \phi_t \leq 1
\label{eq:phit}
\end{equation}
where $H_{h,t}^{st}$ is the stored hydrogen inventory. The energy-state signal regulates the rate of network
energization:
\begin{equation}
\sum_{n\in\mathcal{N}} (x_{n,t}^{bus} - x_{n,t-1}^{bus}) \leq \kappa_0 + \kappa_1 \phi_t
\label{eq:energization}
\end{equation}
When $\phi_t$ is high, the system can energize more buses or loads. As $\phi_t$ declines, the
admissible action rate decreases, forcing conservative restoration and preventing premature
depletion.

The \ac{ESDAR} study demonstrates this concept in a modified islanded \ac{IEEE} 39-bus system,
where restoration scheduling is linked to the available hydrogen fraction so that the recovery pace
self-adjusts as stored energy declines \cite{ref65}. Compared with a fixed-sequence strategy, \ac{ESDAR}
increases \ac{RI} from 54.2\% to 65.2\% and restores 12\%-15\% more cumulative
load, indicating smoother conversion of stored hydrogen into served load \cite{ref65}. Table~\ref{tab:esdar}
summarizes the resilience improvement obtained by replacing energy-blind fixed restoration
with energy-state-driven adaptive restoration.

\begin{table}[!t]
\caption{Fixed and Energy-State-Driven Restoration Performance \cite{ref65}}
\label{tab:esdar}
\centering
\small
\begin{tabular}{p{3.5cm}p{2cm}p{3cm}p{8cm}}
\toprule
\textbf{Restoration logic} & \textbf{RI (\%)} & \textbf{Load-service gain} & \textbf{Operational implication} \\
\midrule
Fixed sequence & 54.2 & Baseline & Energy-blind recovery; higher risk of premature depletion \\
ESDAR & 65.2 & +12\%--15\% & Energy-state-driven recovery; smoother hydrogen-to-load conversion \\
\bottomrule
\end{tabular}
\end{table}

\begin{figure}[H]
\centering
\includegraphics[width=0.7\linewidth]{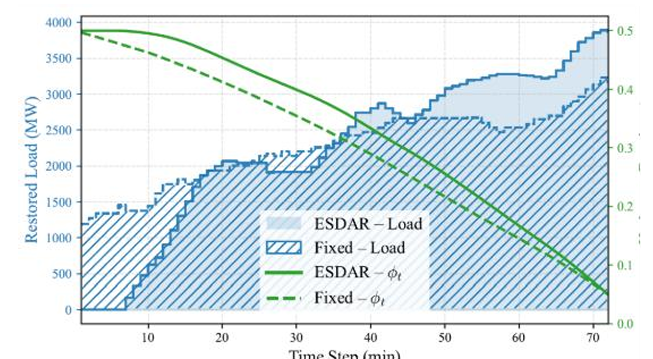}
\caption{Energy-state-driven adaptive resilience operation \cite{ref65}.}
\label{fig:esdar}
\end{figure}

The corresponding recovery trajectory is shown in Fig.~\ref{fig:esdar}, where \ac{ESDAR} maintains a
smoother restored-load increase while coordinating load pickup with the declining hydrogen energy
fraction. These results support the second element of the framework: storage energy should not only
be checked at disturbance onset but should continuously regulate resilience actions. In the broader
\ac{EI} context, $\phi_t$-type energy-state signals can supervise emergency dispatch, islanding, demand
response, critical-load support, and recovery scheduling.

\subsection{Security-Region-Constrained Resilience Feasibility}
This subsection supports the third term of the action gate, $\Delta_t^{sec}$, by showing that energy-ready actions may still be physically inadmissible under inverter and network constraints. Storage
adequacy is necessary but not sufficient. In inverter-rich \ac{EI} systems, a resilience action may be
energetically possible but physically infeasible because inverter current, modulation-voltage,
voltage-stability, or network transfer limits are reached. This is especially important during
restoration, where topology changes stepwise, voltage support is weak, and inverter headroom can
shrink as loading increases. Security-region analysis provides a useful way to
characterize feasible operation for inverter-interfaced systems \cite{ref61}. Grid-forming inverter studies
further show that overcurrent limiting and protection behavior can constrain available support
during disturbances \cite{ref62}.

In this case study, the general physical security margin $\Delta_t^{sec}$ is instantiated as the \ac{SSSR} margin $\Delta_t^{SSSR}$. The \ac{SSSR} margin is defined as:
\begin{equation}
\Delta_t^{SSSR} = \min\{\Delta_t^{solv}, \Delta_t^{OC}, \Delta_t^{MOD}\}
\label{eq:sssr}
\end{equation}
where $\Delta_t^{solv}$ is the margin to the power-flow solvability boundary, $\Delta_t^{OC}$ is the margin to the inverter
overcurrent boundary, and $\Delta_t^{MOD}$ is the margin to the modulation-voltage boundary. The
interpretation is:
\begin{equation}
\Delta_t^{SSSR} > 0: \text{feasible}, \; \Delta_t^{SSSR} \approx 0: \text{boundary}, \; \Delta_t^{SSSR} < 0: \text{infeasible}.
\label{eq:sssr_interp}
\end{equation}

The \ac{SSSR} study operationalizes this feasibility check by embedding inverter overcurrent
and modulation-voltage limits directly into the AC power-flow model using slack-variable
reformulation \cite{ref51}. The resulting formulation separates solvability-, overcurrent-, and modulation-driven boundaries and defines a unified scalar margin for restoration feasibility assessment.

In the \ac{IEEE} 39-bus case with a grid-forming \ac{IBR} at Bus 16, the seven-step restoration
trajectory remains secure in early stages, approaches the \ac{SSSR} boundary at Stage S6, and becomes
infeasible at Stage S7. Sensitivity analysis also shows that reducing the inverter current limit
shrinks the feasible $(P,Q)$ region, confirming that inverter headroom directly controls admissible
restoration actions \cite{ref51}. Table~\ref{tab:sssr_stages} summarizes how the \ac{SSSR} margin translates each restoration
stage into an operational decision condition.

\begin{table}[!t]
\caption{Security-Region Interpretation of Restoration Stages \cite{ref51}}
\label{tab:sssr_stages}
\centering
\small
\begin{tabular}{p{3.5cm}p{3cm}p{10cm}}
\toprule
\textbf{Restoration stage} & \textbf{SSSR margin} & \textbf{Operational interpretation} \\
\midrule
S1--S3 & Large/moderate $>0$ & Secure; sufficient inverter and network headroom \\
S4--S5 & Small $>0$ & Near-limit operation; action should be monitored \\
S6 & $\approx 0$ & Boundary contact; next pickup is critical \\
S7 & $<0$ & Infeasible; action should be blocked \\
\bottomrule
\end{tabular}
\end{table}

\begin{figure}[H]
\centering
\includegraphics[width=0.7\linewidth]{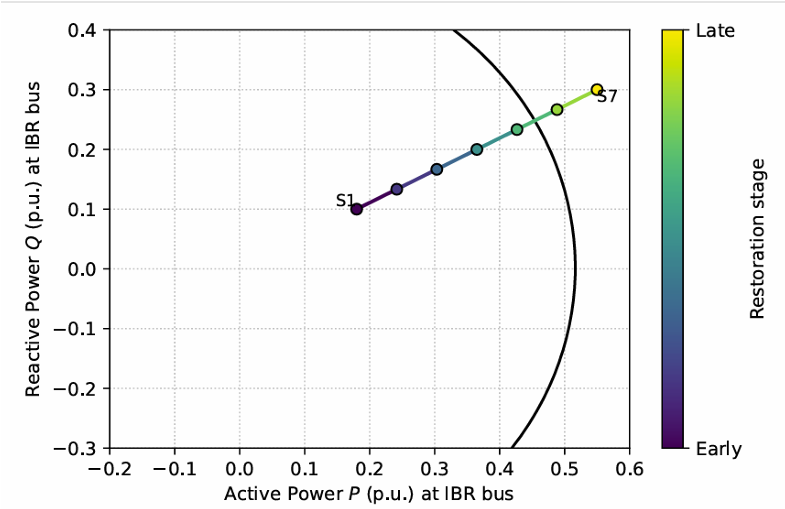}
\caption{Security-region-constrained restoration feasibility \cite{ref51}.}
\label{fig:sssr}
\end{figure}

The corresponding trajectory is shown in Fig.~\ref{fig:sssr}, where the restoration path moves
toward the security-region boundary as additional load is restored. This supports the third
screening condition: $\Delta_t^{SSSR} \geq \Delta^{min}$. Therefore, the action gate must
include physical feasibility in addition to storage adequacy. A restoration
action permitted by storage energy alone may still need to be slowed, blocked, or supported
by additional reactive/inverter headroom if $\Delta_t^{SSSR}$ is small.

\subsection{Integrated Interpretation and Summary}
The preceding subsections show that resilience in storage-integrated \ac{EI} systems is
a \textbf{decision-aware cyber-physical problem}. Storage improves resilience only when three
conditions are jointly satisfied: sufficient usable energy is available at the disturbance onset,
storage-state information is trustworthy, and the selected action is physically feasible under
inverter and network constraints. These conditions are captured by the proposed action gate
in \eqref{eq:actiongate} through $SAI_t$, $DIG_t$, and $\Delta_t^{sec}$.

The storage-readiness results demonstrate that pre-event operation directly affects post-event recovery. In the unified \ac{LDES} study, hydrogen \ac{LDES} improves \ac{RI} to 96.0\% and reduces
restoration time to two hours, compared with 79.5\% RI and seven hours without storage \cite{ref63}. This
shows that installed storage capacity alone is not sufficient resilience indicator; the usable energy
carried into the disturbance period must be explicitly considered.

The adaptive-restoration results further show that storage energy should not only be
checked at the disturbance onset but also used as a real-time supervisory signal. In the \ac{ESDAR}
study, linking restoration pace to the available hydrogen energy state improves \ac{RI}
from 54.2\% to 65.2\% and increases cumulative load served by approximately 12\%-15\% \cite{ref65}.
This supports the role of energy-state feedback in regulating load pickup, islanding, emergency
dispatch, and recovery sequencing. However, this feedback is useful only when the reported storage
state is reliable; otherwise, the decision layer must account for information uncertainty through
the $DIG_t$ condition.

The security-region results establish that energy adequacy does not guarantee action
feasibility. The \ac{SSSR} study shows that restoration trajectories can reach inverter overcurrent,
modulation-voltage, or solvability boundaries before stored energy is exhausted; in the \ac{IEEE} 39-bus case, the trajectory reaches the boundary at Stage S6 and becomes infeasible at Stage S7 \cite{ref51}.
Therefore, physical feasibility screening through $\Delta_t^{sec}$, instantiated as $\Delta_t^{SSSR}$ in the SSSR study, is
an independent requirement, not a consequence of storage availability.

Together, these findings highlight a limitation of conventional resilience metrics. \ac{RI}, \ac{EUE},
\ac{AOC}, and recovery time quantify the outcome after a disturbance, but they do not indicate whether
the system was energy-ready, whether the decision information was trustworthy, or whether the
action was physically admissible. The proposed $SAI_t$, $DIG_t$, and $\Delta_t^{sec}$ checks move resilience
assessment upstream from outcome measurement to decision-point screening.

The \ac{TF} implication is that resilient and secure large-scale \ac{EI} systems require coordinated
treatment of storage operation, cyber-information integrity, and physical feasibility. Storage
investment without energy-aware scheduling may fail to provide usable resilience support.
Cybersecurity without decision-impact assessment may miss resilience degradation caused by
distorted state information. Restoration planning without inverter and network feasibility screening
may produce actions that are energetically possible but physically inadmissible. The proposed
action gate therefore provides the principal contribution of this subsection: a compact cross-layer
condition for deciding whether a resilience action should be allowed, slowed, or blocked. This
contribution links storage scheduling, cyber-information integrity, and inverter/network feasibility
within a unified decision-aware resilience framework. \looseness=-1
\section{Multi-dimensional Resilience Considerations in Energy-Internet Systems}
The transition to decarbonized energy systems has redefined the role of the power grid,
which has evolved from a supply of infrastructure to the foundation of economic, industrial, and
digital activity \cite{ref66}. The accelerated electrification of sectors traditionally dependent on fossil fuels
and the massive integration of variable and distributed renewable generation have significantly
increased the operational complexity of the \ac{EPS}. At the same time, the growing penetration of \ac{IBR},
while essential for decarbonization, introduces new cyber-physical attack surfaces that enable
coordinated attacks to compromise multiple plants and trigger system-wide instability \cite{ref67,ref68}.

In this context, the resilience of the \ac{EPS} has taken on growing importance due to the
increased frequency and intensity of \ac{HILP} events, such as extreme
weather events, cyberattacks, and cascading failures. Unlike reliability, which focuses on managing
foreseeable contingencies, resilience refers to the system's ability to withstand, mitigate, and
quickly recover from severe disturbances, ensuring continuity of supply even under extreme
conditions \cite{ref69,ref70}.

Building on this foundation, the literature characterizes the resilience of \ac{EPS} through five
main dimensions: i) physical, ii) operational, iii) digital-cyber \cite{de2026quic, de2026let}, iv) climate-external, and v) regulatory \cite{ref71}.
Each dimension addresses a specific set of system vulnerabilities and capabilities in response to
disturbances \cite{zografopoulos2021security, topallaj2025impact}. However, most studies have examined these dimensions independently, developing
specific metrics, models, and assessment frameworks for each one without fully considering the
interdependencies among the different domains \cite{ref72}, as shown in Table~\ref{tab:resilience_compare}.

\begin{table}[!t]
\caption{Comparison of Resilience Assessment Approaches}
\label{tab:resilience_compare}
\centering
\small
\begin{tabular}{p{2.7cm}p{2.8cm}p{2.4cm}p{2.8cm}p{2.8cm}p{2.4cm}}
\toprule
\textbf{Reference} & \textbf{Multidim. Metrics} & \textbf{\# Dimensions} & \textbf{Cross-dim. Coupling} & \textbf{Exogenous Factors} & \textbf{Case Study} \\
\midrule
\cite{ref69} & \checkmark & 5 & $\times$ & $\times$ & $\times$ \\
\cite{ref73} & \checkmark & 2 & $\times$ & $\times$ & \checkmark \\
\cite{ref74} & $\times$ & 1 & $\times$ & $\times$ & $\times$ \\
\cite{ref75} & \checkmark & 2 & \checkmark & $\times$ & \checkmark \\
\cite{ref76} & \checkmark & 3 & \checkmark & $\times$ & \checkmark \\
\cite{ref77} & \checkmark & 6 & $\times$ & $\times$ & $\times$ \\
\cite{ref78} & $\times$ & 1 & $\times$ & $\times$ & \checkmark \\
\bottomrule
\end{tabular}
\end{table}

This fragmented perspective hinders the understanding of coupled phenomena, such as
cascading failures, in which an initial disturbance can propagate across the physical infrastructure,
system operations, communication and control networks, as well as external factors, progressively
amplifying its impact. Consequently, a system that is considered resilient from the perspective of a
single dimension may still exhibit critical vulnerabilities when interactions among all dimensions
are considered \cite{ref79,ref80, romero2026resilience}.

\subsection{Multidimensional Resilience Assessment Under Cyber-Physical Attack Scenarios}
This subsection develops a quantitative index that aggregates cross-dimensional degradation under simultaneous stress interactions and then applies it to a case study
of escalating cyber-physical attack scenarios \cite{romero2026multidimensional}.

\subsubsection{System Performance and Resilience Loss}
The system performance function $\phi(t)$ combines frequency deviation and inter-machine
coherency, as expressed in \eqref{eq:phi_t}.
\begin{equation}
\begin{aligned}
\phi(t) &= \omega_f \phi_f(t) + \omega_s \phi_s(t) \\
\phi_f(t) &= \max\left(0, 1 - \frac{f_{COI}(t) - f_{nom}}{f_{nom} - f_{crit}}\right) \\
\phi_s(t) &= \max\left(0, 1 - \frac{\Delta f_{gen}(t)}{\Delta f_{gen}^{coh}}\right)
\end{aligned}
\label{eq:phi_t}
\end{equation}
where $f_{COI}(t)$ is the center-of-inertia frequency, $f_{nom}$ and $f_{crit}$ are the nominal and critical
frequencies, $\Delta f_{gen}(t) = \max_{i,j}|f_i(t) - f_j(t)|$ is the inter-generator frequency spread,
and $\Delta f_{gen}^{coh}$ is the coherency tolerance band. The weights satisfy $\omega_f, \omega_s \in [0,1]$
and $\omega_f + \omega_s = 1$. Equal weights are adopted since frequency deviation and coherence loss are
considered equally important stability phenomena. These two quantities are selected because
they represent the primary indicators of power system stability. Loss of frequency stability or inter-machine coherency typically precedes system collapse, making them the most direct measures of
the physical impact of a cyberattack on the \ac{EPS}.

The resilience loss metric $R_{loss}$ quantifies cumulative performance degradation over a
horizon $T_H$ relative to the pre-disturbance operating point. Integration starts at the disturbance
time $t_{0,i}$ for the scenario $S_i$. To account for collapse, the truncated performance function $\tilde{\phi}_i(t)$ is
defined as:
\begin{equation}
\tilde{\phi}_i(t) =
\begin{cases}
\phi_i(t), & t \leq t_{lim} \\
0, & t > t_{lim}
\end{cases}
\end{equation}
where $t_{lim} = t_f$ if the system recovers within $T_H$, and $t_{lim} = t_{col}$ if collapse occurs at $t_{col}$. The
resilience loss is then computed as:
\begin{equation}
R_{loss,i} = \frac{1}{\phi_{0,i}} \int_{t_{0,i}}^{t_{0,i}+T_H} \max\left(0, \phi_{0,i} - \tilde{\phi}_i(t)\right) dt
\label{eq:rloss}
\end{equation}
where $\phi_{0,i}$ denotes the pre-disturbance performance level for scenario $S_i$.

\subsubsection{Resilience Dimension Definitions}
\textbf{1) Physical Dimension:} The physical disruption index $D_{phy,i}$ in \eqref{eq:dphy} quantifies the weighted
loss of generation capacity caused by a disturbance in scenario $S_i$:
\begin{equation}
D_{phy,i} = \sum_{r\in R} \omega_r \frac{P_{r,lost,i}}{P_{r,total}}
\label{eq:dphy}
\end{equation}
where $P_{r,lost,i}$ is the disconnected or unavailable capacity of resource $r$ in
scenario $S_i$ (MW), $P_{r,total}$ is its installed capacity (MW), $\omega_r \in [0,1]$ is the assigned
weight, and $R = \{$PV, synchronous, storage, substations, ...$\}$ denotes the set of
resource types. The weights reflect the relative importance of affected resources in $S_i$,
while unaffected types are excluded. When multiple resources are impacted, weights are
assigned according to their criticality (i.e., inertia, reserves, or critical-load support). If only
one resource drives the disruption, $\omega_r = 1$.

\textbf{2) Operational Dimension:} The operational disruption index $D_{op,i}$ characterizes the system
dynamic response to a disturbance in scenario $S_i$ by combining frequency variation rate,
performance degradation, and generator coherency, as defined in \eqref{eq:dop}.
\begin{equation}
\begin{aligned}
D_{op,i} &= \frac{X_i}{1+X_i}, \\
X_i &= \left(\frac{|RoCoF|_{max,i}}{RoCoF_{crit}} + \frac{\delta_{\phi,i}}{\delta_\phi^{crit}} + \frac{\Delta f_{gen,i}^{max}}{\Delta f_{gen}^{crit}}\right) \\
\delta_{\phi,i} &= \frac{\phi_{0,i} - \phi_{nadir,i}}{\phi_{0,i}}
\end{aligned}
\label{eq:dop}
\end{equation}
where $|RoCoF|_{max,i}$ is the maximum absolute rate of change of frequency, $\delta_{\phi,i}$ is the
normalized performance drop, $\phi_{nadir,i}$ is the minimum value of $\phi(t)$ during the event,
and $\Delta f_{gen,i}^{max} = \max_t \Delta f_{gen,i}(t)$ is the peak inter-generator frequency spread. Equal
weights are assigned to the three terms due to their complementary role in transient stability
degradation. Critical thresholds are set to $RoCoF_{crit} = 1.0$ Hz/s \cite{ref79}, $\delta_\phi^{crit} = 0.05$,
and $\Delta f_{gen}^{crit} = 2.0$ Hz. The saturating form bounds $D_{op,i}$ within $[0,1)$ while preserving
sensitivity near the critical region ($X_i = 1 \rightarrow D_{op,i} = 0.5$).
As $X_i$ increases, $D_{op,i}$ asymptotically approaches unity.

\textbf{3) Digital-Cyber Dimension:} The digital-cyber disruption index $D_{cyb,i}$ quantifies the impact
of a disturbance on the system during scenario $S_i$, incorporating observability, controllability, integrity, and availability into the normalized index of \eqref{eq:dcyb}:
\begin{equation}
D_{cyb,i} = \sum_{j\in k} \omega_{cy_j} \frac{N_{compr_j,i}}{N_{scope_j}}
\label{eq:dcyb}
\end{equation}
where $k = \{obs, ctrl, int, av\}$ denotes the evaluated cyber aspects, $N_{compr_j,i}$ is the
number of compromised assets in aspect $j$ for scenario $S_i$, $N_{scope_j}$ is the total number of
evaluated assets, and $\omega_{cy_j} \geq 0$ are weighting factors satisfying $\sum_{j\in k} \omega_{cy_j} = 1$.

\textbf{4) Climatic Dimension:} The climatic disruption index $D_{clim,i}$ represents environmental
stressors that exacerbate system degradation during scenario $S_i$. Multiple climatic factors
are aggregated into the normalized index of \eqref{eq:dclim}:
\begin{equation}
D_{clim,i} = \sum_{k\in C} \omega_{c_k} \cdot I_{c_k,i}
\label{eq:dclim}
\end{equation}
where $C = \{$temperature, snow, wind, ice, humidity, extreme weather, ...$\}$ is the set
of climatic stressors considered, $I_{c_k,i} \in [0,1]$ is the normalized intensity of stressor $k$ in
scenario $S_i$, and $\omega_{c_k} \geq 0$ are weighting factors satisfying $\sum_{k\in C} \omega_{c_k} = 1$.

\textbf{5) Regulatory Dimension:} The regulatory dimension quantifies institutional vulnerabilities
by comparing the existing controls against a reference framework. The sub-index is
expressed in \eqref{eq:dreg}.
\begin{equation}
D_{reg} = \frac{1}{N_v^{ref}} \sum_{i=1}^{N_v^{ref}} v_i
\label{eq:dreg}
\end{equation}
where $v_i \in \{0,1\}$ indicate the presence ($v_i=1$) or absence ($v_i=0$) of a regulatory
weakness in the control category $i$, and $N_v^{ref}$ is the total number of reference control
categories, selected independently of the case under study.

\subsection{Multidimensional Resilience Index}
The proposed formulation is based on the premise that evaluating resilience dimensions
independently underestimates system-wide impacts, i.e., their simultaneous compromise and
underlying interdependencies create degradation that no single-dimensional assessment can
capture.

Let $k_{sim} = \{phy, op, cyb\}$ be the set of endogenous dimensions. The endogenous core decomposes
degradation into an additive and a coupling contribution, as given by \eqref{eq:mdri_core}:
\begin{equation}
\mathcal{M}(S_i;\gamma_i) = \underbrace{\frac{1}{|k_{sim}|}\sum_{k\in k_{sim}} D_{k,i}}_{\bar{D}_i \; (additive)} + \gamma_i \underbrace{\prod_{k\in k_{sim}} D_{k,i}}_{\Pi_i \; (coupling)}
\label{eq:mdri_core}
\end{equation}
where $D_{k,i} \in [0,1]$ is the normalized sub-index of dimension $k$ in scenario $S_i$, with equal weights
$1/|k_{sim}|$. The additive term $\bar{D}_i$ measures mean severity across dimensions independently. The
coupling term $\Pi_i = \prod_k D_{k,i}$ captures additional degradation arising from simultaneous cross-dimensional compromise; it collapses to zero whenever any single dimension remains
uncompromised ($D_{k,i}\rightarrow 0$), and reaches its maximum only when all dimensions are jointly and
severely degraded, encoding the cascading failure mechanism whereby an intact dimension
suppresses impact propagation, while simultaneous degradation across all dimensions produces
mutual amplification beyond the additive prediction.

The parameter $\gamma_i$ takes one of two values. Scenarios driven by a single disturbance vector
are assigned to the additive regime ($\gamma_i = 0$), under which $\mathcal{M}(S_i;0) = \bar{D}_i$ and no cross-dimensional
interaction is considered. Scenarios where all endogenous dimensions are simultaneously
compromised are assigned to the coupled regime ($\gamma_i = 1$), activating $\Pi_i$ and amplifying
degradation beyond the additive baseline. The multidimensional resilience index ($\mathcal{MDRI}$) for
scenario $S_i$ is given by \eqref{eq:mdri}:
\begin{equation}
\mathcal{MDRI}_i = \mathcal{M}(S_i;\gamma_i) \cdot \prod_{j\in k_{ext}} (1+D_{j,i})
\label{eq:mdri}
\end{equation}
where $k_{ext} = \{clim, reg\}$ is the set of exogenous dimensions and $D_{j,i} \geq 0$ is the normalized sub-index of exogenous dimension $j$ in scenario $S_i$. Each factor $(1+D_{j,i})$ amplifies the endogenous
core proportionally to the exogenous stress level. When no exogenous stress is present, the factor
reduces to unity, and the index simplifies to $\mathcal{MDRI}_i = \mathcal{M}(S_i;\gamma_i)$. Since the exogenous amplifiers
act multiplicatively on the endogenous core, $\mathcal{MDRI}_i$ is not normalized and serves as a
comparative metric across scenarios.

\subsection{Case Study and Attack Scenarios}
\textbf{System Model:} The proposed framework is validated on the \ac{IEEE} 39-bus test system,
implemented in MATLAB/Simulink. The original system comprises 10 synchronous generators
with a total installed capacity of 10,610 MW. To represent the increasing penetration of \ac{IBR}, 1,500
MW of synchronous generation has been replaced by nine utility-scale grid-forming \ac{PV} plants
distributed across buses 30-38 (14.1\% of total capacity).

\textbf{Threat Model:} Assumes a state-sponsored adversary targeting both the \ac{IT} and \ac{OT}
domains, consistent with the \ac{TTP}s attributed to the
ELECTRUM/Sandworm group \cite{ref81,ref82}. The attacker possesses prior knowledge of grid topology
and \ac{ICS} protocols, exploiting exposed perimeter devices and default credentials for initial access
and lateral movement into \ac{OT} networks. Attack execution involves coordinated, multi-vector
actions across distributed sites, including manipulation of control signals and disruption of
communication.

\textbf{Attack Scenarios:} Two distinct attack scenarios are evaluated, inspired by the coordinated
cyberattack on the Polish energy infrastructure on December 29, 2025 \cite{ref81}.

\textit{Scenario A -- Single-plant baseline attack}: A control input attack manipulates the active power
reference of the \ac{PV} plant at Bus 33 (190 MW), selected for its proximity to the highest load buses
in the system. As a result, a generation loss at this bus produces measurable system-wide frequency
transients while remaining within the single-plant scope for our baseline scenario. Operators retain
full communication and control over all the remaining plants. This represents an isolated
cyberattack without any climatic or regulatory assumptions.

\textit{Scenario B -- Multi-vector cascading attack}: A coordinated attack replicates \ac{TTP}s documented
in the Polish grid incident: i) communication disruption targeting six \ac{PV} plants, eliminating
operator observability and control, and ii) forced disconnection of 1,115 MW of \ac{PV} capacity
(74.3\% of total). The six targeted plants (at Buses 31, 32, 34, 35, 37, and 38) are geographically
dispersed, mirroring the targeting strategy. To reflect the elevated winter demand observed during
the Polish incident, a 25\% load increase is introduced as an operational stress factor, due to the
sub-zero temperatures and snowstorms \cite{ref81,ref83}.

\subsubsection{Attack Impact and Dynamic Response}
\begin{figure}[H]
\centering
\includegraphics[width=0.6\linewidth]{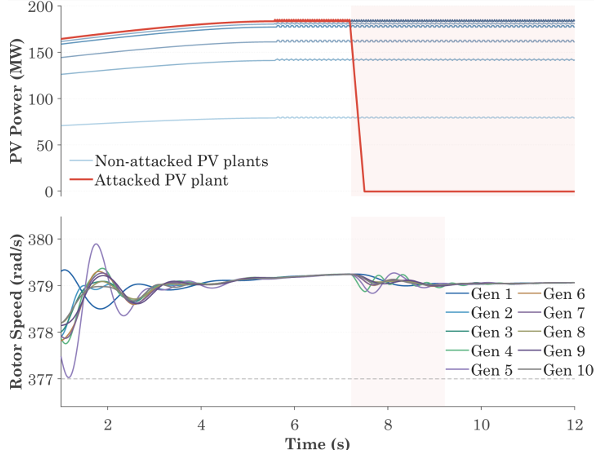}
\caption{Scenario A: PV generation and rotor speed response.}
\label{fig:scenA}
\end{figure}

\begin{figure}[H]
\centering
\includegraphics[width=0.6\linewidth]{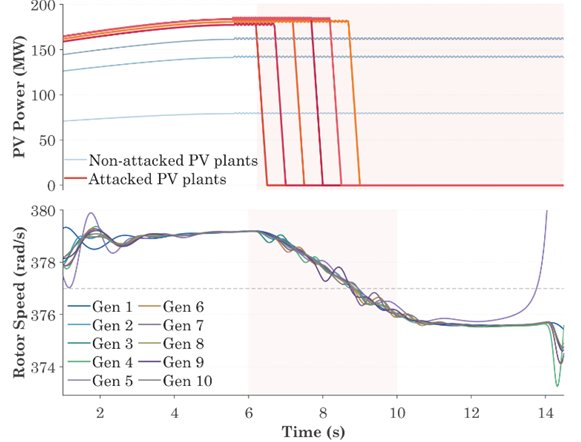}
\caption{Scenario B: PV generation and rotor speed response.}
\label{fig:scenB}
\end{figure}

Fig.~\ref{fig:scenA} and \ref{fig:scenB} present the PV generation output and synchronous generator rotor speeds for both
scenarios, while Table~\ref{tab:freq_metrics} summarizes the frequency response for both scenarios.

\begin{table}[!t]
\caption{Frequency Response Metrics Comparison}
\label{tab:freq_metrics}
\centering
\small
\begin{tabular}{p{6cm}p{5cm}p{5cm}}
\toprule
\textbf{Metric} & \textbf{Scenario A} & \textbf{Scenario B} \\
\midrule
Nadir frequency [Hz] & 60.323 & 59.776 \\
Time to frequency nadir [s] & 8.72 & 12.82 \\
Maximum RoCoF [Hz/s] & -0.055 & -0.089 \\
RoCoF time [s] & 7.5 & 6.5 \\
Steady-state frequency [Hz] & 60.332 & unstable \\
$\Delta f_{gen}$ (Hz) & 0.00012 & 3.10 \\
\bottomrule
\end{tabular}
\end{table}

\textbf{Scenario A:} As demonstrated in Fig.~\ref{fig:scenA}, the attack on the \ac{PV} plant forces its output from 190
MW to zero (at $t=7$s), while non-attacked plants maintain nominal generation. All generators
exhibit brief transient oscillations before converging to a common steady-state speed. During
Scenario A, the system maintains frequency stability with negligible frequency deviations,
indicating full inter-machine coherency throughout the transient.

\textbf{Scenario B:} As demonstrated in Fig.~\ref{fig:scenB}, six PV plants are simultaneously disconnected, removing
1,115 MW from the system. The rotor speed responses reveal dynamic instabilities, i.e., generators
attempt a coordinated response, but beyond $t\approx 8$s their trajectories diverge, leading to loss of
synchronism.

During Scenario B, frequency performance deteriorates, with a 61.8\% increase in
maximum RoCoF compared to Scenario A. The generator frequency fluctuations reach 3.10 Hz,
indicating complete loss of synchronism among synchronous machines.

\begin{figure}[H]
\centering
\includegraphics[width=0.6\linewidth]{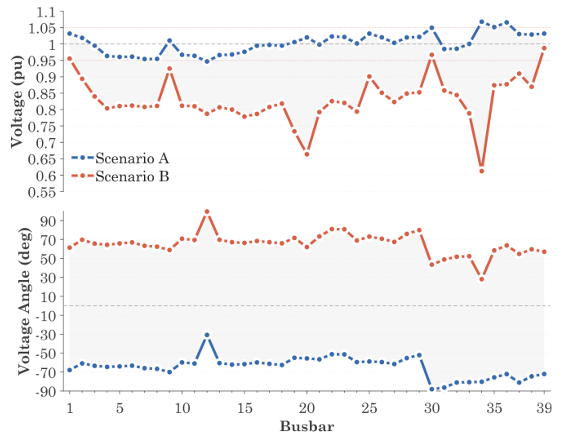}
\caption{Voltage magnitudes and angle profiles across all 39 buses at $t=15$s.}
\label{fig:voltage_profile}
\end{figure}

As shown in the voltage results, Scenario A maintains bus voltages within nominal ranges, whereas Scenario
B exhibits widespread voltage sags and a substantial increase in voltage angle dispersion, reflecting
the loss of angular coherency across the system (Fig.~\ref{fig:voltage_profile}).

\subsubsection{System Performance and Resilience Curves}
\begin{figure}[H]
\centering
\includegraphics[width=0.6\linewidth]{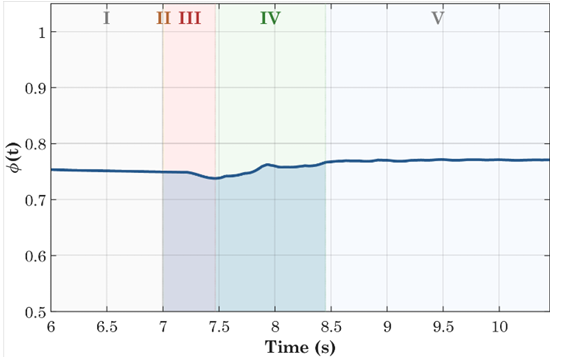}
\caption{System performance function and phases for Scenario A.}
\label{fig:phiA}
\end{figure}

\begin{figure}[H]
\centering
\includegraphics[width=0.6\linewidth]{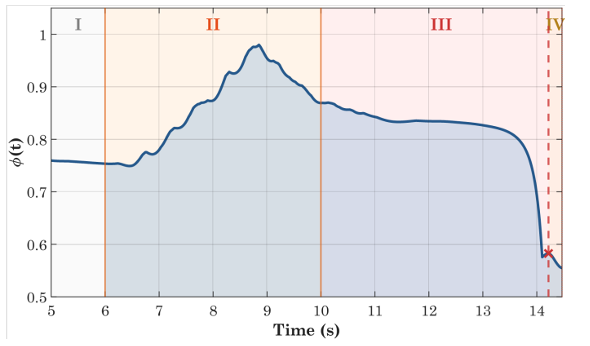}
\caption{System performance function and phases for Scenario B.}
\label{fig:phiB}
\end{figure}

Fig.~\ref{fig:phiA} and \ref{fig:phiB} present $\phi(t)$ for scenarios A and B ($\omega_f=\omega_s=0.5$, $f_{nom}=60$ Hz, $f_{crit}=59$ Hz,
$\Delta f_{gen}^{coh}=1.0$ Hz, $T_H=15$s), where $\phi_0 < 1$ due to residual PV-integration deviations.

\textbf{Scenario A -- Successful Recovery:} Following the attack on a single PV plant, the system
performance degrades from $\phi_0=0.7674$ to a nadir of $0.7379$ at $t=7.47$s, representing a 3.8\%
drop. Recovery to $\phi=0.7670$ (99.9\% of pre-event level) occurs within 1.45s, and
$R_{loss}=0.0305$. The system performance function exhibits five distinct phases: i) pre-event steady
state, ii) absorption (0.02s), iii) degraded operation (0.45s), iv) recovery (0.98s), and v) post-event
equilibrium.

\textbf{Scenario B -- Cascading System Collapse:} The multi-vector attack triggers progressive
degradation from $\phi_0=0.7674$ to a nadir of $0.5549$ at $t=14.57$s, a 27.7\% loss. The system
does not recover, and the collapse is detected at $t=14.22$s with $\phi=0.5824$, and $R_{loss}=1.0080$. The system performance exhibits only four phases (absent post-event equilibrium): i) pre-event steady state, ii) attack propagation (4.00s), during which the system attempts to absorb the
generation loss over approximately 3s before losing stability entirely, iii) degraded operation
(4.22s), iv) failed recovery (0.35s). The short-lived phase iv fails to mitigate the cascading
instability, leading to complete system collapse at $t=14.22$s.

\subsection{MDRI Evaluation}
Table~\ref{tab:endogenous} summarizes the physical, operational, and digital-cyber sub-indices together
with their corresponding average degradation, $\bar{D}_i$, and interaction product, $\Pi_i$.

\textbf{Physical and Operational Dimensions:} The physical sub-index, $D_{phy,i}$, measures the
fraction of inverter-based generation capacity lost. Since all disturbances exclusively affect the PV
fleet, \eqref{eq:dphy} reduces to $D_{phy,i} = P_{PV,lost,i}/P_{PV,total}$, with $\omega_{PV}=1$. These yield $D_{phy,A}=0.127$ and
$D_{phy,B}=0.743$.

The operational sub-index, $D_{op,i}$, in Scenario A, indicators remain below their critical
thresholds, resulting in $D_{op,A}=0.214$. In Scenario B, the simultaneous disconnection of multiple
PV plants causes both $\delta_\Phi$ and $\Delta f_{gen}^{max}$ to exceed their critical limits. Specifically,
$\delta_{\Phi,B}=0.277$ exceeds $\delta_\phi^{crit}$ by a factor of 5.5, while the inter-generator frequency spread reaches
3.097 Hz, surpassing the 2.0 Hz coherency threshold; as a result, $D_{op,B}=0.705$.

\textbf{Digital-Cyber Dimension:} The digital-cyber sub-index, $D_{cyb,i}$, is computed from four
complementary aspects: observability, controllability, integrity, and availability, each assigned an
equal weight of $\omega_{cy_j}=0.25$. Equal weighting reflects their non-redundant contributions to
characterizing operator situational awareness, remote control capability, data trustworthiness, and
asset accessibility during cyber incidents. The assessment scope is defined as $N_{scope}=9$,
corresponding to the nine PV plants comprising the inverter-based fleet.

Under Scenario A, the attack manipulates the active power reference of a single PV plant,
compromising only integrity and availability ($N_{compr,int}=N_{compr,av}=1$), while observability
and controllability remain unaffected. This yields
$D_{cyb,A}=0.056$. In contrast, Scenario B compromises communication, supervisory interfaces,
firmware, and plant availability at the six targeted sites ($N_{compr,int}=N_{compr,av}=N_{compr,obs}=N_{compr,ctrl}=1$),
resulting in $D_{cyb,B}=0.667$, a more than tenfold increase that reflects the broader
cyber impact of the coordinated attack.

\textbf{Climatic-External Dimension:} Scenario A assumes nominal weather conditions, yielding
$D_{clim}^A=0$. In Scenario B, the cyber-physical disturbance coincides with adverse winter conditions
representative of the Polish event. The climatic sub-index considers the latter two exogenous
stressors, which are normalized and equally weighted ($W_{temp}=W_{snow}=0.5$): thermal stress due
to sub-zero ambient temperature, normalized using the IEC 60076 thermal reference
($I_{temp}=0.20$), and additional mechanical loading on non-attacked PV plants caused by snowfall,
normalized according to the EN 1991-1-3 snow load standard ($I_{snow}=0.10$). These values
yield $D_{clim}^B=0.150$.

\textbf{Regulatory Dimension:} Based on the mapping of \ac{IEC} 62443 and \ac{NERC} \ac{CIP}, a reference
set of $N_v^{ref}=10$ control categories is defined. The Polish attack compromised six of these \cite{ref81,ref82,ref83}: i) no multi-factor authentication on FortiGate virtual private networks, ii) unpatched firmware
with known exploitable vulnerabilities, iii) default/reused credentials on Hitachi remote terminal
units and Mikronika controllers, iv) poor \ac{IT}/\ac{OT} segmentation, v) non-compliance with mandatory
\ac{DERs} cybersecurity standards, vi) insufficient \ac{OT} monitoring at remote substations. The remaining
four (i.e., incident response, supply chain risk, physical security, awareness) were unaffected. Thus,
$D_{reg}=0$ for scenario A, and $D_{reg}=0.6$ for scenario B. Based on Table~\ref{tab:endogenous}, Scenario A is
classified under the additive regime ($\gamma_A=0$), whereas Scenario B operates in the coupled
regime ($\gamma_B=1$).

\begin{table}[!t]
\caption{Endogenous Index Values per Scenario}
\label{tab:endogenous}
\centering
\small
\begin{tabular}{p{5cm}p{5.5cm}p{5.5cm}}
\toprule
\textbf{Quantity} & \textbf{$S_A$} & \textbf{$S_B$} \\
\midrule
$D_{phy,i}$ & 0.127 & 0.743 \\
$D_{op,i}$ & 0.214 & 0.705 \\
$D_{cyb,i}$ & 0.056 & 0.667 \\
$\bar{D}_i$ & 0.132 & 0.705 \\
$\Pi_i = \prod_k D_{k,i}$ & 0.0015 & 0.349 \\
Regime & Additive & Coupled \\
$\gamma_i$ & 0 & 1 \\
\bottomrule
\end{tabular}
\end{table}

In Scenario A, the small cyber degradation leads to a negligible interaction product ($\Pi_A=0.0015$),
making the endogenous degradation essentially additive. By contrast, the simultaneous
increase of all three sub-indices in Scenario B yields $\Pi_B=0.349$, causing the coupling term to
account for approximately 33\%. This result highlights the importance of cross-dimensional
interactions under coordinated disturbances. The resulting $\mathcal{MDRI}$ values are reported in
Table~\ref{tab:mdri}.

\begin{table}[!t]
\caption{$\mathcal{MDRI}$ Values per Scenario}
\label{tab:mdri}
\centering
\small
\begin{tabular}{p{6cm}p{5cm}p{5cm}}
\toprule
\textbf{Component} & \textbf{$S_A$} & \textbf{$S_B$} \\
\midrule
$\mathcal{M}(S_i;\gamma_i)$ & 0.132 & 1.054 \\
$(1+D_{clim,i})$ & -- & 1.150 \\
$(1+D_{reg,i})$ & -- & 1.600 \\
$\mathcal{MDRI}_i$ & 0.132 & 1.940 \\
\bottomrule
\end{tabular}
\end{table}

The increase in degradation between scenarios can be expressed through a logarithmic
decomposition as:
\begin{equation}
\ln\left(\frac{\mathcal{MDRI}_B}{\mathcal{MDRI}_A}\right) = \ln\left(\frac{\mathcal{M}_B}{\mathcal{M}_A}\right) + \sum_{k\in k_{ext}} \ln\left(\frac{1+D_{k,B}}{1+D_{k,A}}\right)
\label{eq:decomposition}
\end{equation}

The decomposition shows that 77\% of the increase in $\mathcal{MDRI}$ from Scenario A to Scenario
B is explained by the endogenous core through cross-dimensional coupling, while the remaining
23\% arises from climatic and regulatory effects.

By separating each dimension's independent contribution from its coupled effect, the
$\mathcal{MDRI}$ captures degradation that single-dimensional assessments cannot. The validation
demonstrates that a coordinated multi-vector attack raises the endogenous core roughly 8 times
over a single-vector baseline through cross-dimensional coupling alone, while climatic and
regulatory stressors add a further 84\%, yielding an approximately 15 times overall increase in
resilience loss, confirming that resilience quantification cannot be decoupled from institutional and
environmental contexts.
\section{Electricity Price Forecasting in the Energy Internet Era}
\subsection{Introduction}
The \ac{EPS} is undergoing a fundamental transformation from a centralized infrastructure
dominated by dispatchable generation toward a highly interconnected Energy Internet, in which
electricity, natural gas, hydrogen, transportation electrification, district heating, \ac{DERs}, and
communication networks and digital platforms are increasingly coordinated through sensing,
optimization, and \ac{AI} \cite{ref84}. Unlike traditional power systems that focused primarily on secure
electricity delivery, \ac{EI} systems require continuous coordination across multiple energy carriers,
market participants, intelligent devices, and cyber infrastructures. This transformation creates
important opportunities for flexibility, sustainability, and economic efficiency, but it also
introduces new forms of uncertainty, cyber exposure, and operational interdependence
\cite{ref85}. Electricity markets constitute one of the most important operational layers of this emerging
architecture. In organized markets, \ac{LMP}s, ancillary-service prices, congestion prices, flexibility
prices, and carbon-related signals coordinate the behavior of generators, storage systems, electric
vehicles, aggregators, virtual power plants, industrial loads, and financial participants \cite{ref86}.
Consequently, \ac{EPF} has evolved from a market-participant decision tool into a critical enabling
technology for resilient, secure, and intelligent \ac{EI} operation \cite{ref87}.

Historically, \ac{EPF} focused on improving short-term prediction accuracy in day-ahead and
real-time markets. Classical statistical methods, including autoregressive models, econometric
models, and volatility models, were effective under relatively stable market conditions because
electricity prices exhibit strong seasonality, persistence, and exogenous dependence on load and
fuel prices \cite{ref88}. However, renewable generation, demand response, \ac{DERs}, storage, and market
coupling have increased price volatility, reduced stationarity, and introduced nonlinear interactions
that challenge purely statistical approaches \cite{ref88}.

\ac{ML} and deep learning have improved forecasting capability by learning nonlinear temporal
and spatial relationships from large-scale market and operational data \cite{ref88,ref89}. Recent
architectures, including attention models, temporal convolutional networks, \ac{GNN}s,
and transformers, are particularly relevant to \ac{EI} applications because they
can represent long-range temporal dependencies, heterogeneous variables, and network-driven
spatial correlations \cite{ref89,ref90}. Nevertheless, high accuracy alone is insufficient. Forecasting systems
must quantify uncertainty, remain robust under distribution shift, detect abnormal or malicious data,
and provide interpretable information to market operators and participants \cite{ref91}. Recent work
exemplifies the transition from conventional \ac{EPF} to resilient and trustworthy market
intelligence. A framework was developed to improve short-term \ac{EPF} by enhancing the
representation of price spikes and increasing forecasting robustness under incomplete or noisy
market information \cite{ref92}.

Building on this work, subsequent studies further demonstrated that probabilistic
forecasting can serve not only as a prediction tool but also as an effective anomaly detection
mechanism for identifying abnormal market behavior under cyberattacks \cite{ref93}. These studies
highlight the evolving role of forecasting in resilient and secure large-scale \ac{EI} systems, where
electricity price forecasting, cyber anomaly detection, and operational decision support are
increasingly integrated within a unified cyber-physical intelligence framework.

This subsection of the \ac{TF} report examines \ac{EPF} from the perspective of resilient and secure
\ac{EI} operation. Rather than presenting a purely algorithmic survey, the report connects forecasting to
\ac{ISO}/\ac{RTO} market practice, power-system physics, cyber-physical security, and \ac{AI}-enabled decision
support. The central argument is that \ac{EPF} is evolving into a market intelligence layer that supports
operational reliability, resilience, and security under uncertain and potentially adversarial
conditions.

\subsection{Evolution of Electricity Price Forecasting in the Energy Internet}
\subsubsection{Electricity Price Forecasting as an Operational Function in ISO/RTO Markets}
\ac{EPF} was originally developed to support decision-making in competitive wholesale
electricity markets rather than as an isolated machine-learning problem. In organized markets
operated by \ac{ISO} and \ac{RTO}, such as PJM, MISO, CAISO, NYISO, ISO New England, ERCOT, and
SPP, prices are determined through security-constrained market clearing that simultaneously
enforces economic efficiency and physical reliability. Day-ahead and real-time \ac{LMP}s are computed
through large-scale \ac{SCUC} and \ac{SCED} formulations that represent generator limits, ramping constraints, transmission
constraints, reserve requirements, and contingency criteria \cite{ref94}. Therefore, electricity prices are
not direct functions of demand alone; they are shadow prices arising from constrained network
optimization. From the perspective of market participants, price forecasts support almost
every operational and financial decision. Generation companies use them for bidding, fuel
procurement, maintenance planning, and unit commitment. Load-serving entities use them for
procurement and hedging. Storage operators use anticipated intertemporal spreads to schedule
charging and discharging. Demand response providers and virtual power plants rely on price
forecasts to coordinate flexible loads and distributed resources. Financial participants use forecasts
to value \ac{FTR}s, \ac{CRR}s, and derivatives. For \ac{ISO}s/\ac{RTO}s,
forecasts also support situational awareness, resilience studies, scarcity assessment, and evaluation
of future congestion and reserve conditions. \looseness=-1

\begin{figure}[H]
\centering
\includegraphics[width=0.7\linewidth]{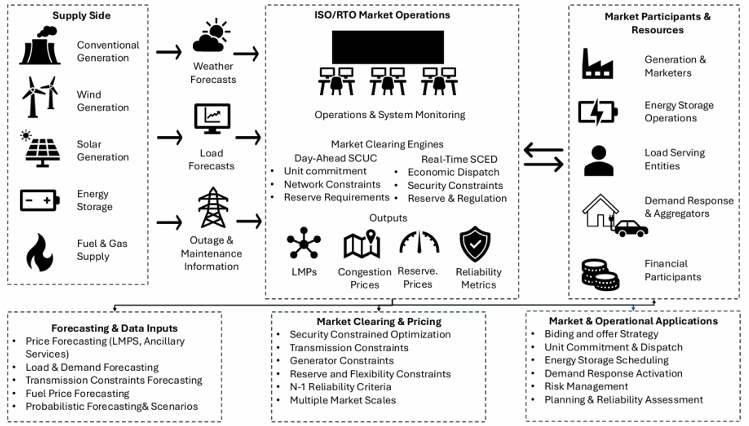}
\caption{EPF is embedded within ISO/RTO operational workflows and supports market participation, resource scheduling, and system reliability.}
\label{fig:epf_isorto}
\end{figure}

Fig.~\ref{fig:epf_isorto} illustrates how \ac{EPF} is embedded within \ac{ISO}/\ac{RTO} market operations. The figure
emphasizes that price forecasts are not independent products; they are shaped by weather forecasts,
load forecasts, renewable forecasts, generator availability, fuel prices, and transmission topology,
and they feed directly into bidding strategies, storage scheduling, demand response, reserve
procurement, and reliability assessment. The forecasting errors propagate into market bids,
dispatch schedules, reserve procurement, battery operation, and congestion management, while
market outcomes subsequently modify future system states. Thus, the operational value of a
forecast cannot be assessed only by mean absolute error or root mean square error; it must also be
judged by its impact on market decisions, risk exposure, and reliability.

Table~\ref{tab:epf_roles} summarizes the operational roles of \ac{EPF} in organized electricity markets. The
table is included not to list stakeholders, but to emphasize that forecasting errors affect different
actors through different mechanisms. A storage operator may be harmed by errors in price spreads,
while an \ac{ISO}/\ac{RTO} may be more concerned with missed scarcity events or congestion
conditions. As summarized in the table, the forecasting problem is inherently operational.
Different stakeholders require different forecast horizons, spatial resolutions,
and uncertainty information. This motivates forecasting frameworks that are not only accurate but
also interpretable, uncertainty-aware, and robust during abnormal operating conditions.

\begin{table}[!t]
\caption{Operational Roles of EPF in Organized Electricity Markets}
\label{tab:epf_roles}
\centering
\small
\begin{tabular}{p{3.5cm}p{6.5cm}p{6.5cm}}
\toprule
\textbf{Stakeholder} & \textbf{Operational Decision} & \textbf{Role of Price Forecast} \\
\midrule
ISO/RTO & Reliability assessment, reserve procurement, congestion monitoring & Anticipate stressed market conditions and operating risk \\
Generation companies & Unit commitment, bidding, fuel procurement, maintenance & Maximize expected revenue while managing operational constraints \\
Storage operators & Charging/discharging and ancillary-service participation & Identify intertemporal spreads and scarcity opportunities \\
Virtual power plants & Aggregation and DER scheduling & Coordinate distributed flexibility across market intervals \\
Load-serving entities & Energy procurement and retail hedging & Minimize procurement cost and manage exposure \\
Financial participants & FTR/CRR valuation and derivatives & Quantify congestion risk and market uncertainty \\
\bottomrule
\end{tabular}
\end{table}

\subsubsection{Evolution Toward the Energy Internet}
Electricity price formation is changing because the physical and market mechanisms
that determine prices are changing. In conventional power systems, prices were largely driven by
predictable demand patterns, thermal generation costs, and transmission congestion. In the Energy
Internet, the price formation process is shaped by renewable uncertainty, distributed resources,
flexible loads, energy storage, transportation electrification, multi-energy coupling, and cyber-physical coordination \cite{ref95}. Prices increasingly reflect the aggregated behavior of both centralized
market participants and millions of distributed devices.

In \ac{ISO}/\ac{RTO} markets, \ac{LMP}s remain endogenous outputs of security-constrained
optimization. However, the inputs to that optimization are increasingly uncertain and distributed.
Wind and photovoltaic resources have near-zero marginal costs and displace thermal generation
through the merit-order effect, suppressing average prices while increasing negative-price events
and volatility under oversupply, congestion, and inflexibility \cite{ref96}. Conversely, renewable forecast
errors, generator outages, gas-delivery constraints, or reserve shortages can produce scarcity prices
and extreme real-time spikes. Extreme weather events reinforce this challenge because they
simultaneously affect load, renewable generation, fuel availability, outages, and transmission
capability \cite{ref97,ref98}.

\begin{figure}[H]
\centering
\includegraphics[width=0.7\linewidth]{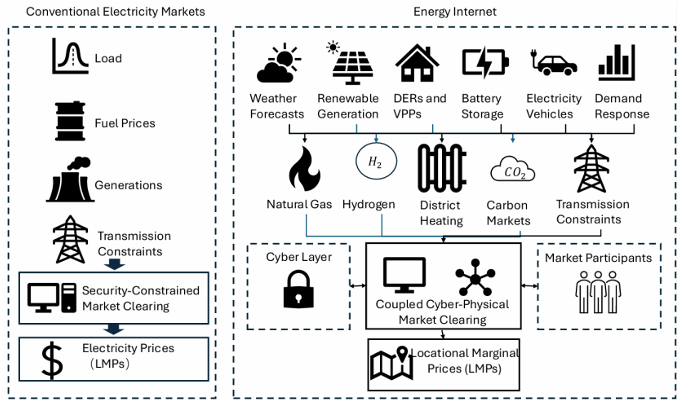}
\caption{The EI fundamentally changes electricity price formation by introducing distributed intelligence, renewable uncertainty, cross-energy coupling, and cyber-physical interactions.}
\label{fig:ei_market}
\end{figure}

Fig.~\ref{fig:ei_market} compares conventional market price formation with the emerging Energy
Internet. The figure highlights that the \ac{EI} introduces additional information streams and feedback
paths into market clearing, including \ac{DERs}, battery storage, electric vehicles, demand response,
natural gas, hydrogen, carbon markets, and cyber layers.

The increased complexity changes the forecasting task. Traditional
models often combined historical prices with load and fuel-price regressors. Modern \ac{EI} forecasting
must integrate variables with different temporal resolutions, spatial footprints, and reliability
characteristics. Numerical weather prediction affects renewable uncertainty; \ac{DERs} modify net load;
storage and demand response create strategic intertemporal behavior; and communication
networks determine data quality. Forecasting therefore becomes a cyber-physical learning problem
rather than a low-dimensional time-series problem. Table~\ref{tab:price_drivers} summarizes this transition,
comparing price-formation drivers rather than algorithms, because methodological advances in
forecasting are responses to changes in the underlying \ac{EPS}.

\begin{table}[!t]
\caption{Price-Formation Drivers in Conventional Electricity Markets and EI Markets}
\label{tab:price_drivers}
\centering
\small
\begin{tabular}{p{3cm}p{5cm}p{8cm}}
\toprule
\textbf{Characteristic} & \textbf{Conventional Markets} & \textbf{Energy Internet Markets} \\
\midrule
Generation mix & Dispatchable thermal generation dominant & High penetration of inverter-based renewable generation \\
Demand & Mostly passive consumers & Active prosumers, flexible demand, and aggregators \\
Storage & Limited system-level storage & Distributed batteries, grid storage, and EV \\
Market participants & Relatively centralized & Millions of distributed and automated participants \\
Energy coupling & Electricity-centered & Electricity, natural gas, hydrogen, heating, transportation, and carbon coupling \\
Data layer & SCADA and market data & SCADA, PMU, AMI, IoT, weather, market, edge, and cloud data \\
Forecasting challenge & Temporal prediction under moderate stationarity & Cyber-physical, multimodal, spatio-temporal learning under uncertainty \\
Primary objective & Market forecasting & Resilient, secure, and intelligent decision support \\
\bottomrule
\end{tabular}
\end{table}

As shown in Table~\ref{tab:price_drivers}, the \ac{EI} expands forecasting complexity across generation, demand,
storage, market participation, data availability, and energy coupling. This explains why purely
statistical or single-market forecasting models are increasingly insufficient. Future models must be
able to learn nonlinear relationships, represent spatial topology, quantify uncertainty,
and maintain reliable performance under distribution shifts.

\subsubsection{From Accurate Forecasting to Resilient Market Intelligence}
The evolution described above suggests that future \ac{EPF} should no longer be evaluated
solely by statistical accuracy. Metrics such as mean absolute error, root mean square error, and
mean absolute percentage error remain useful, but they provide limited information about
operational value \cite{ref99}. A model that slightly improves average error may still be operationally
weak if it misses price spikes, underestimates scarcity risk, fails during extreme weather, or
produces overconfident forecasts when data are corrupted. Therefore, \ac{EPF} is evolving into resilient
market intelligence: a decision-support capability that combines prediction, uncertainty
quantification, anomaly detection, and operational feedback.

\begin{figure}[H]
\centering
\includegraphics[width=0.7\linewidth]{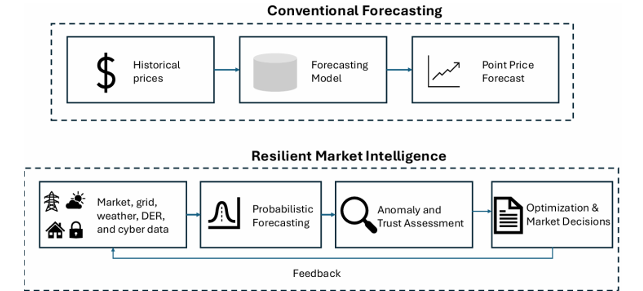}
\caption{Evolution from conventional point forecasting to resilient market intelligence. Future systems combine forecasting, uncertainty quantification, anomaly detection, optimization, and operational feedback.}
\label{fig:resilient_intel}
\end{figure}

Fig.~\ref{fig:resilient_intel} illustrates this conceptual shift. Conventional forecasting follows a mostly linear
workflow from historical data to deterministic price prediction. Resilient market intelligence
is closed-loop: forecasting assimilates heterogeneous cyber-physical
data, generates probabilistic forecasts, checks data trustworthiness, supports optimization,
and learns from realized market outcomes. Forecasting models must distinguish ordinary
uncertainty from abnormal behavior caused by sensor failures, missing data, communication
outages, or malicious cyberattacks. This distinction is becoming essential as market operations rely
more heavily on \ac{AMI}, \ac{PMU}s, cloud computing, \ac{IoT} devices, and distributed data streams. Recent
advances in \ac{EPF} have increasingly shifted the research focus from improving prediction accuracy
alone to enhancing robustness, uncertainty quantification, and cyber resilience. Modern forecasting
frameworks are designed to better represent price spikes, maintain reliable performance under
incomplete or noisy market information, and quantify predictive uncertainty under rapidly
changing operating conditions. More importantly, forecasting is no longer viewed solely as a
prediction task. By integrating probabilistic forecasting with anomaly detection and confidence
assessment, forecasting systems can distinguish normal market behavior from anomalies caused by
cyberattacks, corrupted measurements, communication failures, or abnormal operating conditions.
As a result, \ac{EPF} evolves from a passive prediction tool into an active cyber-physical intelligence
capability that simultaneously supports market prediction, anomaly detection, and resilient
operational decision making. Table~\ref{tab:objective_shift} summarizes the corresponding change in
forecasting objectives.

\begin{table}[!t]
\caption{Transition from Conventional Forecasting Objectives to Resilient EI Objectives}
\label{tab:objective_shift}
\centering
\small
\begin{tabular}{p{5.3cm}p{10.5cm}}
\toprule
\textbf{Conventional Objective} & \textbf{Resilient EI Objective} \\
\midrule
Minimize average forecasting error & Support reliable operational decision-making \\
Produce deterministic point forecasts & Provide probabilistic forecasts with calibrated uncertainty \\
Fit historical data & Adapt to changing operating conditions and extreme events \\
Forecast a single market or pricing node & Generalize across multi-region and multi-energy systems \\
Operate as a standalone prediction tool & Integrate forecasting, optimization, and control \\
Update models offline & Support continuous online learning, monitoring, and feedback \\
Rely primarily on historical data & Incorporate physical constraints, network topology, and cyber trust \\
Evaluate forecasting accuracy alone & Evaluate reliability, robustness, resilience, trustworthiness, and operational value \\
\bottomrule
\end{tabular}
\end{table}

The transition summarized in Table~\ref{tab:objective_shift} establishes the foundation for the remainder of this
report. The key question is no longer which model produces the lowest average error in a static
benchmark. The more important question is which forecasting architecture can support reliable,
secure, and adaptive market operation under uncertainty.

\subsection{AI for Resilient Electricity Price Forecasting}
\begin{figure}[H]
\centering
\includegraphics[width=0.65\linewidth]{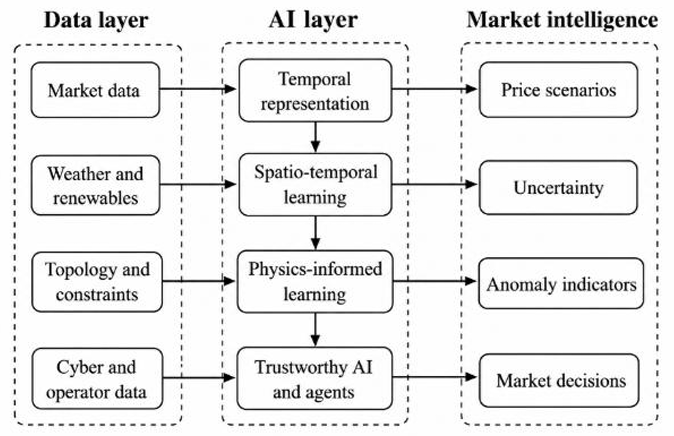}
\caption{Capability-based AI architecture for resilient electricity price forecasting. AI methods should be selected according to the EI functions they support: temporal learning, topology awareness, uncertainty quantification, anomaly detection, and decision support.}
\label{fig:capability_arch}
\end{figure}

\ac{AI} has transformed \ac{EPF}, but its role should not be interpreted as a simple replacement of
classical time-series methods with larger neural networks. Its deeper contribution is representation
learning: the ability to map heterogeneous data streams into latent variables that capture market
states, temporal patterns, topology-dependent congestion, uncertainty, and abnormal behavior
\cite{ref100}. In \ac{EI} systems, this capability is essential because prices emerge from interactions among
network constraints, renewable forecasts, storage decisions, fuel systems, market offers, and cyber
infrastructure. Fig.~\ref{fig:capability_arch} presents a capability-based architecture for \ac{AI}-enabled \ac{EPF}. The
architecture begins with heterogeneous data ingestion, passes through representation-learning
modules, and produces market intelligence outputs for operational applications.

As shown in Fig.~\ref{fig:capability_arch}, \ac{AI} forecasting systems must process market, weather, topology,
and cyber data simultaneously. Deep learning architectures such as recurrent networks, temporal
convolutional networks, attention models, and transformers are useful because they capture
temporal dependencies across hours, days, seasons, and market cycles \cite{ref101,ref102}. However,
temporal learning alone is insufficient. In \ac{LMP} markets, congestion propagates through
transmission networks according to electrical connectivity and binding constraints. \ac{GNN}s and
\ac{GAT}s can exploit this structure by representing buses or pricing nodes as graph
vertices and transmission lines or learned correlations as graph edges \cite{ref103,ref104}. Physics-informed models can further constrain learning to respect power-flow relationships, operational
limits, and market-clearing logic \cite{ref105}.

Table~\ref{tab:ai_capabilities} summarizes the operational capabilities of \ac{AI} forecasting methods,
grouped by function rather than chronology, emphasizing that modern forecasting platforms will
likely combine multiple methods rather than rely on a single model class.

\begin{table}[!t]
\caption{AI Capabilities for Resilient EPF in EI Systems}
\label{tab:ai_capabilities}
\centering
\small
\begin{tabular}{p{3cm}p{4cm}p{6.5cm}p{2.5cm}}
\toprule
\textbf{Capability} & \textbf{Representative Technologies} & \textbf{Forecasting Contribution} & \textbf{EI Relevance} \\
\midrule
Nonlinear Temporal Learning & DNNs, LSTM, TCNs & Captures nonlinear price dynamics, seasonality, and temporal dependencies & High \\
Long-Range Temporal Reasoning & Attention, Transformers & Learns long-term dependencies, market regimes, and multi-horizon dynamics & Very High \\
Topology-Aware Learning & GNNs, GATs & Captures transmission topology, congestion propagation, LMP spatial correlations & Very High \\
Physics-Informed Learning & PINNs, constrained learning, differentiable optimization & Embeds physical laws, market constraints, and power system knowledge & Critical \\
Uncertainty Modeling & Quantile regression, Bayesian learning, deep ensembles, conformal prediction & Generates calibrated probabilistic forecasts for risk-aware decisions & Critical \\
Cyber-Resilient Learning & Robust learning, anomaly detection, adversarial learning & Detects corrupted measurements, cyberattacks, abnormal behavior & Critical \\
Foundation Model Intelligence & Multimodal foundation models, LLMs, AI agents & Integrates numerical data, text, operator knowledge for decision support & Emerging \\
\bottomrule
\end{tabular}
\end{table}

Table~\ref{tab:ai_capabilities} also clarifies why the future of \ac{EPF} is not simply ``larger models''. A model with
excellent temporal accuracy but no topology awareness may fail during congestion events. A model
with low average error but poor calibration may mislead storage or reserve decisions. A model that
cannot detect corrupted inputs may amplify cyberattacks. Therefore, resilient forecasting requires
combining temporal representation learning, physical consistency, uncertainty quantification, and
cyber trust.

\subsubsection{Representation Learning and Market Dynamics}
Representation learning addresses the fact that electricity prices are driven by hidden
market states rather than by observable prices alone. Recurrent neural networks and \ac{LSTM}
models represent sequential dependence, making them useful for short-term market forecasting and
abnormal event detection \cite{ref106,ref107}. \ac{TCN}s provide stable long-receptive-field modelling and can be more computationally efficient for large-scale datasets \cite{ref108}.
Attention mechanisms and transformers allow models to assign varying importance to historical
intervals, exogenous variables, and market contexts, making them attractive for multi-horizon
forecasting \cite{ref109,ref110}. From a power-system perspective, the value of representation learning is
not merely that it improves benchmark accuracy. Its value is that it can learn latent drivers such as
scarcity conditions, renewable ramping events, congestion regimes, and price-spike precursors that
are difficult to encode manually. However, representation learning must be coupled with
operational insight. For example, a model may learn that evening net-load ramps in high-solar
systems often precede price volatility, but an \ac{ISO}/\ac{RTO} user also needs to understand whether the
forecast is driven by load uncertainty, renewable uncertainty, transmission congestion, or reserve
scarcity. This motivates interpretable attention, feature attribution, and physics-informed model
diagnostics.

\subsubsection{Physics-Informed and Topology-Aware Learning}
Electricity prices differ from many commodity prices because they are produced by
constrained network optimization. The \ac{LMP} at a pricing node can be decomposed conceptually
into an energy component, congestion component, and loss component, and congestion patterns
are determined by network topology, power-flow constraints, and dispatch economics \cite{ref111}.
Therefore, topology-aware forecasting is essential for \ac{EI} applications, especially when forecasting
nodal prices across thousands of buses.

\ac{GNN}s provide a natural modeling framework because they propagate information along graph
structures. In \ac{LMP} forecasting, graph edges may represent physical transmission connectivity,
electrical distance, power-transfer distribution factors, or learned price correlations. This allows the
model to learn how a renewable forecast error, generator outage, or transmission constraint in one
area can affect prices in neighboring or electrically coupled areas. Physics-informed learning
further improves robustness by embedding power-flow relationships, market constraints, or
differentiable optimization layers into the model \cite{ref103,ref112}. For \ac{ISO}/\ac{RTO} practice, these methods
are promising because they align model structure with grid structure rather than treating each node
as an independent time series.

\subsubsection{Probabilistic Forecasting and Uncertainty Quantification}
Deterministic price forecasts are insufficient for resilient \ac{EI} operation because the most
consequential events are often rare, high-impact, and difficult to predict exactly. Probabilistic
forecasting provides predictive distributions, quantiles, or scenarios that support risk-aware
bidding, storage scheduling, reserve procurement, and resilience analysis \cite{ref113}. This is especially
important for price spikes and scarcity events, where the cost of underestimating risk can be much
larger than the cost of ordinary forecast errors.

Renewable uncertainty has become one of the dominant drivers of electricity price
volatility in modern power systems. Consequently, forecasting methodologies have evolved from
deterministic point prediction toward probabilistic frameworks capable of quantifying forecast
uncertainty and supporting risk-aware operational decisions. Recent advances have incorporated
spatial-temporal dependence, weather scenario generation, ensemble learning, predictive
distribution optimization, and adaptive reserve estimation to better characterize the uncertainty
associated with renewable generation. These developments are directly relevant to \ac{EPF} because
uncertainty in wind and solar generation propagates through net load, reserve requirements,
transmission congestion, and ultimately market clearing. More recently, robust forecasting
frameworks have been developed to improve forecasting reliability under highly volatile market
conditions by explicitly addressing price spikes, incomplete market information, and data
uncertainty. Collectively, these approaches represent an important step toward uncertainty-aware
forecasting that supports resilient operation of future \ac{EI} systems.

Modern probabilistic methods include quantile regression \cite{ref114}, Bayesian neural networks
\cite{ref115}, deep ensembles \cite{ref116}, scenario generation \cite{ref117}, and conformal prediction \cite{ref118}. For
\ac{ISO}/\ac{RTO} applications, calibration is as important as sharpness. A narrow-forecast interval
that frequently misses price spikes is operationally dangerous, while an overly conservative
interval may lead to inefficient reserve or hedging decisions. Therefore, probabilistic \ac{EPF} should
be evaluated using proper scoring rules, reliability diagrams, prediction-interval coverage, and
downstream operational value.

\subsubsection{Cyber-Resilient Forecasting and Anomaly Detection}
Digitalization improves observability but also expands the cyberattack surface of
electricity markets. Forecasting systems may rely on market data, \ac{SCADA} measurements, \ac{PMU}
streams, \ac{AMI} data, weather feeds, \ac{DERs} telemetry, and cloud services \cite{ref119}. These inputs can be
affected by missing data, communication failures, \ac{FDIA} attacks, adversarial manipulation, or
compromised forecasting services. If such inputs are trusted blindly, forecasting models may
produce misleading prices and propagate errors into bidding, scheduling, and reliability decisions.

Cyber-resilient \ac{EPF} requires models that not only predict future market outcomes but also
continuously assess the trustworthiness of incoming data and the reliability of their own
predictions. Rather than treating forecasting and cyber security as independent functions, recent
research has increasingly integrated probabilistic forecasting with anomaly detection to distinguish
normal market behavior from abnormalities caused by \ac{FDIA} attacks, communication failures,
compromised renewable forecasts, or corrupted measurements. By combining deterministic
forecasting, probabilistic prediction, and uncertainty-aware anomaly detection, these frameworks
enable forecasting systems to identify abnormal market conditions while providing calibrated
confidence measures for operational decision making. Consequently, forecast distributions become
valuable cyber-physical security indicators, supporting both market prediction and real-time
monitoring of data integrity and system resilience.

\begin{figure}[H]
\centering
\includegraphics[width=0.7\linewidth]{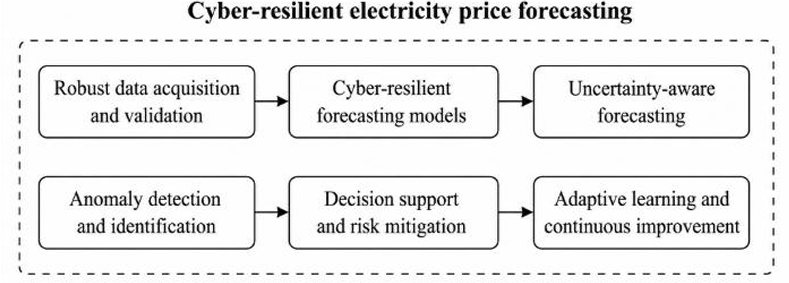}
\caption{Cyber-resilient EPF framework. Probabilistic forecasts provide both market predictions and anomaly indicators for detecting corrupted or abnormal cyber-physical inputs.}
\label{fig:cyber_resilient}
\end{figure}

Fig.~\ref{fig:cyber_resilient} summarizes the cyber-resilient forecasting concept. The framework separates
data assimilation, probabilistic forecasting, anomaly scoring, and decision support. When the
anomaly score is high, the system can trigger operator review, robust optimization, fallback models,
or additional cyber diagnostics. This highlights an important design principle: forecasting and
cyber monitoring should not be separate afterthoughts. A resilient forecasting platform must
understand when it does not trust its inputs or its own predictions. This requirement is central to
secure \ac{EI} operation, where automated market decisions may be made faster than manual operator
review.

\subsubsection{Foundation Models and Autonomous Market Intelligence}
Foundation models and \ac{LLM}s create new possibilities for
electricity market intelligence because they can integrate numerical data, textual information,
market rules, operator procedures, outage reports, weather narratives, and historical cases
\cite{ref120,ref121,ref122}. For \ac{EPF}, foundation models may support transfer learning across markets, few-shot adaptation to new pricing nodes, multimodal reasoning over weather and market data, and
retrieval-augmented decision support using market manuals and historical event reports
\cite{ref123,ref124}.

However, foundation models also introduce new risks. They may hallucinate explanations,
fail under out-of-distribution events, or be vulnerable to prompt injection and data poisoning.
Therefore, their use in \ac{ISO}/\ac{RTO} environments must be constrained by verified data sources,
retrieval systems, uncertainty estimates, human-in-the-loop workflows, and physics-informed
checks. A practical near-term role is not autonomous market clearing, but decision support:
summarizing forecast drivers, explaining congestion scenarios, generating operator-facing
diagnostics, and assisting analysts in comparing historical analogs.

\subsection{Electricity Price Forecasting for Secure Market Operation}
The preceding subsections establish that \ac{EPF} is no longer only a statistical exercise. In
large-scale \ac{EI} systems, price forecasts influence how flexible resources are scheduled, how market
participants hedge risk, how operators anticipate congestion and scarcity, and how cyber-physical
abnormalities are detected. Therefore, the final role of \ac{EPF} is operational: it must support secure
and resilient market operation under uncertainty, digitalization, and potentially adversarial
conditions. This subsection connects forecasting capability to grid resilience, cyber monitoring,
market security, and operator-facing intelligence.

\subsubsection{Forecasting for Secure Market Operation}
Secure market operation requires reliable decisions under physical constraints, market
rules, and incomplete information. In day-ahead markets, price expectations influence generator
offers, virtual bids, storage schedules, and demand response commitments. In real-time markets,
short-horizon price forecasts are relevant to storage dispatch, flexible load control, reserve
deployment, and congestion management. For \ac{ISO}s/\ac{RTO}s, forecasting also supports market
surveillance by identifying unexpected price patterns, abnormal congestion, or deviations between
expected and realized scarcity conditions.

This motivates a decision-centric view of \ac{EPF}. Conventional model development often
minimizes statistical loss functions such as mean absolute error. However, the operational cost of
a forecasting error is asymmetric. Underestimating a scarcity event may lead to insufficient reserve
positioning or missed demand response, while overestimating ordinary prices may cause inefficient
commitment or unnecessary hedging. Similarly, missing congestion at a critical interface may be
more consequential than small errors at uncongested nodes. Therefore,
forecasting objectives should be aligned with downstream applications, including storage revenue,
reserve adequacy, congestion exposure, and risk-adjusted operating cost. Table~\ref{tab:secure_market}
maps forecasting outputs to secure market-operation functions.

\begin{table}[!t]
\caption{Forecasting Outputs for Secure Market Operation}
\label{tab:secure_market}
\centering
\small
\begin{tabular}{p{3.5cm}p{6.5cm}p{6.5cm}}
\toprule
\textbf{Market Function} & \textbf{Useful Forecasting Outputs} & \textbf{Security and Resilience Contribution} \\
\midrule
SCUC and SCED support & Price scenarios, scarcity probabilities, congestion forecasts & Anticipates stressed dispatch conditions and reserve shortages \\
Battery and flexible load scheduling & Quantiles of price spreads, event duration, uncertainty & Improves deployment of storage and demand response \\
VPP and DER coordination & Locational price forecasts, uncertainty intervals, flexibility estimates & Coordinates DER while reducing impact of uncertainty and noise \\
Congestion management & Nodal LMP forecasts, congestion-component forecasts, risk indicators & Supports proactive congestion mitigation and transmission risk management \\
FTR/CRR analysis & Congestion-price forecasts, scenarios, risk metrics & Enhances congestion hedging and financial risk management \\
Market surveillance & Anomaly scores, confidence measures, explanatory indicators & Detects abnormal market behavior, data corruption, potential manipulation \\
Operator decision support & Forecast narratives, confidence levels, event analogs, recommendations & Improves situational awareness and informed decision-making \\
\bottomrule
\end{tabular}
\end{table}

Table~\ref{tab:secure_market} also demonstrates why secure market operation requires more than a single
forecast trajectory. Operators and market participants need information about confidence, drivers,
and plausibility. For example, a high price forecast caused by forecasted scarcity should be
interpreted differently from a high price forecast caused by suspicious telemetry or
a model extrapolation outside its training distribution. Thus, explanation and trust scoring should
accompany forecasts whenever they are used in security-relevant decisions.

\subsubsection{Toward Operational Market Intelligence}
The long-term trajectory of \ac{EPF} is toward operational market intelligence: systems that
combine forecasting, uncertainty quantification, cyber monitoring, optimization, and human-AI
interaction. Foundation models and \ac{LLM}s may contribute to this trajectory by integrating
numerical time series with textual market manuals, outage reports, operator logs, weather
discussions, reliability assessments, and historical event reports. Retrieval-augmented generation
can help ensure that operator-facing explanations are grounded in verified market rules and
historical evidence rather than unsupported model outputs. In a practical \ac{ISO}/\ac{RTO} environment,
the near-term role of foundation models is likely to be decision support rather than autonomous
control. A market-intelligence assistant could summarize why a forecasted congestion pattern
resembles a previous outage event, identify which weather drivers are contributing to price-spike
probability, retrieve relevant market-manual procedures, or compare forecast scenarios against
historical analogs.

Digital twins can provide a complementary environment in which forecasting models,
optimization solvers, and operator workflows are tested before deployment in real systems.
However, operational market intelligence must remain constrained by power-system physics,
market rules, cybersecurity requirements, and human accountability. Foundation models can
hallucinate, extrapolate incorrectly, or be manipulated through data poisoning or prompt injection.
Therefore, \ac{EI} applications require verified data pipelines, model validation, uncertainty calibration,
physics-informed checks, role-based access control, and human-in-the-loop review. The most
credible path forward is not fully autonomous market operation, but trustworthy augmentation of
analysts and operators.

\ac{EPF} is therefore evolving from a statistical prediction tool into an intelligent cyber-physical
reasoning layer for \ac{EI} markets. Its value will increasingly be measured by whether it improves
operational decisions, detects abnormal conditions, supports resilient flexibility deployment, and
helps operators maintain secure market operation under uncertainty and cyber risk.
\section{Adversarial Risks from AI Integration to Energy Systems}
\subsection{Introduction}
The use of \ac{AI}, particularly in the energy sector, is seen as a key technology to enable the
net-zero goals set up several nations. \ac{AI}-based techniques are rapidly evolving as powerful tools
for real-time grid control, operational decision support, predictive maintenance analysis, regulatory
compliance monitoring, etc. At the same time, power grids are also increasingly witnessing cyber
threats targeting their operation. The decentralization of power systems (such as \ac{DERs}) has
significantly expanded the attack surface, and impact on the grid operation is also increasing due
to tight coupling of cyber and power grid and interaction at faster time scale. In this context, the
use of \ac{AI} technologies can be a double-edged sword. Despite the significant capabilities they bring
out, \ac{AI}-based systems can themselves become attractive targets for cyber attackers. Recent studies
have shown that these \ac{ML} models are also vulnerable to adversarial attacks, in which carefully
crafted perturbations are added to the original attack measurements to manipulate the prediction of
the deployed \ac{AI} model while remaining difficult to distinguish from legitimate inputs \cite{ref125}. The
problem is further exacerbated by the adoption of frontier \ac{AI} models, such as \ac{LLM}s, for critical
applications including operator decision support. In such settings, vulnerabilities to attacks such as
prompt injection may compromise model reliability, potentially leading to unsafe or misleading
actions that threaten power-grid security and resilience \cite{ref126,ref127}. Table~\ref{tab:ml_apps} lists some popular
\ac{ML} applications in \ac{EPS}, and the potential consequence of adversarial attacks targeting these
applications.

\begin{table}[!t]
\caption{ML Applications in Power Systems and Adversarial Attack Consequences}
\label{tab:ml_apps}
\centering
\small
\begin{tabular}{p{5cm}p{11cm}}
\toprule
\textbf{ML Application} & \textbf{Potential Adversarial Consequence} \\
\midrule
Forecasting & Incorrect generation scheduling and reserve planning \\
Predictive Maintenance & Premature maintenance or unexpected equipment failures \\
Cyber security & Cyberattacks evade detection \\
Autonomous Grid Control & Unsafe or destabilizing control actions \\
Customer Support Chatbots (LLMs) & Customers receive incorrect billing, service information, or credential theft \\
Operator Decision Support Assistants (LLMs) & Engineers and operators receive incorrect operational recommendations \\
\bottomrule
\end{tabular}
\end{table}

\subsection{Adversarial Attacks Against Machine Learning in Power Systems}
Early studies on adversarial attacks against power systems largely adapted attack
generation techniques developed for computer vision, such as \ac{FGSM},
to construct perturbations capable of deceiving \ac{ML}-based detectors \cite{ref128,ref129}.

In image-processing tasks, the objective is to generate visually imperceptible perturbations
that alter the model prediction while preserving the semantic content of the image. While these
techniques are broadly applicable to \ac{ML} in power systems, there lies an important distinction.
Unlike adversarial attacks in image processing, adversarial attacks in power systems are
constrained not only to bypass the \ac{ML} model's detection, but also by the underlying physical laws
governing power system operation \cite{ref130}. Specifically, the perturbation must remain within the
feasible attack space defined by the physics of the system's measurement model so that the resulting
attack continues to bypass the conventional detection methods (such as the power grid \ac{BDD}),
while simultaneously misleading the \ac{ML}-based detector.

\begin{figure}[H]
\centering
\includegraphics[width=1.0\linewidth]{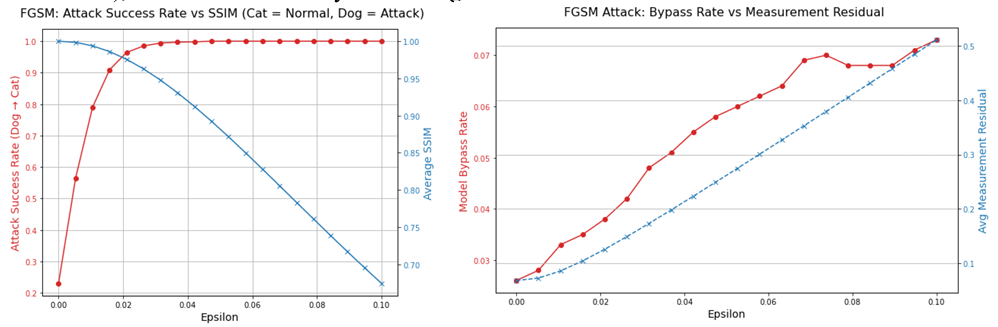}
\caption{Comparison of the impact of FGSM adversarial perturbations on conventional image classification and power system FDIA detection (Fig. Source \cite{ref130}).}
\label{fig:fgsm}
\end{figure}

Fig.~\ref{fig:fgsm} provides a comparison of the impact of \ac{FGSM} adversarial perturbations on conventional
image classification and power system \ac{FDIA} detection. In the image classification task, increasing
the perturbation strength rapidly improves the attack success rate while only slightly reducing the
\ac{SSIM}, enabling effective attacks with imperceptible visual distortion. In
contrast, for \ac{FDIA} detection, the attack success rate remains at a low level, while the measurement
residual increases steadily with the perturbation magnitude. Before a sufficiently high attack
success rate can be achieved, the measurement residual exceeds the \ac{BDD} threshold, causing the
malicious measurements to be detected. Consequently, the adversarial attack cannot simultaneously
achieve high effectiveness and maintain stealthiness. This comparison highlights the fundamental
difference between the two domains. These results indicate that gradient-based perturbations that
are highly effective for attacking image classifiers may no longer remain effective against \ac{FDIA}
detectors in power systems because they must simultaneously satisfy the physical constraints and
stealthiness requirements of the power system.

Thus, straightforward application of adversarial techniques to \ac{ML} in power systems is not
effective. Nevertheless, researchers have designed adversarial attacks that remain undetected by
the \ac{BDD}, by modifying the \ac{CW} attacks. Under this method, an \ac{A-FDIA}
is generated by adding carefully designed perturbations to an
existing \ac{BDD}-bypassing \ac{FDIA}, enabling it to evade both the AI detector and the underlying
physics-based detection mechanism simultaneously.

Beyond satisfying the stealthiness constraints imposed by the power system and \ac{AI}-based
detector, \ac{A-FDIA}s must also preserve the objective of the original \ac{BDD}-bypassing \ac{FDIA}. Unlike
image classification, where the attack is considered successful once the classifier is misled,
adversarial perturbations in power systems should not significantly alter the physical impact of the
underlying attack. Otherwise, the perturbation may reduce the effectiveness of the original \ac{FDIA},
defeating the purpose of constructing the adversarial example. To characterize this effect,
the \ac{CAI} metric has been introduced, which quantifies the extent to
which the adversarial perturbation modifies the original attack vector. A \ac{CAI} value close to
one indicates that the original attack objective is largely preserved while simultaneously bypassing
both \ac{AI}-based and physics-based detection mechanisms. Existing studies have shown that
improving the success rate of \ac{A-FDIA}s generally comes at the cost of a lower \ac{CAI}, indicating that
stronger adversarial perturbations tend to compromise the effectiveness of the original \ac{FDIA}. This
fundamental trade-off further distinguishes adversarial attacks in power systems from those in
conventional image classification tasks.

\subsection{Defending Against Adversarial Attacks}
To improve the robustness of \ac{AI}-based detectors against adversarial attacks, a wide range
of adversarial defense techniques have been proposed in the \ac{ML} literature. For example, adversarial
training incorporates adversarial examples into the training process to improve model
robustness \cite{ref131}. Gradient masking obscures the gradient information exploited by gradient-based
attacks. Input transformation techniques preprocess the input measurements to suppress adversarial
perturbations before they reach the detector \cite{ref132}. These approaches have subsequently been
adapted to power system applications to strengthen \ac{ML}-based detectors against \ac{A-FDIA}s \cite{ref133}.
However, despite their effectiveness against known attack strategies, these
defenses remain fundamentally static, as the deployed \ac{AI} model and its decision boundary remain
unchanged after deployment.

The static nature of these defenses makes them vulnerable to adaptive adversaries. By
continuously probing the deployed detector and incorporating knowledge of the defense
mechanism into the attack generation process, adaptive attackers can progressively construct
adversarial examples that bypass the protected model. Representative optimization-based attacks,
such as the \ac{CW} attack, explicitly exploit the characteristics of the deployed
defense during attack generation and have been shown to bypass several existing static defense
mechanisms. Consequently, many existing defenses provide robustness primarily against
previously observed attack patterns rather than adaptive attacks that explicitly optimize against the
deployed detector \cite{ref134}.

The limitations of static adversarial defenses are particularly pronounced in power system
applications. Since power grid monitoring and \ac{SE} operate continuously over long-time horizons,
attackers often have sufficient opportunities to conduct prolonged
reconnaissance, observe the behavior of the deployed \ac{AI} model, and accumulate knowledge of
both the detection algorithm and the underlying physical system. This allows attackers to jointly
exploit vulnerabilities in the \ac{AI} model and the cyber-physical characteristics of the power grid to
construct increasingly sophisticated adversarial \ac{FDIA}s. Consequently, improving the robustness of
\ac{AI} models through one-time model hardening alone is generally insufficient for providing long-term protection, highlighting the need for defense mechanisms that continuously evolve together
with the attacker.

A promising approach to addressing adaptive adversarial attacks is \ac{MTD}. Unlike
conventional adversarial defense techniques that attempt to harden a fixed \ac{AI} model, \ac{MTD} adopts
a fundamentally different philosophy by continuously invalidating the attacker's knowledge rather
than attempting to identify every possible attack. The key idea is to introduce periodic or event-triggered controlled changes to either the cyber or physical components of the power grid, such that
the information collected during the attacker's reconnaissance phase becomes obsolete before it can
be exploited to construct stealthy attacks. As a result, adversarial perturbations generated using
outdated system knowledge or learned model characteristics become ineffective, substantially
increasing the attacker's cost and reducing the probability of successful attacks.

In contrast to traditional defense mechanisms that primarily focus on attack detection,
\ac{MTD} can simultaneously provide prevention, detection, and mitigation capabilities. If attackers
become aware that \ac{MTD} is deployed, the uncertainty introduced by the continuously changing
system configuration significantly increases the complexity and cost of launching successful
attacks, thereby discouraging malicious activities. If attackers remain unaware of the \ac{MTD}
activation, they will construct attacks using outdated knowledge, causing the resulting attacks to
become detectable by existing protection mechanisms. Furthermore, \ac{MTD} can also be employed
to mitigate the impact of ongoing attacks by dynamically reconfiguring system resources after an
attack is identified. \looseness=-1

The effectiveness of \ac{MTD} against adversarial attacks is particularly evident in power
systems because both stealthy FDI attacks and adversarial \ac{FDIA}s fundamentally depend
on accurate knowledge of the system measurement model \cite{ref135}. As discussed previously,
adversarial perturbations must satisfy the physical constraints imposed by the power system
Jacobian matrix in order to simultaneously evade the \ac{BDD} and \ac{AI}-based detection models.
Consequently, once the system topology, transmission line reactances, measurement configuration,
or \ac{AI} detection model is dynamically changed, the previously learned attack space and decision
boundary become invalid. This prevents attackers from directly transferring previously generated
adversarial perturbations to the updated system, thereby significantly reducing the transferability
and effectiveness of adversarial attacks. \looseness=-1

Existing studies have developed several categories of \ac{MTD} for power grid security \cite{ref135}.
Physics-based \ac{MTD} periodically perturbs physical system parameters, such as transmission
line reactances, measurement configurations, or distributed energy resource \ac{DERs} control signals,
thereby invalidating the attacker's knowledge of the underlying power system model \cite{ref136}.
Network-based \ac{MTD} dynamically reconfigures communication paths, \ac{IP} addresses, or \ac{SDN} configurations to increase uncertainty during cyber reconnaissance.
More recently, ML-based \ac{MTD} has emerged to specifically defend against adversarial \ac{ML} attacks
by dynamically generating multiple \ac{AI} detectors with different decision boundaries. Since
adversarial examples generated for one detector exhibit limited transferability to another,
continuously changing the deployed \ac{AI} model substantially reduces the success probability of
adversarial attacks while maintaining satisfactory detection performance. \looseness=-1

\begin{figure}[H]
\centering
\includegraphics[width=0.7\linewidth]{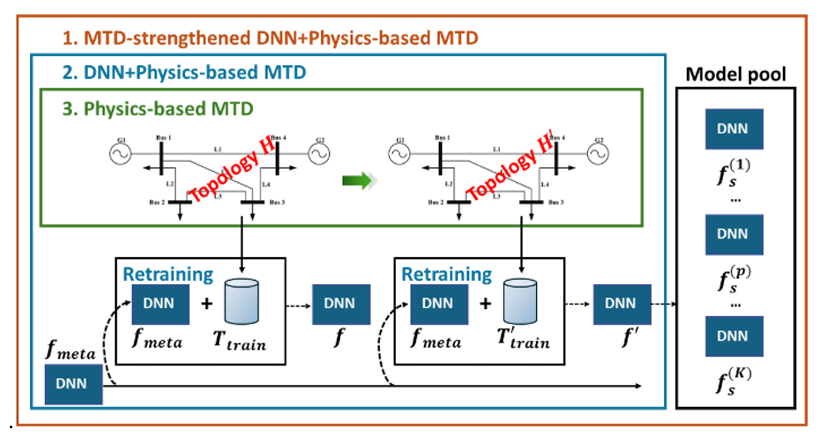}
\caption{MTD Implementation for power systems (Fig. Source \cite{ref137}).}
\label{fig:mtd}
\end{figure}

Fig.~\ref{fig:mtd} presents a representative framework that integrates physics-based \ac{MTD} with
ML-based \ac{MTD} to defend against adversarial attacks in power systems. The ML-based component deploys a pool of diverse detection models and combines their outputs through
majority voting, exploiting the limited transferability of adversarial examples across models with
different decision boundaries. Meanwhile, the physics-based \ac{MTD} perturbs the power system
topology to invalidate the attacker's knowledge of the underlying system model and further
improve detection accuracy. By combining these two mechanisms, the framework can achieve
strong defense performance with a smaller model pool and less adversarial training, while requiring
only a relatively low reactance perturbation and therefore reducing the operational cost associated
with physics-based \ac{MTD}.

In conclusion, compared with conventional static adversarial defenses, \ac{MTD} adopts a
fundamentally different strategy by continuously changing the attack surface to invalidate the
attacker's knowledge. This dynamic defense paradigm makes \ac{MTD} a promising approach for
protecting AI-enabled power grids against adaptive adversaries.

This subsection focused on identifying and mitigating threats from \ac{AI} to power grid
operations. It also outlined adversarial-resilient \ac{AI} frameworks designed to detect cyber-attacks,
ensuring robust and trustworthy \ac{AI} deployment.
\section{Trustworthy AI-Driven Cyber-Physical Security for a Resilient Energy Grid}
\subsection{Scope}
This subsection examines \ac{AI} as an enabling capability for cyber-physical security in
Energy Internet-enabled electric grid systems. The focus is not on \ac{AI} as a generic analytics tool,
but on \ac{AI} as part of the operational security architecture of a distributed power system in which
sensing, communication, control, and actuation are tightly coupled. The objective is to identify the
main cyber-physical security challenges introduced by \ac{EI} operation of the grid, describe the classes
of \ac{AI} methods that are technically relevant to those challenges, and outline the architectural
principles required for secure, explainable, and resilient deployment in power system
environments.

The technical scope includes \ac{DERs}, inverter-based assets, microgrids, networked control
systems, transmission-distribution coordination interfaces, and digital platforms that mediate grid
operation through measurement, communication, and automation. Within this scope, particular
emphasis is placed on cyber-physical monitoring, anomaly detection, secure data exchange, model
interpretability, and the role of unsupervised and hybrid \ac{AI} in environments with sparse or evolving
attack data that affect grid observability, control integrity, and operational trust.

\subsection{Background \& Technical Context}
The \ac{EPS} is evolving from a largely centralized network with relatively simple supervisory
coordination into a distributed cyber-physical infrastructure supported by communication
networks, software-defined control, and digitally connected field assets \cite{ref16}. This evolution is
driven by the increasing penetration of \ac{DERs}, advanced sensing, \ac{IBR} control, transactive
coordination, distributed automation, and cloud- or edge-enabled decision platforms. The resulting
system does not operate only through physical power flow. It operates through the continuous
interaction of energy transfer, information exchange, and control action across the grid \cite{ref138}.

The term Energy Internet is useful in this setting because it captures that interaction
explicitly. Physical state, cyber state, and control state are now interdependent. Voltage, frequency,
current, and \ac{PF} remain central to system behavior, but they no longer define that behavior by
themselves. Controller logic, communication timing, telemetry integrity, software state, device
interoperability, and platform trustworthiness now influence how the system responds to
disturbances, how it coordinates distributed assets, and how quickly it can restore service after
degradation.

A useful technical abstraction is to describe the system as a coupled cyber-physical process.
Let the grid state at time $t$ be represented as:
\begin{equation}
x(t) = \begin{bmatrix} x_p(t) \\ x_c(t) \end{bmatrix},
\label{eq:xstate}
\end{equation}
where $x_p(t)$ denotes the physical state variables and $x_c(t)$ denotes the cyber state variables. The
physical portion may include variables such as voltage magnitude, voltage angle, current,
frequency, real power, reactive power, \ac{DERs} operating state, breaker state, or protection status. The
cyber portion may include communication delay, packet retransmission count, protocol commands,
authentication states, controller states, message paths, configuration attributes, and event logs.
Under normal operation, the combined state evolves within a constrained region defined by
physical laws, control objectives, timing requirements, communication quality, and system
topology. Disturbance, failure, or malicious action may force the system outside that region and
thereby alter both grid behavior and cyber observability.

This interpretation has several implications. First, cyber infrastructure is not merely a
support layer. It is part of the operational process of the grid. Second, the cyber and physical layers
cannot be defended effectively in isolation. Third, anomaly detection in such systems requires
reasoning over the combined structure of $x_p(t)$ and $x_c(t)$, not just over one of them \cite{ref139}. These
observations motivate the need for \ac{AI}-driven cyber-physical monitoring that can interpret both
physical response and cyber conditions as part of the same system state.

\subsection{Cyber-Physical Threat Environment in Energy Internet-Enabled Grids Systems}
\subsubsection{Cross-Domain Propagation of Cyber-Physical Attacks}
The defining security property of Energy Internet-enabled grid systems is that cyber events
can cause physical effects, and physical disturbances can generate cyber symptoms. This creates a
threat environment in which attacks are often cross-domain rather than purely digital or purely
physical and in which the operational impact of an event cannot be assessed from one domain alone.

A \ac{FDIA} may alter measurements, protocol fields, or control-state information. If those
corrupted values are consumed by a \ac{DERs} management system, a controller, or a supervisory
application, the resulting action may produce incorrect dispatch, distorted volt-var behavior,
degraded frequency support, or unsafe operating conditions. A \ac{DoS} attack may first appear as a
communication availability problem, but its consequence may be delayed control, stale situational
awareness, or missed protective action. A man-in-the-middle attack may preserve message syntax
while altering payload values, thereby creating an operationally plausible but false view of the
system. A replay attack may reintroduce previously valid measurements or commands into a
context where they are no longer valid for the actual physical state of the grid \cite{ref140}.

More advanced attacks target operational logic directly. An adversary may alter grid-support functions, disable autonomous modes, change setpoint priorities, modify inverter response,
or manipulate edge gateway logic \cite{ref141}. These scenarios are difficult to detect with protocol
compliance checks alone because the messages may remain well-formed. The anomaly becomes
meaningful only when the physical system response no longer matches the expected effect of the
cyber action or when the reported cyber state no longer explains the measured grid response.

The reverse direction is also important. A genuine physical disturbance may induce traffic
bursts, control retries, unusual event timing, or transient inconsistencies in measurements. If these
effects are interpreted without physical context, the system may attribute them to cyber
compromise. That ambiguity delays diagnosis and increases the chance of incorrect
response during time-critical grid operations.

\subsubsection{Heterogeneity, trust boundaries, and system exposure}
The challenge is intensified by heterogeneity. \ac{EI} architectures include utility-owned
infrastructure, customer-owned \ac{DERs}, aggregator-managed portfolios, legacy operational
technology, protocol translators, remote management interfaces, and cloud-connected
services \cite{ref142}. These components differ in communication protocol, authentication support, update
mechanism, telemetry fidelity, and access control maturity. They also differ in who owns them,
who manages them, and what level of trust can be assumed across their interfaces during
monitoring, coordination, and control of the grid.

Security exposure is therefore not determined only by the properties of individual devices.
It is strongly influenced by the communication and control paths that connect them. Measurements
and control actions often pass through gateways, middleware services, message brokers, protocol
translation layers, and cross-domain administrative interfaces. A message may be syntactically
valid but operationally untrustworthy. A controller may authenticate correctly and still act on stale,
manipulated, or context-inconsistent data. A field device may report a nominal status while its
firmware, configuration, or logic has already been altered through a separate access path. This
means that endpoint validation alone is insufficient. Security must be established as an end-to-end
property of sensing, communication, control, and actuation throughout the cyber-physical control
path.

\subsubsection{Limits of conventional detection and monitoring}
Conventional detection methods remain useful, but they are insufficient for this
environment when used in isolation. Signature-based intrusion detection can identify known
exploits and protocol misuse, but it does not generalize well to attacks that remain protocol-compliant while altering physical behavior \cite{ref143}. Physics-based monitoring can identify abnormal
voltage or frequency behavior, unexpected \ac{DERs} response, or protection miscoordination, but it
often cannot determine whether the cause is attack, failure, misconfiguration, or environmental
stress.

The most relevant anomaly often appears as a cyber-physical inconsistency. This includes
cases where: i) a valid command produces an implausible physical response, ii) communications
appear healthy while control performance degrades, iii) reported device state does not match
measured operating behavior, iv) controller actions are consistent with cyber inputs but
inconsistent with power system conditions. These conditions are difficult to capture with single-domain monitoring because the security-relevant signal lies in the mismatch between cyber
coordination and physical response rather than in either one independently.

Such conditions motivate the use of cyber-physical data fusion and AI-based inference,
which are discussed next as mechanisms for learning the normal joint structure of grid behavior
and for detecting that conventional monitoring may miss.

\subsection{AI For Cyber-Physical Security}
\subsubsection{Why AI is necessary}
The technical need for \ac{AI} arises from the structure of the monitoring problem. The
available data are high-dimensional, multimodal, multi-rate, and strongly time-dependent. Relevant
evidence may be contained in \ac{PMU} streams, \ac{SCADA} measurements, relay records, \ac{DERs} telemetry,
protocol exchanges, controller states, network timing statistics, and configuration changes. The
system must not only observe these sources but infer whether their joint
behavior remains consistent with expected operation of the grid under normal, disturbed, and
potentially compromised conditions.

The required decision function can be viewed as:
\begin{equation}
y(t) = f(x_p(t-\tau:t), x_c(t-\tau:t)),
\label{eq:decisionfn}
\end{equation}
where $f(\cdot)$ maps a recent temporal window of physical and cyber observations into an
estimate $y(t)$ of system condition, anomaly state, attack class, or control recommendation. Because
the mapping is nonlinear, context-dependent, and often changes with operating mode, it is difficult
to define with static thresholds or isolated rules alone without losing important information about
the coupling between the cyber and physical layers.

\ac{AI} provides a mechanism to learn such mappings from data. This learning can support
anomaly detection, attack classification, cross-domain state consistency checking, alert
prioritization, local and system-level situational awareness, and response support. Supervised
learning is useful for known patterns. Unsupervised learning is more natural when attack labels are
sparse. Sequence models are important when anomalies unfold over time. Physics-informed and
causal approaches improve operational meaning and model trust by aligning the inference process
with grid dynamics, control behavior, and cyber-physical dependencies.

\subsubsection{Supervised Learning}
Supervised learning is appropriate when representative labeled datasets exist. In the
simplest form, given input-output pairs $(x_i, y_i)$, the model learns a mapping $f_\theta(x)\rightarrow y$ by
minimizing a loss such as:
\begin{equation}
\min_\theta \sum_{i=1}^N \mathcal{L}(f_\theta(x_i), y_i).
\label{eq:supervised}
\end{equation}
Here, $x_i$ may contain fused cyber-physical features, $y_i$ may denote a binary label, a multi-class attack type, or control decision and $\theta$ denotes the model weights. Support vector machines,
random forests, gradient boosting, multilayer perceptrons, and convolutional classifiers all fall into
this category.

Supervised learning is well suited for recurring attack patterns, known protocol misuse,
and benchmarked detection tasks in the grid \cite{ref144}. It is also useful when the goal is to distinguish
among several previously characterized cyber-physical scenarios, for example, separating \ac{FDIA}
behavior from \ac{DoS} effects or distinguishing communication-layer anomalies from physically
induced control anomalies. However, its main limitation is that the training set defines the
recognition space. If unseen attacks, new configurations, or new operating states differ substantially
from the training distribution, the classifier may not generalize well in real-time production
systems. This limitation is serious in \ac{EI} systems because both operating conditions and attack
surfaces evolve with changing \ac{DERs} participation, controller settings, communication paths, and
operating models of the grid.

\subsubsection{Unsupervised Learning \& Semi-supervised Learning}
Unsupervised and semi-supervised learning are especially important in cyber-physical grid
security because normal operating data are abundant, while labeled attack data are sparse and
incomplete. These methods learn the structure of nominal operation and detect deviations from
it without requiring the threat space to be fully enumerated in advance.

A reconstruction-based autoencoder provides a common formulation \cite{ref145}. An
encoder $g_\phi(\cdot)$ maps input $x$ to a latent representation $z$, and a decoder $h_\psi(\cdot)$ reconstructs $x$ as $\hat{x}$.
Training is typically performed on normal data only using:
\begin{equation}
\min_{\phi,\psi} \sum_{i=1}^N \lVert x_i - \hat{x}_i \rVert^2.
\label{eq:autoencoder}
\end{equation}
At inference time, the anomaly score is:
\begin{equation}
e(x^*) = \lVert x^* - \hat{x}^* \rVert^2.
\label{eq:anomalyscore}
\end{equation}
If $e(x^*)$ exceeds a threshold derived from the distribution of training errors, the observation
is treated as anomalous. This formulation is especially effective when the cyber and physical feature
sets are jointly represented in the input. The model then learns the normal relation between the two
and flags cases where that relation breaks down under anomalous grid conditions or adversarial
manipulation.

This is particularly valuable in Energy Internet-enabled grid systems because cyber and
physical abnormalities often appear in combination. A physical signal may remain within
acceptable bounds when viewed alone. A cyber signal may also appear benign when viewed alone.
The abnormality only becomes visible when the normal relation between the two is
violated through inconsistent control response, stale or manipulated telemetry, or degraded
communication performance. In this sense, the autoencoder is not merely a compression model. It
is a cyber-physical structure learner.

Clustering methods, one-class support vector machines, isolation forests, latent density
estimators, and graph-based unsupervised methods are also relevant. Their common strength is that
they detect structure-breaking events without requiring explicit labels for each attack type and can
therefore provide a more practical basis for deployment in evolving grid environments.

\subsubsection{Temporal \& Sequence-Based Models}
Many cyber-physical anomalies are not point anomalies. They are sequence anomalies.
The relevant information lies in the order, persistence, and timing of changes rather than in an
instantaneous value. This is especially true for denial-of-service, replay, timing manipulation,
stealthy false data injection, and coordinated control interference in which the effect on the
grid emerges over a time horizon rather than at a single observation input.

Let a temporal observation window be represented as:
\begin{equation}
X_{t,T} = \{x(t-T+1), x(t-T+2), \dots, x(t)\}.
\label{eq:window}
\end{equation}
A sequence model learns a representation of $X_{t,T}$ that captures both intra-window
evolution and cross-domain relations. \ac{LSTM}-based autoencoders, recurrent neural networks,
transformers, and temporal convolutional models are especially useful for this purpose because
they preserve temporal dependencies between cyber and physical response \cite{ref146}. The anomaly
decision can then be written as:
\begin{equation}
a(t) = \begin{cases} 1, & \text{if } e(X_{t,T}) > \gamma, \\ 0, & \text{otherwise}, \end{cases}
\label{eq:seqanomaly}
\end{equation}
where $e(X_{t,T})$ is the sequence anomaly score and $\gamma$ is the threshold.

These models are well suited to cases where the physical response to a cyber event is
delayed or where an attack manifests through gradual divergence rather than abrupt change. They
can capture situations in which command timing appears valid, but the subsequent physical
response deviates from the learned pattern. They can also identify sequence-level mismatches
between repeated cyber actions and evolving system state such as repeated control requests,
delayed actuator response, or progressive divergence between reported and measured grid behavior.

\subsubsection{Physics-Informed and Hybrid AI Approaches}
Purely data-driven models can perform well, but they often lack awareness of physical
consequence in the grid. Physics-informed \ac{AI} addresses this problem by incorporating engineering
knowledge into the learning process. This may take the form of physical constraints in the loss
function, state-estimation-informed features, digital-twin-based residuals, topology-aware graph
representations, or control-loop-aware feature design that preserves the relation between cyber
observations and physically plausible system response \cite{ref147}.

Hybrid detection architectures are particularly promising. One part of the system may use
model-based checks such as state estimation, \ac{PF} consistency, or relay logic validation. Another
part may use \ac{ML} to detect subtle or distributed anomalies that do not manifest clearly in those
models. The outputs can then be fused to improve both sensitivity and reliability for cyber-physical
threat detection in the grid.

This approach reduces the risk of purely statistical false alarms while preserving the ability
to capture nonlinear and context-dependent cyber-physical behavior that emerges across
measurement, communication, and control layers.

\subsubsection{Generative AI and Operator Support}
Generative \ac{AI} and large language model-based systems should be used with appropriate
operational guardrails in cyber-physical grid security. Even though Generative \ac{AI} is extremely
underexplored for cyber-physical security in the grid, their strongest uses include alert
summarization, structured incident description, support for analyst queries, and synthesis of
evidence across multiple detection modules. They may also assist in embedding or representation
learning for structured system descriptions of grid assets, events, or cyber-physical
dependencies and may increasingly support higher-level reasoning over heterogeneous operational
data \cite{ref148}.

At present, however, their use in power system cybersecurity should remain grounded in
physically meaningful constraints, validated cyber-physical evidence, and controlled decision
workflows. Generative \ac{AI} outputs should be checked against physical rules, operating limits,
protection logic, and trusted model outputs before being used in operational decision-making. In
this sense, generative \ac{AI} is most effective when it is coupled with trustworthy detection, estimation,
and data fusion pipelines rather than used in isolation. Its role is to improve the interaction between
technical security analytics and the human users who must interpret them during monitoring,
diagnosis, and response in power system operations.

This area is evolving rapidly. Recent efforts in power-system-focused \ac{AI} agents, such
as PowerAgent, indicate that \ac{LLM}-based systems may eventually support more
active roles in grid analysis, operator assistance, and cyber-physical decision support, provided they
are constrained by domain knowledge, physical validation, and secure governance mechanisms.

Table~\ref{tab:ai_methods_cps} summarizes each \ac{AI} class and the part of the cyber-physical
security problem it addresses. The implication is that no single \ac{AI} method is sufficient for all
Energy Internet-enabled grid security tasks. A practical architecture will combine several of them
under a layered design in which generative \ac{AI} complements, rather than replaces, physically
grounded cyber-physical security analytics.

\begin{table}[!t]
\caption{AI Methods for the Cyber-Physical Security of the Energy Internet-enabled Grid}
\label{tab:ai_methods_cps}
\centering
\small
\begin{tabular}{p{3cm}p{3.5cm}p{5cm}p{5cm}}
\toprule
\textbf{AI method class} & \textbf{Primary function} & \textbf{Main strength} & \textbf{Main limitation} \\
\midrule
Supervised learning & Attack classification & High accuracy on known classes & Weak generalization to unseen attacks \\
Unsupervised learning & Anomaly detection & Detects unseen deviations from nominal behavior & Limited semantic interpretation without explanation \\
Semi-supervised learning & Normality-driven detection with limited labels & Balances normal-data abundance with sparse attack evidence & Depends on reliable nominal baseline \\
Temporal models & Sequence anomaly detection & Captures delayed and evolving attack effects & Higher training complexity and data demands \\
Physics-informed / hybrid AI & Operationally grounded detection & Preserves physical meaning and reduces false alarms & Requires model and feature engineering \\
Generative AI & Analyst and operator support & Evidence synthesis and context-aware reasoning & Requires operational guardrails and secure workflow constraints \\
\bottomrule
\end{tabular}
\end{table}

\subsection{Explainability, Trustworthiness \& Model Governance}
The increasing use of \ac{AI} in cyber-physical grid security introduces a second challenge
beyond detection accuracy. The model must not only detect abnormal behavior. It must also
provide outputs that are technically interpretable, operationally useful, and sufficiently trustworthy
for deployment in critical infrastructure environments \cite{ref149}. In Energy Internet-enabled grid
systems, this requirement is particularly important because the output of a cybersecurity model may
influence protection logic, operator response, \ac{DERs} control decisions, or broader situational
awareness functions that directly affect grid operation and resilience.

\subsubsection{Explainable AI as a cyber-physical diagnostic layer}
A conventional anomaly score is rarely sufficient on its own. An operator
must determine whether the abnormality is likely associated with communications, controller
behavior, data integrity, protocol misuse, or physical system stress. The operator must
also determine whether the model is reacting to a localized feature deviation, a coordinated cross-domain pattern, or an inconsistency between cyber inputs and physical response in the grid.

Explainable \ac{AI} helps expose why a model produced a given output. In cyber-physical
security, this means revealing which features, time windows, or cross-domain relations contributed
most strongly to a detection. That information supports interpretation, triage, engineering review,
and model debugging while linking the model output back to observable power system behavior.

A generalized explanation can be written as:
\begin{equation}
\mathcal{E}(x) = \{(f_1,w_1), (f_2,w_2), \dots, (f_m,w_m)\},
\label{eq:explanation}
\end{equation}
where $f_i$ denotes a feature or feature group and $w_i$ denotes its local contribution to the model output
for the observation $x$. In grid cyber-physical security, the features may include both cyber and
physical quantities, allowing the explanation to show whether the anomaly is cyber-driven,
physically driven, or rooted in an inconsistency between the two.

\subsubsection{SHAP-based explanation}
One important explainability approach is \ac{SHAP} \cite{ref150}. It provides additive feature
attribution by expressing the model output locally as:
\begin{equation}
f(x) \approx \phi_0 + \sum_{i=1}^M \phi_i,
\label{eq:shap1}
\end{equation}
where $\phi_0$ is the baseline contribution and $\phi_i$ is the attribution associated with feature $i$. The
\ac{SHAP} value $\phi_i$ can be defined as:
\begin{equation}
\phi_i = \sum_{S\subseteq F\setminus\{i\}} \frac{|S|!(|F|-|S|-1)!}{|F|!} \left(f_{S\cup\{i\}}(x) - f_S(x)\right),
\label{eq:shap2}
\end{equation}
where $F$ is the full feature set and $S$ is a subset of features. This formulation is useful because it
quantifies the marginal effect of each feature on the prediction.

In a cyber-physical context, \ac{SHAP} supports separation of cyber and physical evidence. A
high anomaly score driven mainly by communication timing and protocol features suggests a
cyber-centric event. A score driven mainly by physical variables suggests a disturbance or
physically implausible response. Jointly high contributions indicate a genuinely cyber-physical
anomaly and can therefore support root-cause reasoning and response prioritization in grid
operations.

\subsubsection{Custom-made explanations for unsupervised models}
\begin{figure}[H]
\centering
\includegraphics[width=0.7\linewidth]{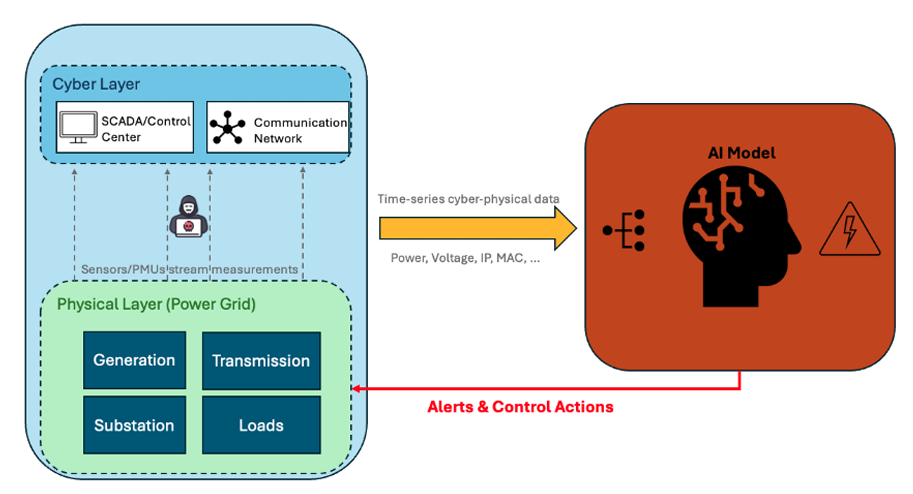}
\caption{Custom-made per-feature explainability \& misclassification detection for LSTM Autoencoders.}
\label{fig:lstm_explain}
\end{figure}

Explainability must also be adapted to unsupervised models such as autoencoders. For a
reconstruction-based model, the anomaly score is:
\begin{equation}
e(x) = \lVert x - \hat{x} \rVert^2,
\label{eq:erecon}
\end{equation}
which can be decomposed feature-wise as:
\begin{equation}
e(x) = \sum_{i=1}^M (x_i - \hat{x}_i)^2.
\label{eq:erecon2}
\end{equation}
This makes it possible to define a per-feature anomaly contribution:
\begin{equation}
e_i(x) = (x_i - \hat{x}_i)^2.
\label{eq:eifeat}
\end{equation}
A normalized comparison to the baseline feature-wise error from normal training data can then be
written as:
\begin{equation}
r_i(x) = \frac{e_i(x)}{\bar{e}_i + \epsilon},
\label{eq:rifeat}
\end{equation}
where $\bar{e}_i$ is the average reconstruction error of feature $i$ under normal conditions and $\epsilon$ is a small
stabilization constant. High values of $r_i(x)$ indicate features that contribute disproportionately to
the anomaly relative to normal behavior. This is especially valuable in cyber-physical detection
because it helps localize whether the anomaly is dominated by cyber features, physical features, or
both and whether the deviation is likely associated with the cyber part (e.g., communications) or
the physical part (e.g., control and sensing). Additionally, this technique helps security
operators and defenders identify potential misclassifications of the AI models, as illustrated in
Fig.~\ref{fig:lstm_explain} in the case of a Long Short-Term Memory Autoencoder that utilizes the per-feature reconstruction
error for explainability \cite{ref151}.

\subsubsection{Trustworthy AI, Uncertainty, and Lifecycle Governance}
Explainability is necessary for \ac{AI}-enabled cyber-physical security, but it is not sufficient
for operational trust. A model may be interpretable and still be unsuitable for deployment if it is
brittle under changing operating conditions, poorly calibrated, sensitive to degraded data quality,
or exposed to insecure update and retraining processes. In Energy Internet-enabled grid systems,
trustworthy \ac{AI} must therefore be treated as a full system property rather than as a characteristic of
model accuracy alone and must be evaluated in terms of its effect on secure and reliable grid
operation.

From a technical standpoint, trustworthy \ac{AI} requires consistent behavior within the
intended operating domain, explicit handling of uncertainty, resilience to corrupted or drifting data,
and secure control of the model lifecycle. This is particularly important in cyber-physical security
because the underlying data distribution is not stationary. Network conditions change. \ac{DERs}
operating modes change. Control configurations evolve. System topology may also change. A
model that performs well during offline evaluation may become unreliable if the deployed
environment no longer matches the assumptions of training or if the cyber-physical relations
learned from past operating data no longer reflect the present grid state.

The trust problem is therefore inseparable from governance. The \ac{AI} component must be
supported by secure data provenance, controlled training and retraining, threshold validation, drift
detection, secure logging, and access control over model assets. The detector itself becomes part of
the cyber-physical attack surface. If training data are manipulated, if model parameters are altered,
or if thresholds drift silently over time, the \ac{AI} system may continue to produce plausible outputs
while no longer providing reliable security value for the grid.

A useful operational view is to represent the deployed detector as:
\begin{equation}
y(t) = f_{\theta(t)}(x(t)),
\label{eq:deployed}
\end{equation}
where the model parameters $\theta(t)$ may evolve through retraining, recalibration, or controlled
updates. Trustworthy deployment requires that such changes remain governed, observable, and
auditable. In practice, this means that model versions, explanation behavior, anomaly-score
distributions, and input data quality should all be monitored as part of the operational security
posture of the power system.

Uncertainty must also be represented explicitly. In critical infrastructure, the model should
not present every inference as equally reliable. Confidence estimates, ensemble disagreement,
explanation consistency, or calibrated anomaly distributions can help distinguish between results
that are strongly supported, results that require cross-checking, and results that fall outside the
model's reliable operating region \cite{ref152}. This distinction is operationally important because false
positives and false negatives have very different consequences in grid cybersecurity and may lead
to either unnecessary intervention or missed mitigation of a cyber-physical threat.

These governance requirements should be aligned with existing cybersecurity and
operational standards. In North America, the discipline imposed by \ac{NERC} \ac{CIP} is directly relevant
for configuration management, access control, logging, recovery, and change control for AI-enabled security functions that influence grid operations. More broadly, the \ac{NIST} Cybersecurity
Framework and NIST SP 800-82 \cite{ref153} provide a useful structure for integrating \ac{AI} into \ac{OT}
security processes across the identify, protect, detect, respond, and recover lifecycle. At the
information governance layer, ISO/IEC 27001 remains useful for defining auditable management
and risk processes \cite{ref154}. For \ac{DERs}-specific environments, \ac{IEEE} 1547 and
emerging \ac{IEEE} 1547.3 guidance are important because they connect interoperability requirements
with communication security, access control, and operational trust in distributed grid
environments \cite{ref155}.

Accordingly, trustworthy \ac{AI} in the electric grid should not be treated as an isolated
analytics capability. It should be managed as an operational cyber asset, governed under the same
discipline applied to other security-critical functions, and continuously evaluated against the
evolving cyber-physical environment in which it operates so that AI-driven cybersecurity remains
technically reliable, operationally interpretable, and aligned with the resilience needs of the Energy
Internet-enabled grid.

\begin{figure}[H]
\centering
\includegraphics[width=0.7\linewidth]{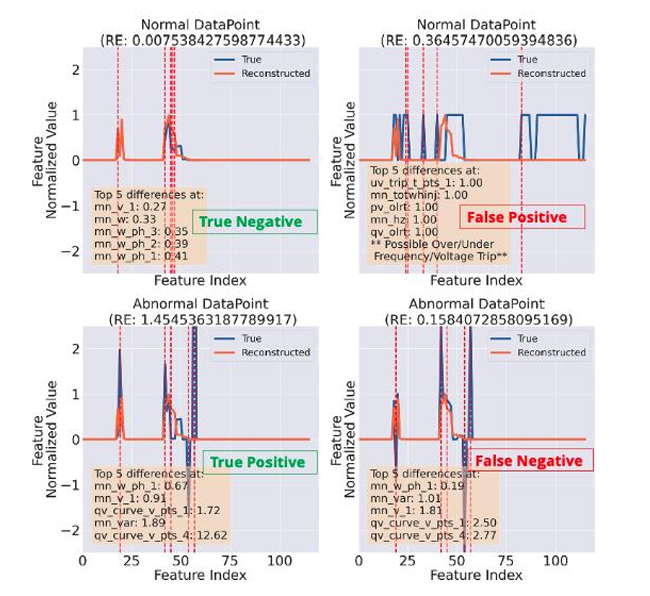}
\caption{AI-driven cyber-physical security pillars for the Energy Internet-enabled grid.}
\label{fig:pillars}
\end{figure}

\subsection{Conclusions \& Recommendations}
Energy Internet-enabled grid systems create a cybersecurity problem that is fundamentally
cyber-physical. Their operating state is shaped jointly by electrical behavior, digital coordination,
and distributed control. As a result, conventional single-domain monitoring is no longer sufficient.
The most relevant anomalies often appear as inconsistencies between cyber inputs, control actions,
and physical response, rather than as isolated cyber events or isolated physical deviations in the
grid.

\ac{AI} is technically important in this environment because it can learn structure across
heterogeneous, time-dependent, and cross-domain data streams. Different classes of \ac{AI} provide
different benefits. Supervised methods are useful for known attack patterns and benchmarked
operating conditions. Unsupervised and semi-supervised methods are especially important for
unknown, evolving, or site-specific threats. Temporal models capture delayed and sequence-based
anomalies. Physics-informed and hybrid methods improve operational relevance by aligning
inference with system behavior. Causal \ac{AI} improves explanation and model orchestration, while
generative \ac{AI} can enhance operator-facing workflows when used in constrained and well-grounded
roles within cyber-physical grid security workflows.

At the same time, several technical needs remain before \ac{AI}-driven cyber-physical security
can be deployed broadly and confidently in \ac{EI} systems. First, the field needs stronger cyber-physical resilience metrics. Detection accuracy alone is not sufficient. Evaluation must also reflect
service continuity, control quality, containment effectiveness, and trusted recovery. Second, there
is a need for stronger \ac{AI} assurance under changing conditions. Models must
be validated under distribution shift, sparse attack data, site-specific operating variation, and
adversarial manipulation. Third, distributed architectures require scalable and secure data
fusion methods that can support the trusted exchange of cyber-physical data, derived features,
explanations, and alerts across multiple trust boundaries. Fourth, operational explainability must
move beyond post hoc interpretation and support online triage and misclassification
identification. Finally, generative \ac{AI} should be explored further for structured summarization,
reasoning support, and operator interaction, but only when grounded in validated cyber-physical
evidence and constrained by secure workflows.

Based on these observations, several recommendations follow. Cybersecurity for Energy
Internet-enabled grid systems should be built around joint cyber-physical inference rather than
isolated cyber or physical monitoring. Unsupervised, temporal, and hybrid physics-informed \ac{AI}
should be prioritized for operational deployment because they are better matched to the structure
of the real problem. Explainability, uncertainty awareness, and trustworthy lifecycle governance
should be treated as core design requirements rather than auxiliary features. \ac{AI} should be deployed
in layered architectures that support local, enclave, and system-level situational awareness. Finally,
standards and governance frameworks should treat AI-enabled cybersecurity functions as
operational infrastructure, not as external analytics add-ons but as integral elements of secure grid
operation and resilience management.

Taken together, these conclusions and recommendations define a practical path toward AI-driven cyber-physical security that is technically grounded, operationally relevant, and appropriate
for the evolving architecture of Energy Internet-enabled electric grid systems with increasing
dependence on distributed intelligence, digital coordination, and cyber-physical observability.

\subsection*{\underline{Acknowledgment:}}
\noindent\textit{Sandia National Laboratories is a multi-mission laboratory managed and operated by National Technology \& Engineering Solutions of Sandia, LLC (NTESS), a wholly
owned subsidiary of Honeywell International Inc., for the U.S. Department of Energy's National Nuclear Security Administration (DOE/NNSA) under contract DE-NA0003525. This written work is authored by an employee of NTESS. The employee, not NTESS, owns the right, title and interest in and to the written work and is responsible for its contents. Any subjective views or opinions that might be expressed in the written work do not necessarily represent the views of the U.S.
Government. The publisher acknowledges that the U.S. Government retains a non-exclusive, paid-up, irrevocable, world-wide license to publish or reproduce the published form of this written work or allow others to do so, for U.S. Government purposes. The DOE will provide public access to results of federally sponsored research in accordance with the DOE Public Access Plan. This material is based upon work supported by the U.S Department of Energy's Office of Cybersecurity, Energy Security, and Emergency Response (CESER), Award Numbers TL0101010-05900-4219057-41059, TL0103030-05900-4219069-52603 and TL0101010-05900-4219057-55697.}
\section{Graph Algorithms for Attack-Resilient Information Routing in the Energy Internet}

Modern large-scale \ac{EI} systems consist of \ac{DERs}, inverter-based generation, energy storage,
and several other heterogeneous entities. The constituents of the \ac{EI} are coordinated through layered
control frameworks that can extend from local microgrids to the wide-area operators \cite{ref156}. These
assets in the grid are numerous and geographically dispersed. Hence, their coordination rests on a
dense communication fabric. \ac{IED}, \ac{PMU}, and \ac{DERs} controllers often exchange measurements and
commands on a continuous basis. Hierarchical as well as distributed control schemes, transactive
energy markets, and edge intelligence all depend on such information reaching their destinations
on time and unaltered. Standards such as IEC 61850 were developed primarily to carry this traffic
across substations and feeders.

While communication is advantageous, dependence on it also exposes the system to one of
its biggest attack surfaces \cite{ref157}. An adversary capable of reading, delaying, or tampering (creating
or modifying) messages can steer the physical system without ever touching a breaker or inverter
directly \cite{ref4,ref57}. While encryption and access control are certainly very helpful security measures,
yet they (both) rest on an assumption that the endpoints and the paths connecting them stay
trustworthy. However, once a device on the path is compromised, it can emit messages that are
cryptographically valid and yet factually wrong. In a general sense, communication networks
supporting the entire \ac{EI} can essentially be represented as a graph. In this graph, information can be
steered along whichever paths are most likely to be clean (`clean' here meaning free from attacked
or compromised elements). This is called graph-based, trust-aware routing. It is a promising
approach for keeping information flowing when parts of the communication network have been
compromised. This subsection explains how networks can be represented as a weighted graph, and
how a combination of intrusion detectors and graph routing algorithms can help achieve cyber-resilience even when an attacker is trying to actively manipulate the system.

\subsection{Graph Representation of Energy Internet Systems}
\begin{figure}[H]
\centering
\includegraphics[width=0.7\linewidth]{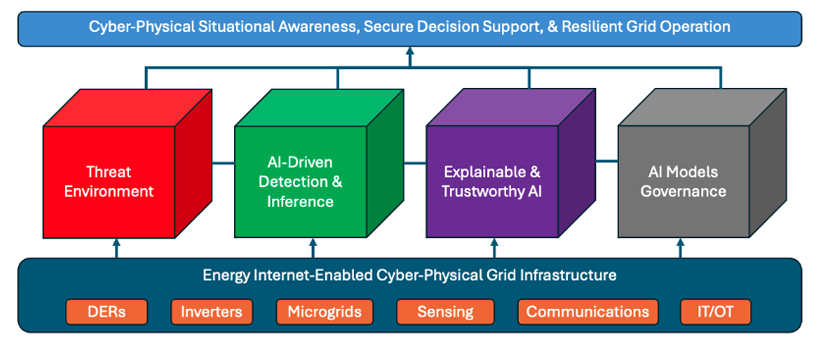}
\caption{Graph representation of a symbolic EI network.}
\label{fig:graph_ei}
\end{figure}

The information exchange framework in the \ac{EI} can inherently be represented as a graph
with nodes, intermediate nodes, and edges (or links, see Fig.~\ref{fig:graph_ei}). In this representation, the
endpoints that produce or consume information become the nodes. Hence, nodes can include
distributed generators, inverters, \ac{IED}, \ac{PMU}, and supervisory controllers, all of which send and
receive information. Between the nodes are the repeaters which can be considered as the
intermediate nodes. Examples of intermediate nodes include routers, gateways, protocol
converters, and communication relays that simply forward traffic without either being its source or
destination. As may be obvious here, the communication lines that join nodes and repeaters are the
edges of the graph.

Every edge can be considered as carrying a weight that expresses some quantity of interest,
such as latency, bandwidth cost, or, in the security setting (as explained below), the degree of
suspected manipulation. Links that carry traffic in both directions are modeled as undirected edges.
One-way telemetry (data exchange) can be modeled as directed edges. Note that modeling here
does not change the routing logic that follows.

Any message traveling from a source-end point to a destination-end point follows a path,
which is simply an ordered sequence of edges through zero or more repeaters. When every edge is
trustworthy, the choice of path is governed only by cost, such as latency or bandwidth. Once certain
links pass through compromised devices, path selection becomes a matter of security rather than
efficiency. This is because in a communication-based critical power grid, security takes precedence
over efficiency for the simple fact that a security failure can corrupt the control traffic itself and
cause physical damage.

This contrasts with an efficiency problem that can typically only degrade performance
(without threatening safety). A key benefit of graph representations is that they provide a basis
for identifying paths that bypass compromised devices or links. For example, if a rootkit-based,
device-level cyberattack \cite{ref158} compromises a particular gateway, the resulting impact is often
limited to the edges (i.e., the communication links) that pass through that device. The edges that do
not rely on it remain typically unaffected. This localization (of the
attacker's immediately manipulable pathway) is what makes rerouting an effective defensive
framework. This rests on the simple fact that if an alternative path that avoids the compromised
region can still deliver information successfully, the network is not required to depend on the
corrupted device to maintain normal operation \cite{ref157}.

Graphs don't just help avoid compromised devices but rather they also provide a concrete
way to measure network resilience. A network in which every pair of endpoints (nodes or vertices)
is connected by only a single pathway is very fragile in terms of resilience. This is because the loss
or compromise of one communication link (edge) can disrupt the communication entirely. On a
similar note, a network offering multiple independent paths proves substantially more robust due
to the possible existence of alternative pathways avoiding tainted/compromised network devices as
described in the preceding paragraph. This property is captured by a term called `connectivity'. A
graph is described as two-connected when at least two nodes must be removed before it becomes
disconnected. Similarly, a graph is called $k$-connected when at least $k$ nodes are required to
achieve the disconnectedness. Higher connectivity provides more independent routes, giving
defenders greater flexibility to steer traffic around a compromised device/region.

Deliberately adding redundant links can increase connectivity. This, in turn, directly
strengthens the network's resilience against attack. Paths are considered edge-disjoint when they
share no common communication line, and node-disjoint when they share no intermediate device, and
both forms carry unique value in terms of security. When two node-disjoint paths carry the
same message, an adversary would need to compromise a device on each path simultaneously to
corrupt every copy. Therefore, gaining a foothold on (access to) a single device is no longer
sufficient to defeat the system. A major point to note here is that the graph in an \ac{EI} system is not
fixed. It is very dynamic in nature. Links go up and down, devices join and leave, and,
most importantly for this discussion, edge weights change every sampling period as the detectors
update their scores. The graph is therefore time-varying, and the routing layer operates on a fresh
snapshot at each sampling instant. The message routing algorithms below are described on a single
snapshot, with the understanding that the computation repeats as the snapshot changes.

\subsection{Edge-Level Intrusion Detection in Trusted Execution Environments}
One essential condition for rerouting to be trustworthy is that the operator must receive a
reliable per-edge estimate of how corrupted each edge is. This estimate can be supplied by intrusion
detectors placed carefully within the network itself. A hybrid set of detectors, including physics-based and data-driven, is well suited to this task. The first one draws on the physics of the power
system, while the second one draws on learned models of normal behavior. Used together, they can
cover threats that neither would catch alone. More information on these detectors is provided below
for clarity on their complementary nature.

Physics-informed detectors exploit the fact that power-system quantities obey well-established physical laws and known dynamics. Measurements crossing a communication edge can
be checked against a physical (estimated) model of the system. A general example is residual-based
checks comparing received values against those predicted by \ac{SE} or the governing equations of the
local subsystem. When a measurement stream is manipulated, it tends to violate these physical
constraints. These violations can result in an increase in the residual magnitude \cite{ref159}. Because this
logic does not depend on data or experience-based learning, a physics-informed
detector remains effective even against attacks that have never been observed before. A more
specific example is the residual test used in weighted least-squares state estimation. The
operator maintains an estimate of the internal state of the system, such as voltage magnitudes and
angles. From that estimate, it predicts what each measurement should (generally) read. The
difference between the predicted and the received reading is the residual magnitude. Under
normal conditions residuals are small and follow a known statistical distribution. When a
measurement is altered during transmission under adversarial manipulation, the resulting residual
exceeds its expected statistical range. This deviation allows the test to identify the communication
edge associated with the manipulated signal. As stated above, this method remains effective even
for previously unseen attacks, as detection is based only on the physical model of the system rather
than on historical records of prior intrusions.

\begin{figure}[H]
\centering
\includegraphics[width=0.7\linewidth]{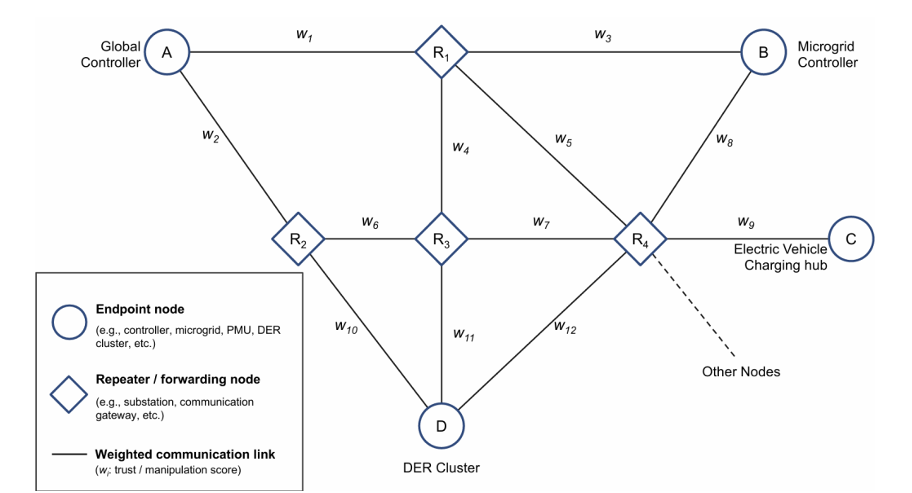}
\caption{An intrusion detector placed within a repeater. Host repeaters must have enough processing power to allow reasonably fast decision-making.}
\label{fig:tee}
\end{figure}

Despite the robustness of physics-informed detectors, machine learning-based intrusion
detector models are also important as they address a very distinct problem. They characterize
and identify patterns within the statistical structure of normal traffic and measurements so that any
outliers not conforming to established patterns can be identified. This remains valid even when no
explicit physical relationship captures those attacks \cite{ref160}. A representative example is a recurrent
autoencoder. Autoencoders structured with long short-term memory units learn to reconstruct
standard time-series data, naturally resulting in a high reconstruction error when the incoming
signals deviate from established baseline patterns. Hence, both physics-informed and data-driven
intrusion detectors have unique characteristics that essentially complement each other. The former
provides guaranteed coverage, and the latter contributes by being sensitive to fine-grained
anomalies, that the former may not catch.

An aspect not often discussed in power grid cybersecurity literature is the safety of the
intrusion detectors or monitoring entities that screen for cyberattacks. It is imperative to understand
that the monitoring/intrusion-detecting infrastructure must also be considered part of any system's
attack surface. If an adversary compromises the \ac{AI} detector (for example), malicious activity may
be suppressed, misclassified, or removed from the reported results. One possible
protection mechanism (against sophisticated attacks directly targeting intrusion detectors) is to
execute security-sensitive parts of the detector within a \ac{TEE}. TEE-enabling technologies such as Intel SGX (which creates enclaves) and Arm TrustZone isolate
selected code and data from other software running on the host helping enable isolated
environments that have higher security. This may isolate even from a compromised operating
system in some threat models \cite{ref161}, which protects the detector's model parameters and scoring
logic from tampering. Even malware such as rootkits sitting outside the enclave are typically
unable to read or modify what happens inside it. However, the enclave's own interface
does remain exposed to the outside world. An attacker who cannot break the isolation directly may
still attempt to overwhelm the enclave with requests. Attackers may try to extract information
through timing or power measurements. Attackers can also try to exploit bugs directly in the
enclave's own code. Pairing the enclave with higher and more powerful monitoring frameworks at
the network and device level often closes vulnerabilities like these that it alone cannot address.

A suitable deployment location (for detectors) often depends on the network architecture
and the visibility needed, as illustrated in Fig.~\ref{fig:tee}. A viable option can be the enclave region on
an intermediate node that already handles bidirectional traffic. This option is feasible when the
node has sufficient processing capacity and can reliably host the enclave. When that isn't the case,
the detector can instead be installed at a gateway, a substation aggregation node, or the destination
device itself.

The main role of each detector in graph-based defense is to generate scores for the network
edge it observes. That score is then used to update the total edge weight. This score typically takes
values on a continuous scale, where larger values signal stronger evidence of manipulation.
However, it can also take a discrete or binary form when the routing mechanism only needs a
limited number of security states. In every case, a clean edge keeps a low weight, and the weight
climbs as the detector gathers more evidence that the edge may be compromised. Detector outputs
also need to be converted into comparable edge weights for effective understandability. Since
different detection methods produce scores on different scales, detector outputs would need to be
normalized first. The normalized value often reflects how strong or reliable the evidence (of
manipulation) is, providing a higher degree of insight as compared to just the raw number a detector
reports. A weak (manipulation) signal that repeats consistently can carry more weight than a single
sharp spike that never recurs. When several detectors watch the same edge (communication link),
their scores are combined through a fusion rule. An example of one such rule can be weighted
averaging. Weighted averaging typically favors the detector that has proven more reliable over a
particular operation cycle of the energy internet. An appropriate fusion method should ensure that
stronger evidence must always increase the weight, never reduce.

The edge weight should also change when the evidence (of manipulation) is no
longer observed. Without such an update mechanism, an edge that was briefly associated with an
anomaly could remain penalized long after normal operation has resumed. A decay function can
gradually reduce the weight when no new suspicious observations are reported. Fresh
measurements obtained through the exploration mechanism can then determine whether the edge
has recovered or whether the elevated weight should be maintained. In this way, an edge can return
to normal routing status after sustained clean observations, while continued anomalous evidence
keeps its cost high. These updates produce a weighted graph in which each edge represents the
most recent security estimate available from the deployed detectors. The weights are refreshed at
each sampling interval and supplied to the routing layer as part of its path-selection process. The
graph should therefore be interpreted as a continuously updated estimate of network risk rather than
as a complete representation of the network's true security state.

\subsection{Attack-Resilient Routing via Epsilon-Greedy Graph Search}
Once the weights are formulated resulting in a weighted graph, the routing objective would
be to move information from sources to destinations along paths that accumulate as
little manipulation weight as possible. Minimizing total weight corresponds to high trustworthiness
as weights are indicative of malicious manipulation (as described above). Thus, a routing method
that consistently prefers low-weight edges will naturally avoid compromised devices and lean on
the parts of the network that the detectors judge to be clean. A purely greedy policy always takes
the lowest-weight route it currently knows, but this does have a limitation. The limitation is that it
never revisits paths it believes to be poor/compromised. Suppose an attacker abandons an edge, or
an earlier suspicion by the detector turns out to be mistaken. A strictly greedy router will never
discover that the edge has become safe again, because it has stopped sending traffic along it.
However, the threat landscape in an \ac{EI} system is not static. This is because compromises
continually appear and disappear as attackers move and defenders take corrective steps. So, a router
that cannot notice recovery will keep on wasting healthy (information-carrying) capacity
indefinitely.

The epsilon-greedy strategy can resolve this tension between using what is known and
discovering what has changed. Most of the time the router exploits its current best knowledge. It
chooses the lowest-weight path with high probability. This probability is equal to one minus a small
valued hyperparameter called epsilon ($\varepsilon$). With the remaining $\varepsilon$ magnitude, it instead selects an
alternative path for exploration. Exploration occasionally favors a higher-weight route instead of
the current minimum-weight path \cite{ref162}. Routing a small fraction of traffic along less-preferred
paths provides fresh observations for updating edge trust estimates, revealing when a suspect edge
has quietly recovered or when a trusted one has started to slip. Getting this right depends heavily
on the value of $\varepsilon$. A small $\varepsilon$ value keeps almost all traffic on trusted paths and probes alternatives
only cautiously. Conversely, pushing $\varepsilon$ values higher speeds up adaptation to change. However,
this does come at the cost of sending more traffic down paths whose reliability is still uncertain.

Path computations can be performed by classical graph algorithms working on the
weighted graph \cite{ref163}. Three of such classical graph algorithms are described here. Prim's
algorithm and Kruskal's algorithm both work towards building a \ac{MST}.
MSTs are useful when information must reach many endpoints, as in a broadcast of set points or a
synchronization message. This is typically seen in the case of microgrids \cite{ref157,ref162}. Dijkstra's
algorithm finds the lowest-weight path between a single source and a single destination, making it
the standard choice for point-to-point control and measurement traffic. In every case the edge
weights represent manipulation scores, so minimizing accumulated weight is equivalent to
avoiding manipulation. Note that the three algorithms described below are as
per their classical versions. All the algorithms run deterministically on whatever weighted graph
they are given and provide an optimal result based on that snapshot (of the graph). Exploration here
can be considered as a separate outer layer. This does not change the algorithms themselves, which
continue to operate as per their classical procedures. In any given period, the routing policy (not
the algorithms) may enforce exploration by temporarily biasing the decision toward a currently
distrusted element before the algorithm runs. For the traffic, this means occasionally using a route
that the following algorithms may not determine as the best. Hence, the graph algorithm
itself remains unchanged, and only its input or its selected output may be adjusted to incorporate
an exploratory behavior.

\begin{itemize}
\item \textbf{Prim's Algorithm}: This algorithm builds an \ac{MST} by growing it outward from a single
starting node. Often this node is conveniently a trusted (global) supervisory controller. At
each step, it looks at the edges connecting the current tree to nodes not yet attached and
adds whichever one carries the smallest weight. The smallest weight typically corresponds
to the link the detectors currently trust the most. Repeating this process until every node
has joined leaves a tree built from the safest connections available at each stage. The
running total simply adds up manipulation scores as the tree grows, so the finished tree
ends up carrying the least accumulated evidence of tampering. So, Prim's algorithm can
naturally help information avoid compromised repeaters whenever a cleaner detour exists.

\item \textbf{Kruskal's Algorithm}: This algorithm also constructs an \ac{MST} using an edge-driven
approach rather than growing from a single starting node. It first sorts every edge from
most trusted to least trusted. Then it moves through that sorted list and adds each edge only
if it joins two parts of the network that are not yet connected. It skips any edge that would
close a loop. Once the separate fragments have merged into one tree, every node is
connected through the lowest-weight edges that avoided creating cycles. This makes
Kruskal's algorithm convenient when the security scores are already available as a ranked
list, since the first step is simply to sort by trust.

\item \textbf{Dijkstra's Algorithm}: This algorithm finds the single lowest-weight path from a source
to a destination. This would be the ideal routing decision for most point-to-point traffic. It
works outward from the source, keeping a running cost to every node and repeatedly
settling the unsettled node of the smallest running cost. Each time a node is settled, the
running costs of its neighbors are lowered whenever reaching them through that node is
cheaper, and the path itself is recovered afterward by tracing these choices back from the
destination to the source. Because the running cost accumulates edge weights, and those
weights are manipulation scores, the path it returns is the one with the least total evidence
of tampering. This helps route around (avoid) compromised repeaters whenever a cleaner
detour exists.
\end{itemize}

It is important to be noted that the three algorithms serve different communication patterns,
and the routing layer chooses among them by the nature of the traffic. When a single source must
reach a single destination, Dijkstra's algorithm is the right tool. This is because it returns one least-weight path, and this covers most control commands and measurement reports. When one source
must reach many destinations at once, a spanning tree is more economical than computing a
separate path to each destination. A broadcast of set points, a global synchronization signal, and/or
a firmware update all fit this pattern. Prim's and Kruskal's algorithms both compute a minimum-weight tree, so the choice between them makes no difference to the routes that get selected. The
difference shows up in implementation. Prim's runs faster on denser graphs, and Kruskal's
underlying data structure handles sparser graphs more efficiently.

\subsubsection{Multi-Path Delivery for Critical Traffic}
Protective trip commands and critical alarm notifications often require strong delivery
guarantees in grid communication systems. In these cases, loss or corruption of a message can
trigger immediate and potentially unsafe actions in physical equipment. For this kind of critical
traffic, the routing layer can compute multiple node-disjoint paths and send an identical copy of the
message down each one. The receiver accepts the first valid copy it receives
and subsequently examines any additional copies for evidence of tampering.

The effectiveness of this scheme depends on the paths being truly independent. Node-disjoint routes share no intermediate device. This is why an adversary trying to suppress or alter
the message must compromise at least one device on every path simultaneously (to be successful).
Adding a third disjoint path forces the attacker to control a third device, adding a fourth forces a
fourth, and the required foothold (representing the number of devices that need to be controlled for
manipulation) keeps growing one device at a time as more paths are added. The disjoint paths
themselves come from running Dijkstra's algorithm repeatedly. A shortest path is computed, and
its intermediate devices are removed before recomputing. This can be done until no further disjoint
path remains. Note that network connectivity often sets the limit on how many independent paths
are available in the first place. Hence, operators who need this level of protection for critical traffic
must build in redundant links to enable rerouting.

\subsubsection{Scheduling Exploration}
Exploration requirements will typically vary over the lifetime of a deployed system. Hence,
a fixed exploration rate is not equally effective across all operational stages in the energy internet.
During early deployment (for example) or following a suspected attack, higher exploration rates
are necessary to improve visibility of edge conditions and identify reliable communication paths.
Once more stability is achieved, exploration can be reduced. This will enable more traffic to be
routed on more trusted paths. A decaying schedule captures this behavior by starting the exploration
rate higher and letting it fall gradually. The rates can then be raised again whenever the detectors
report a change. Exploration can also be performed in a targeted manner rather than random. For
example, the router can be configured to prefer edges whose scores are old or uncertain rather than
probing any arbitrary alternatives. This may help spend the exploration budget more judiciously
and only where it can yield the most information.

\subsubsection{Behavior When No Clean Path Exists}
It is sometimes possible that every path between a source and a destination passes through
a suspect (possibly compromised) edge. In this case the routing layer must behave sensibly rather
than failing silently. Two responses are appropriate here. The first is that the router selects the least-weight path available (which is the least-bad option). Then it continues to deliver traffic (through
the path) while flagging the elevated risk to the operator. The second applies when the accumulated
weight of even the best path exceeds a safety threshold, at which point the router raises an alarm
and must invoke a fallback such as switching to an out-of-band channel or entering a safe local
mode of operation \cite{ref156}. The graph view makes this decision practical because the total weight of
the best path is a direct measure of the residual risk in delivering the message.

\subsubsection{Resilience Under Active Manipulation}
Resilient information flow under an active manipulation requires complete coordination
between the intrusion detectors and routing mechanisms. At each sampling period, the detectors
translate the network's current state into edge weights based on which information can be routed
and the level of manipulation. The routing algorithms then use those weights to select paths that
avoid manipulated devices. Epsilon-greedy policies as described above can keep weight estimates
fresh (meaning representative of current conditions). Keeping weights `fresh' helps the system
track a possibly shifting or evolving threat landscape \cite{ref160}. Once an adversary
compromises a device, the affected edges are assigned more weight. Then the next routing pass
reroutes traffic away from them. Following remediation of the compromised device, exploration
eventually starts detecting the improved conditions. This can allow the affected edges to gradually
return to normal service.

The system does not need to depend on a compromised device to keep functioning. It just
needs another path to exist somewhere in the network for information rerouting in the presence of
one or more adversarial entities. An attacker actively trying to corrupt parts of communication lines
(or edges) still cannot stop traffic. This is because no single device carries the responsibility of
traffic routing alone. Thus, resilience is brought to the entire graph (as a whole), rather than to any
single link within it. This rerouting also pairs naturally with \ac{MTD} principles \cite{ref164}. Moving target
defenses periodically shuffle observable features of the network including addresses and active
paths, so an attacker's earlier reconnaissance is rendered useless. Trust-aware routing also provides
an independent reason to shift paths. This is typically triggered by evidence of compromise rather
than a fixed schedule. Thus, an attacker facing both defenses simultaneously has to deal with a
network whose layout not only keeps changing on its own timer but whose traffic also keeps getting
(actively) rerouted to abandon any part that an attacker actually manages to compromise \cite{ref164}.

\subsubsection{An Example to Aid Understanding}
\begin{figure}[h]
\centering
\includegraphics[width=0.75\linewidth]{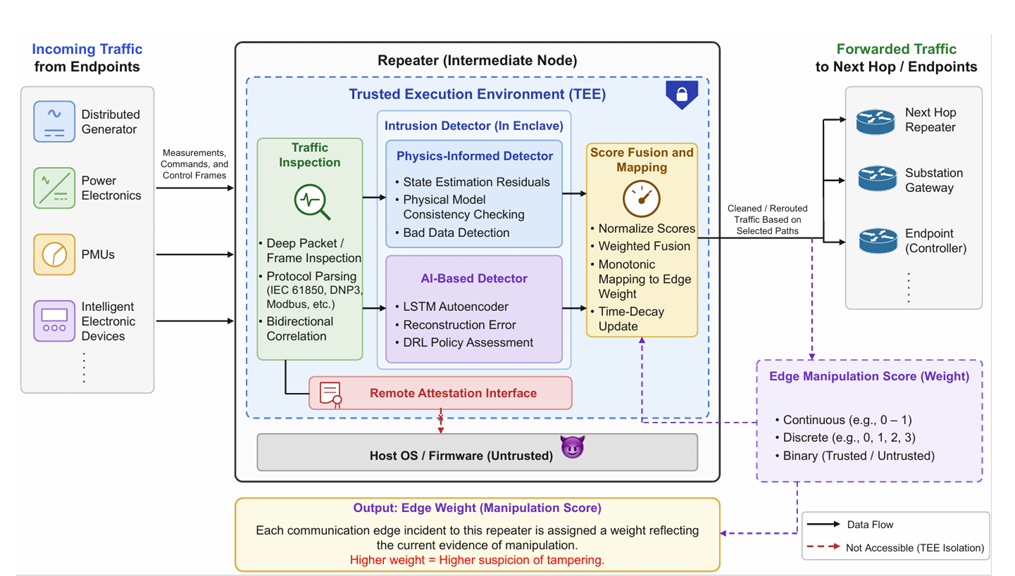}
\caption{Epsilon-greedy rerouting around a compromised network region.}
\label{fig:epsgreedy}
\end{figure}

This subsection presents an illustrative walkthrough of attack detection, routing adaptation,
and recovery in the proposed communication framework. As shown in Fig.~\ref{fig:epsgreedy}, consider a
controller that transmits routine set points to an inverter over a three-repeater chain. In normal
operation, all edges in this path carry low manipulation scores. Hence, Dijkstra's algorithm selects
this route as the minimum-weight path. The inverter receives and applies the set points without
delay.

The scenario changes once an adversary breaches the second repeater and installs a rootkit.
The rootkit tries to alter the set-point values passing through the repeater \cite{ref158}. As these altered
values no longer match what the inverter's physical model predicts, the physics-informed detector
inside the repeater's enclave picks up a growing residual. In parallel, the learning-based
detector identifies a deviation from the established traffic profile. Within a single sampling period,
both detectors raise a flag by increasing the risk scores of edges associated with the compromised
repeater. At the next routing update, the increased total weight of the compromised chain causes
Dijkstra's algorithm to exclude it from selection. The algorithm instead computes an alternative
path to the inverter through another set of repeaters. In this state, no device on the compromised
chain is relied upon for message delivery or validation. Set points are subsequently routed along
this alternative path, and the inverter continues to receive correct commands through this route.

The operator later removes the rootkit and restores the repeater \cite{ref162}. The edges through
it are clean again, but their weights are still elevated from the earlier attack. With low probability,
the router occasionally sends traffic back down the recovered path. The detectors report clean
readings. Subsequently, the decay rule brings the weights back down. Finally, the recovered path
returns to service within a few sampling cycles, all without an operator adjusting a single route by
hand.

\subsection{Relationship to Existing Communication-Resilience Approaches}
Graph-based trust-aware routing is not a unitary defense. Operational networks also use
several other (complementary) communication-resilience techniques. One of them is \ac{SDN}, which
separates the `control of forwarding' from `the devices that forward'. They allow a controller to
install paths across the network from a central vantage point. \ac{SDN} as an approach has been studied
extensively for smart-grid communication \cite{ref165}. Because the trust-aware routes computed here are
just another set of forwarding rules, an \ac{SDN} controller offers a ready-made mechanism for pushing
them out across the network. Time-sensitive networking and related deterministic-networking
standards guarantee bounded latency for critical streams \cite{ref166}, but they address timing rather than
trust. Hence, \ac{SDN} complements (rather than replaces) a routing layer that chooses paths by their
security state. Path-control mechanisms such as segment routing and multiprotocol label switching
can likewise carry the specific, possibly disjoint, paths that the algorithms above select.

The approach also fits within broader security frameworks. Industrial security standards
such as IEC 62443 organize defenses into zones and conduits and call for defense in depth. Trust-aware routing is one control that helps meet that goal at the network layer. Zero-trust networking,
with its principle of never trusting and always verifying, is also closely related, because where a
zero-trust architecture continuously re-evaluates the trust of endpoints and sessions, the routing
described here continuously re-evaluates the trust of paths, using detector evidence as the metric.
Seen this way, graph-based trust-aware routing is not an alternative to these
established approaches but a security-aware routing metric and decision layer that can be realized
on top of them.

\subsection{Centralized and Distributed Route Computation}
Routing decisions can either be computed in one place or distributed across many nodes.
In a centralized arrangement, a single operator collects the edge weights from all enclaves,
assembles the complete weighted graph, and runs the algorithms for the whole network. Centralized
route computation often produces the most consistent result because every route is computed
against the same global picture. However, it also needs a very capable and well-protected control
center. This is because, in centralization, every routing decision depends on one single node and its
connections to the rest of the network. If the central node itself is compromised (single-point failure
\cite{ref159}), it can fail the system by compromising entire routing arrangements.

In distributed route computation, each repeater computes routes for its own traffic. Route
computation relies on locally observed edge weights together with weight information exchanged
among neighboring repeaters. This eliminates the need for any node to maintain a complete view
of the communication graph. Hence, losing part of the network under this setup does not bring the
entire system down. Instead, the remaining repeaters keep routing around the gap while relying
only on the local information available to them. Consistency suffers in exchange, as two repeaters
may end up holding slightly different views of the same region. Neighbor-exchange protocols
reduce this gap by letting repeaters periodically compare and reconcile their estimates. However,
even then, small discrepancies can persist between updates. Whether to prioritize consistency or
independence in this exchange is a decision that shapes how the energy internet's information
exchange network responds under stress, and it should be made based on how quickly the
deployment expects conditions to change.

It is worth specifying where these computations actually run in an \ac{EI} system, since the
natural placement follows the existing control hierarchy. Fast, local rerouting can be performed at
edge gateways, feeder controllers, and microgrid controllers \cite{ref156}. In these locations, routing
decisions can be made close to the traffic source. This placement enables routing decisions to be
completed within a sampling period without relying on wide-area communication. Network-wide
optimization operates on a longer timescale and is therefore better suited to the utility control center
or a \ac{SDN} controller. These entities maintain a network-wide view and can install consistent
forwarding rules, although they operate on a longer timescale. In practice, both approaches are
deployed together. Local controllers respond immediately to newly detected compromises, while
the control center periodically recomputes globally consistent routes and reconciles local routing
decisions. The resulting separation between local and network-wide functions mirrors existing
grid protection practices and preserves the independence of time-critical routing from wide-area
communication.

Exploration must also be coordinated in the distributed case. If every repeater explores
independently at the same time, an excessive fraction of traffic may be redirected onto uncertain
paths simultaneously. One safeguard is to limit the total amount of exploration allowed across the
network at any given moment, keeping only a small fraction of traffic on unproven paths. Repeaters
can also stagger their exploration over time instead of probing simultaneously. Spreading probing
out this way keeps the network gathering fresh information without disrupting normal traffic in the
process.

It is important to note that, whichever arrangement is chosen, the routing layer sits below
the physical control functions and serves them, since it does not decide what commands to send,
only how those commands travel. This separation is deliberate, because it lets the routing layer
change paths freely in response to the threat landscape without altering any control logic that
governs the power system. The control functions thus have a robust communication service that
always stays available and trustworthy. Controllers can remain unaware of rerouting changes that
preserve communicated continuity, keeping security exposure limited and only on a need-to-know
basis.
\section{Conclusions and Recommendations}
With the increasing penetration of renewable energy in power grids and the deepening
digitization brought by \ac{ICT}-coupled, interoperable components, the \ac{EI} creates new opportunities
for power system operation, control, and market participation, but also new challenges stemming
from the growing dependencies between the physical, cyber, and business layers of the system.
This \ac{TF} report provides a comprehensive analysis of the challenges and opportunities arising at the
confluence of the Energy and Internet domains. Consistent with the scope outlined in Section 1, it
emphasizes the need to understand and model the multi-network, multi-level, and multi-layer
interactions of \ac{EI} systems to ensure their reliability, security, and resilience, given the scale and
complexity of modern, internet-dependent energy infrastructures. To this end, the report examines
the \ac{EI} threat landscape and assurance mechanisms, modeling and control of \ac{TnD} systems through
\ac{EI} architectures, decision-aware and storage-integrated resilience, multi-dimensional resilience
assessment, price forecasting in the \ac{EI} era, the adversarial risks and trustworthy deployment of \ac{AI},
and attack-resilient information routing, collectively highlighting the different layers of cyber-physical interdependence that shape \ac{EI} operation. These considerations are rendered even more
pressing by the rapid transformation of the demand side itself, the proliferation of controllable and
flexible loads, ranging from smart buildings and electrified transportation to industrial demand
response, and, most notably, the unprecedented load growth driven by \ac{AI} datacenters, whose large,
fast-ramping, and geographically concentrated consumption profiles introduce new stability,
planning, and security pressures. Paradoxically, these same loads embody the \ac{EI} vision, as they are
software-defined, network-connected, and inherently controllable, making them simultaneously a
stressor on the grid, a potential attack surface, and one of the most powerful flexibility resources
available for resilient operation.

Based on these findings, the \ac{TF} puts forward the following recommendations. First,
security and resilience should be treated as coupled, cross-layer design objectives rather than
isolated attributes, since cyber intrusions can masquerade as physical faults and single-dimension
assessments fail to capture the degradation produced by simultaneous cross-dimensional failures.
Second, the community should invest in realistic validation infrastructure, including real-time co-simulation, controller-in-the-loop and \ac{HIL} testbeds, and digital twins, to
evaluate detection, mitigation, and recovery mechanisms under conditions that faithfully reproduce
\ac{EI} cyber-physical coupling. Third, the flexibility inherent to \ac{DERs}, energy storage, controllable
loads, and adaptive control schemes, such as event-triggered islanding, should be systematically
leveraged as a resilience resource against both cyber and physical disturbances. This includes
formalizing the participation of large flexible consumers, particularly \ac{AI} datacenters, in grid
services, while ensuring that the control and communication channels enabling such participation
are secured against manipulation, since compromised load flexibility at datacenter scale could itself
become a destabilizing attack vector. Fourth, \ac{AI} and \ac{ML} applications in \ac{EI} systems must be
deployed with explicit consideration of adversarial risks, favoring trustworthy, explainable, and
physics-informed designs. Finally, given the \ac{EI}'s dependence on communication and network
infrastructures, a coordinated approach to operation and stability control is essential, supported by
secure communication protocols, resilient information routing, and the extension of existing
standards and regulatory frameworks to the decentralized, transactive nature of the \ac{EI}, including
interconnection and flexibility requirements for hyperscale loads. The \ac{TF} views this report as a
foundation for continued collaboration among academia, industry, system operators, and
regulators, and recommends that future work refine quantitative cross-layer resilience metrics,
mature validation methodologies, and translate the lessons of the Internet vertical into standardized
practice for secure and resilient EI deployments.

\bibliographystyle{IEEEtran}
\bibliography{refs_manual}
\end{document}